\documentclass[iop]{emulateapj}
 
\usepackage{wasysym}
\usepackage{natbib}
\usepackage{longtable}
\usepackage{graphicx}
\usepackage{fancyref}
\usepackage{gensymb}
\usepackage{adjustbox}
\usepackage{booktabs}
\usepackage{scrextend}

\def\mgii{\ifmmode {\rm Mg{\sc ii}} \else Mg~{\sc ii}\fi}
\newcommand{\ra}[1]{\renewcommand{\arraystretch}{#1}}

\shortauthors{Fischer, Smith, Kraemer, Schmitt, et al.}
\shorttitle{AGN Feedback across the Electromagnetic Spectrum}

\begin{document}

\title{A Dissection of Spatially Resolved AGN Feedback across the Electromagnetic Spectrum}

\author{Travis Fischer\altaffilmark{1,2,3}, Krista Lynne Smith\altaffilmark{4}, Steve Kraemer\altaffilmark{2}, Henrique Schmitt\altaffilmark{5}, D. Michael Crenshaw \altaffilmark{6}, Michael Koss\altaffilmark{7}, Richard Mushotzky\altaffilmark{8}, Kirsten Larson\altaffilmark{9}, Vivian U \altaffilmark{10}, Jane Rigby\altaffilmark{3}}

\altaffiltext{1}{U.S. Naval Observatory, 3450 Massachusetts Avenue NW, Washington, DC 20392, USA}
\altaffiltext{2}{Institute for Astrophysics and Computational Sciences, Department of Physics, The Catholic University of America, Washington, DC 20064, USA}
\altaffiltext{3}{Observational Cosmology Lab, Goddard Space Flight Center, Code 665, Greenbelt, MD 20771, USA}
\altaffiltext{4}{KIPAC at SLAC, Stanford University, Menlo Park CA 94025u}
\altaffiltext{5}{Naval Research Laboratory, Washington, DC 20375, USA}
\altaffiltext{6}{Department of Physics and Astronomy, Georgia State University, Astronomy Offices, 25 Park Place, Suite 600, Atlanta, GA 30303, USA}
\altaffiltext{7}{Eureka Scientific, 2452 Delmer Street Suite 100, Oakland, CA 94602-3017, USA}
\altaffiltext{8}{Department of Astronomy and Joint Space-Science Institute, University of Maryland, College Park, MD 20742, USA}
\altaffiltext{9}{Infrared Processing and Analysis Center, MC 314-6, Caltech, 1200 E. California Blvd., Pasadena, CA 91125}
\altaffiltext{10}{Department of Physics and Astronomy, 4129 Frederick Reines Hall, University of California, Irvine, CA 92697}

\begin{abstract}

We present optical SuperNova Integral Field Spectrograph (SNIFS) integral field spectroscopy, {\it Hubble Space Telescope} optical imaging, {\it Chandra} X-ray imaging, and {\it Very Large Array} radio interferometry of the merging galaxy 2MASX~J04234080+0408017, which hosts a Seyfert 2 active galactic nucleus (AGN) at z = 0.046. Our observations reveal that radiatively driven, ionized gas outflows are successful to distances $>$ 10\,kpc due to the low mass of the host system, encompassing the entirety of the observed optical emission. We also find that at large radii, where observed velocities cannot be reproduced by radiative driving models, high velocity kinematics are likely due to mechanical driving from AGN winds impacting high density host material. This impacting deposits sufficient energy to shock the host material, producing thermal X-ray emission and cosmic rays, which in turn promote the formation of in situ radio structure in a pseudo-jet morphology along the high density lanes. 

\end{abstract}


\section{Introduction}
\label{sec:intro}

There is clear and demonstrable evidence for a relationship between supermassive black holes and the physical properties of their host galaxies. 
Tight correlations exist between the black hole mass and the stellar bulge mass, bulge luminosity, and stellar velocity dispersion 
\citep{geb00,fer00,gul09}. The probable mechanism for such correlations is feedback from an active galactic nucleus (AGN) 
\citep{Beg04} on the host galaxy. The existence of an AGN could either quench star formation in the host galaxy by expelling or heating the necessary molecular gas 
(e.g., \citealt{DiM05,Hop06}), or trigger star formation by instigating outflows which generate turbulence and shocks, 
causing the collapse of giant molecular clouds \citep{Kla04}. These effects may be one of the main drivers of evolution in massive galaxies: 
AGN feedback is necessary in cosmological simulations to suppress the formation of massive blue galaxies \citep{Hop08,Cat09}. 
Further, AGN host galaxies tend to lie in the otherwise sparsely-populated region between the blue, star forming sequence 
and red ellipticals on color-mass plots \citep{Nan07,Sch09,Kos11}, and appear to have suppressed star formation rates 
for a given total stellar mass compared to normal galaxies \citep{Sal07,Shi15,Ell16}. However, the observed correlations between the 
black hole and the host galaxy could also be a result of co-evolution, in which galaxy mergers instigate 
both the fueling of the black hole and star formation.

Many surveys have made progress in understanding the effects of feedback on galaxy evolution writ large (e.g. \citet{Mul18,Rif18,Neu19}). An important complement to this 
approach is studying the detailed interaction of an AGN and its host galaxy in individual objects. Recently, we analyzed kinematics in 
the luminous, nearby (z=0.017) Seyfert 2 galaxy Mrk 573 \citep{Fis17}, where we found the AGN-ionized gas morphology to be consistent 
with an intersection between spiral arms in the host disk and ionizing radiation from the central engine. Dust lanes rotating in the 
host disk, traced by molecular hydrogen emission, were found to connect regions of ionized gas from outside of the biconical AGN 
radiation field, suggesting that kinematics of the outflowing ionized gas in the narrow-line region (NLR) of Mrk 573 are due to in 
situ acceleration of material originating in the host disk via radiative driving by the AGN, until radiative driving is no longer 
able to disrupt the rotational motion of the spiral arms, forming the Extended NLR (ENLR).  

\begin{figure}[!htbp]
\centering

\includegraphics[width=0.48\textwidth]{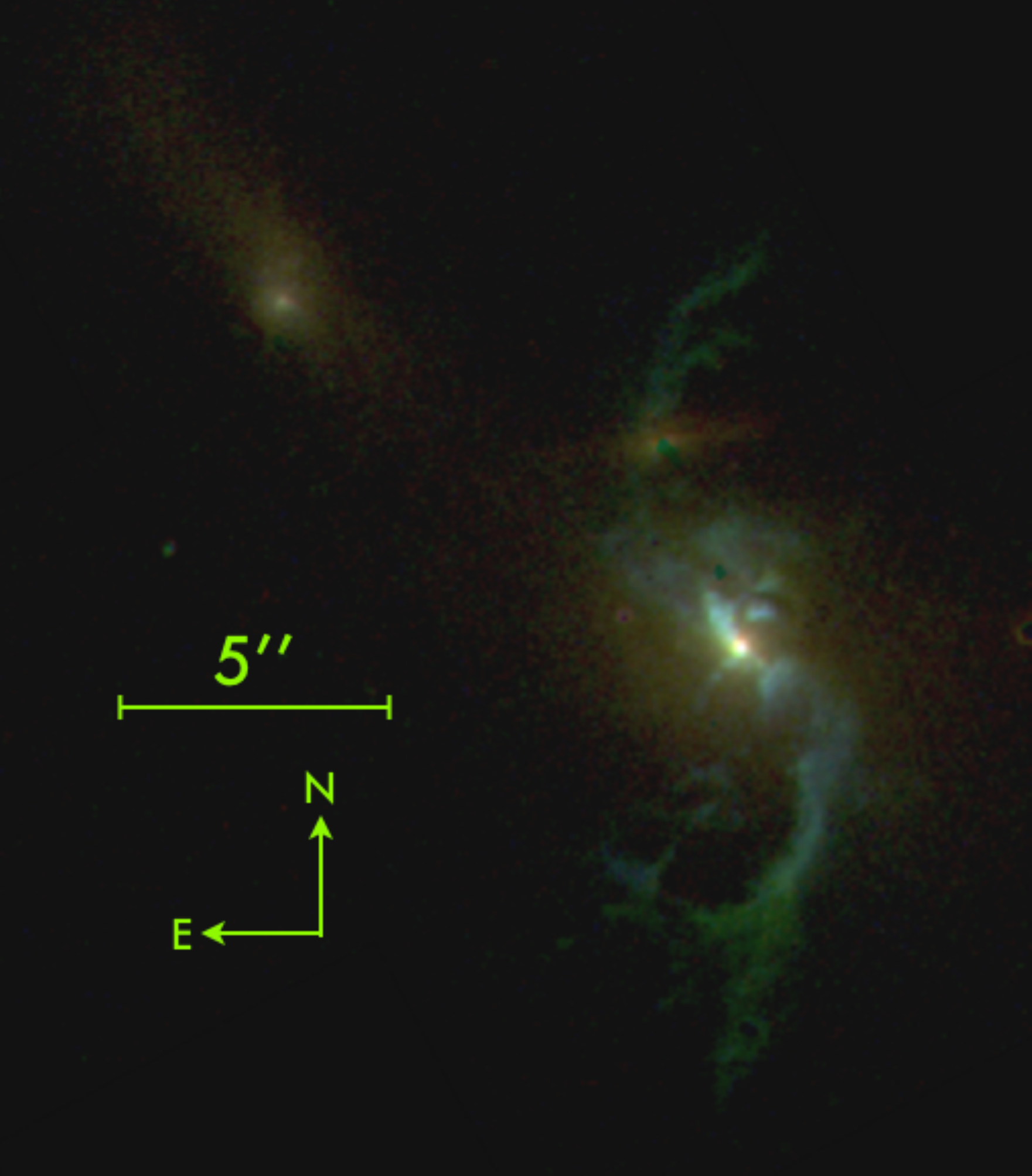}
\caption{\emph{Hubble Space Telescope} imaging of the merging AGN 2MASX~J0423 (right) with companion 
(left). A third, background galaxy at a different redshift is in the northern limb of 2MASX~J0423 at approximately 4.1$''$ from the 
continuum peak. Color rendition composed of \emph{HST}/WFPC2 F814W (red); F675W (green); and F547M (blue) filters. 5$''$ corresponds 
to a spatial scale of $\sim$4.5 kpc.}
\label{fig:fig1}

\end{figure} 

Another individual galaxy potentially capable of providing similar insights is 
2MASX~J04234080+0408017, henceforth 2MASX~J0423, shown in Figure \ref{fig:fig1}. A merging system which houses a Type 2 AGN, 
2MASX~J0423 exhibits bright optical ionized gas extents on scales of several kpcs, and potentially 
provides an excellent opportunity to measure high-resolution ionized gas outflows to much larger distances.

Alongside extensive optical ionized-gas structures, the object has large radio lobes that extend well outside the 
host galaxy, spanning approximately 30 kpc, as first reported by \citet{Beichman1985}. At the time, the host was thought 
to be a spiral galaxy. However, optical spectroscopy by \citet{Hill1988} found that the spiral structure was not produced by 
starlight or scattered nuclear light. \citet{Steffen1996} analyzed the kinematics in more detail and found the radio structure 
to be bent along the optical ionized spiral. Using long-slit spectroscopy of the radio lobes, \citet{Holloway1996} found that 
the radio lobes have optical emission lines with velocity differentials of $\sim1000$~km~s$^{-1}$, and suggest that the radio 
jet expands dramatically into the observed lobes after emerging from the dense interstellar medium. 

After this, the galaxy remained unstudied for many years before having the good fortune to be detected as part of the ultra-hard X-ray 
selected \emph{Swift}-BAT AGN sample, which has been the subject of a large multi-wavelength follow-up effort \citep{Kos17}. A 22~GHz high-resolution 
radio imaging survey of the BAT AGN \citep{Smi16} obtained a much more sensitive radio image of 2MASX~J0423 than those 
previously available. 
A {\it Chandra} survey of merging galaxies \citep{Kos12} included this system, although only a single AGN was detected, with 
complementary optical spectroscopy to also identify dual AGN using SuperNova Integral Field Spectrograph (SNIFS) integral 
field unit (IFU) observations. The new data present an opportunity to learn more about the AGN/host galaxy interaction in this 
unique object. In this paper, we synthesize all of these observations and use them to scrutinize the conclusions of the work 
in the 1980s and 90s and add our own insights. 

In our analysis of 2MASX~J0423, we use a redshift of z = 0.046099 \citep{Spr05}, a Hubble constant of H$_0$ = 71 km s$^{-1}$, 
$\Omega_m$ = 0.3, and $\Omega_{\lambda}$ = 0.7 for a Hubble distance of 201.5 Mpc and a scale of 0.893 kpc arcsec$^{-1}$ \citep{Wri06}.

\section{Observations and Reduction}
\label{sec:obs}

\subsection{\emph{HST} Imaging Observations}
Archival \emph{HST} Wide Field and Planetary Camera 2 (WFPC2)
imaging was retrieved from the Mikulski Archive at the Space
Telescope Science Institute. The observations
were obtained on 1995 January 31 as a part of Program ID 5746
(PI: F. D. Macchetto) with a plate scale of
0.05$''$ pixel$^{-1}$. 2 300\,s exposures were taken using 
the F547M filter and individual 600\,s exposures were taken using F675W 
and F814W filters. Cosmic rays were removed from the latter images using 
the L.A. Cosmic algorithm \citep{Dok01}. F547M and F814W filters cover 
largely featureless continuum, while the F675W bandwidth covers 
[O~I]$\lambda$6300, [N~II]$\lambda\lambda$6548,6584, H$\alpha$, and 
[S~II] $\lambda\lambda$6716,6731 emission lines. 

\subsection{{\it UH 2.2m}/SNIFS Optical IFU Observations}
\begin{figure*}[h]
\centering
\includegraphics[width=0.9\textwidth]{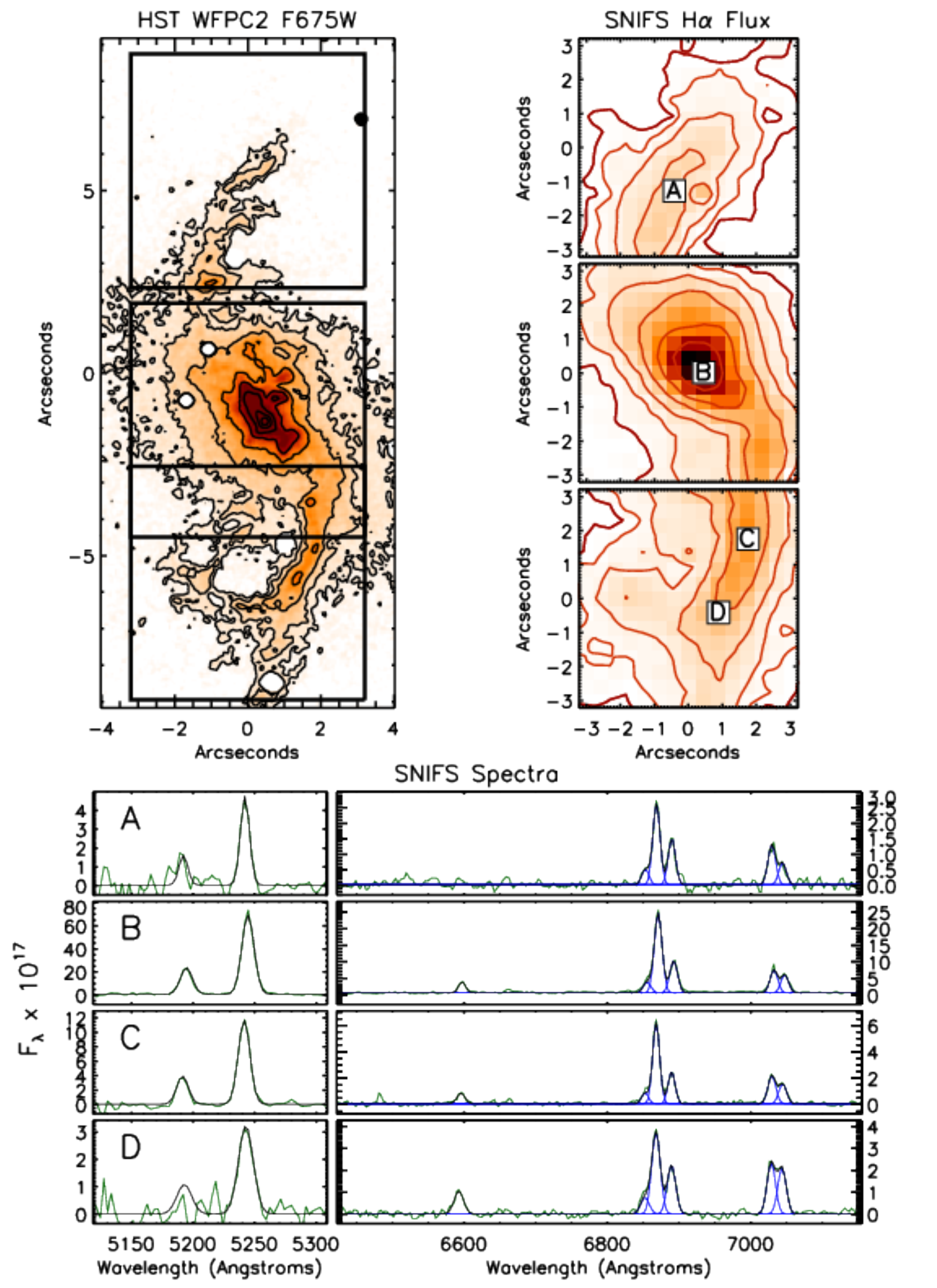}
\caption{Top left: Black squares depict the distribution of $6.4'' \times 6.4''$ 
SNIFS fields of view of 2MASX~J0423 overlaid on an \emph{HST}/WFPC2 
F675W image. The top, center, and bottom fields are unevenly distributed,
with the top field offset from the center field by approximately 0.4$''$ and 
the center and bottom fields overlapping by approximately 2$''$.
White circles in the center SNIFS field box are artifacts. 
Top right: Continuum-subtracted H$\alpha$ flux distributions obtained 
from the top, center, and bottom $6.4'' \times 6.4''$ SNIFS datacubes. Outer, dark 
red contours represent a 3$\sigma$ S/N lower flux limit. Bottom: Representative spectra 
across 2MASX~J0423, overplotted with their best fitting model. Approximate spaxel positions 
are labeled in the H$\alpha$ image. Left boxes exhibit [O~III]~$\lambda\lambda$4959,5007, right boxes 
exhibit [O~I]~$\lambda$6300 (when present), H$\alpha~+$ [N~II]~$\lambda\lambda$6549,6585, 
and [S~II]~$\lambda\lambda$6718,6732. Green represents data, blue represents individual 
Gaussians in blended emission lines, and black represents the combined model. The optical 
continuum flux peak of 2MASX~J0423 is sampled in spectrum B. North is up and east is left 
in all images.}
\label{fig:fig2}

\end{figure*}

Spectroscopic observations of 2MASX~J0423 were obtained using SNIFS \citep{Ald02,Lan04},
mounted at the University of Hawaii 2.2 m telescope (UH88) at Mauna Kea in 
UT 2012 Nov 11 for a total of 7 hrs of integration with an effective seeing FWHM 
for science observations of 0.97$''$ across three fields of view. The top left portion 
of Figure \ref{fig:fig2} depicts the positions of the SNIFS fields overlaid on a 
cropped F675W {\it HST} image. The center and bottom fields overlap by 2$''$, while the top 
field is offset from the center field by approximately 0.4$''$. Individual H$\alpha$ flux 
distributions in the top, center, and bottom SNIFS fields illustrate the correspondence between 
morphologies observed in our SNIFS and {\it HST} observations.

SNIFS employs a fully filled 15$\times$15 lenslet array in the IFU, covering
a 6.4$''$ $\times$  6.4$''$ field of view. The corresponding spatial
resolution is 0.43 arcseconds per spatial pixel (spaxel).
The spectrograph consists of two arms operating simultaneously
to cover the entire optical wavelength range at R $\sim$ 1000:
the blue channel covers 3000-5200 \AA~and the red channel covers
5200-9500 \AA.  The SNIFS reduction pipeline, SNURP, was used for 
wavelength calibration, spectro-spatial flat-fielding, cosmic ray 
removal, and flux calibration \citet{Bac01,Ald06}. A sky image was taken 
after each source image and subtracted from each IFU observation. The 
extraction aperture was 2.4$"$ in diameter. 

\subsection{{\it Chandra} and NuSTAR X-ray Observations}

On 2012 October 20, 2MASX~J0423 was imaged with the ACIS-S detector on \emph{Chandra} for 20\,ks. 
We reprocessed the image using the \emph{Chandra} Interactive 
Analysis of Observations (CIAO) software version 4.11 \citep{Fru06}. After restricting the energy 
to 0.3-7~keV with the task \texttt{dmcopy} to avoid the sharply rising high-energy particle background outside 
this range, we normalize by the exposure map and correct for bad pixels using the task \texttt{fluximage}.

In July 2012, 2MASX-J0423 was observed three times by NuSTAR as part of the BAT legacy survey. 
Each observation is approximately 6\,ks in duration. Although not 
sufficiently high S/N for imaging, the hard energy range of NuSTAR ($\leq$ 78\,keV) allows us to constrain 
the power law component of the spectral fits from \emph{Chandra}.

\begin{figure*}[!htbp]
\centering
\includegraphics[width=0.31\textwidth]{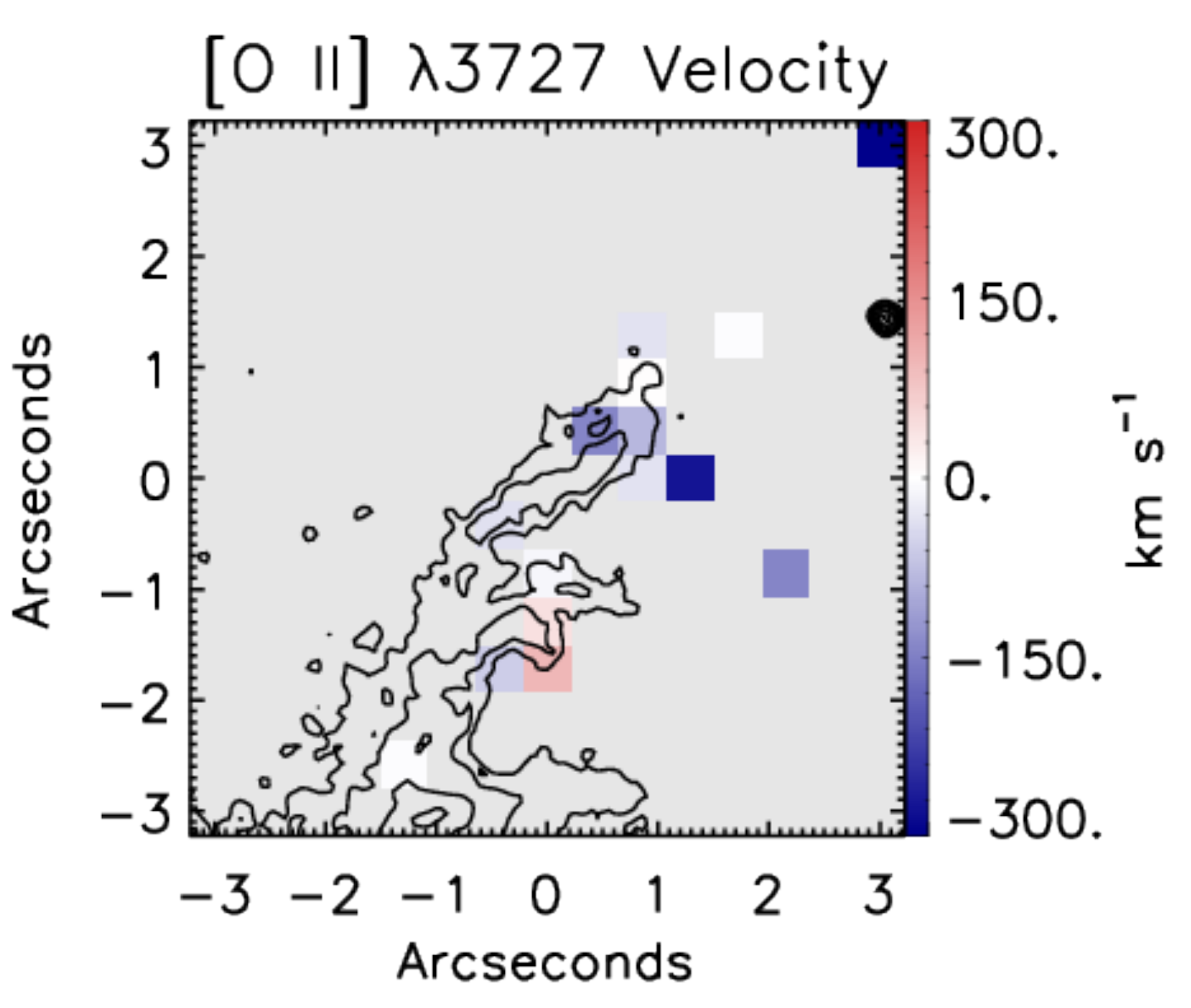}
\includegraphics[width=0.31\textwidth]{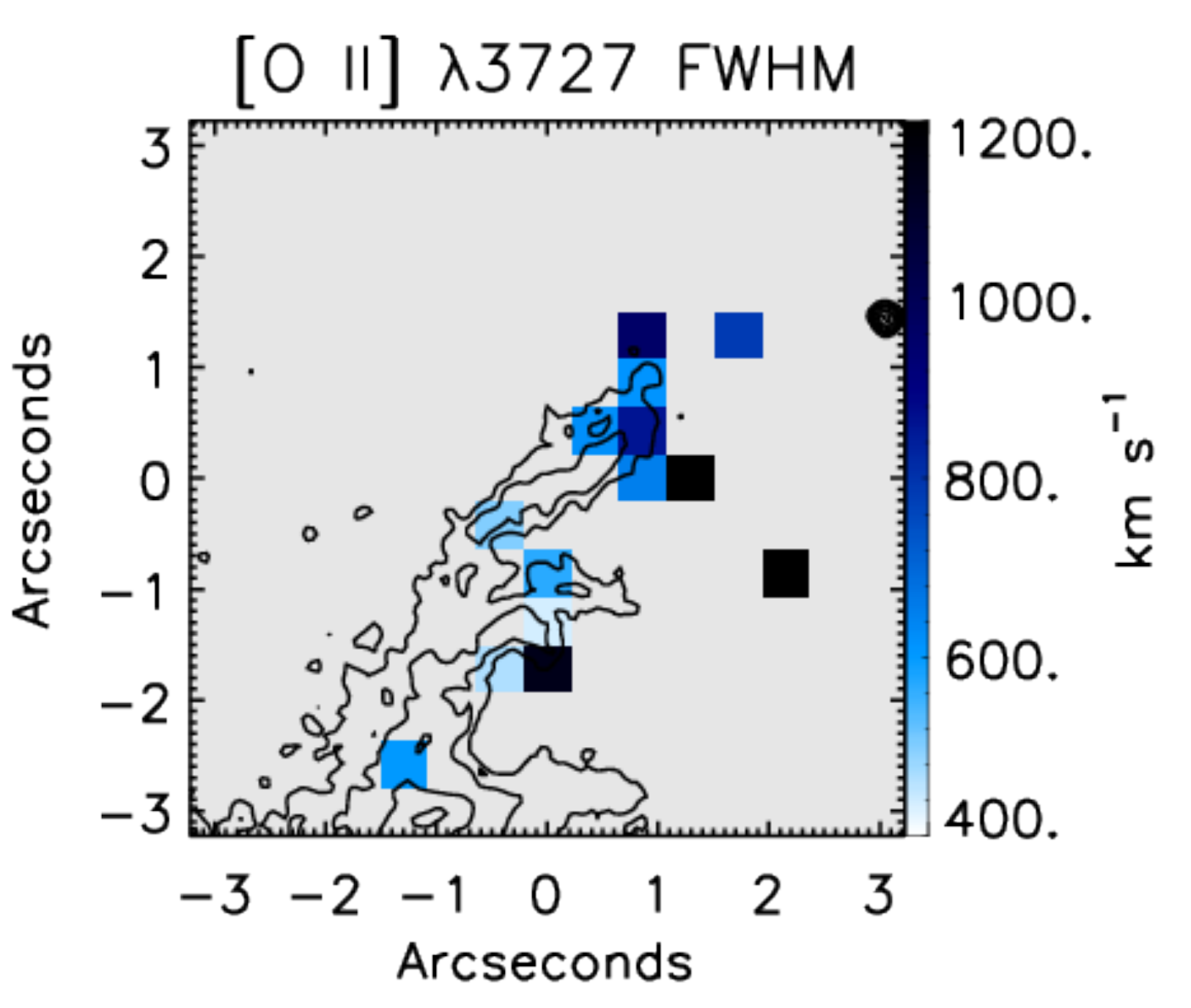}
\includegraphics[width=0.31\textwidth]{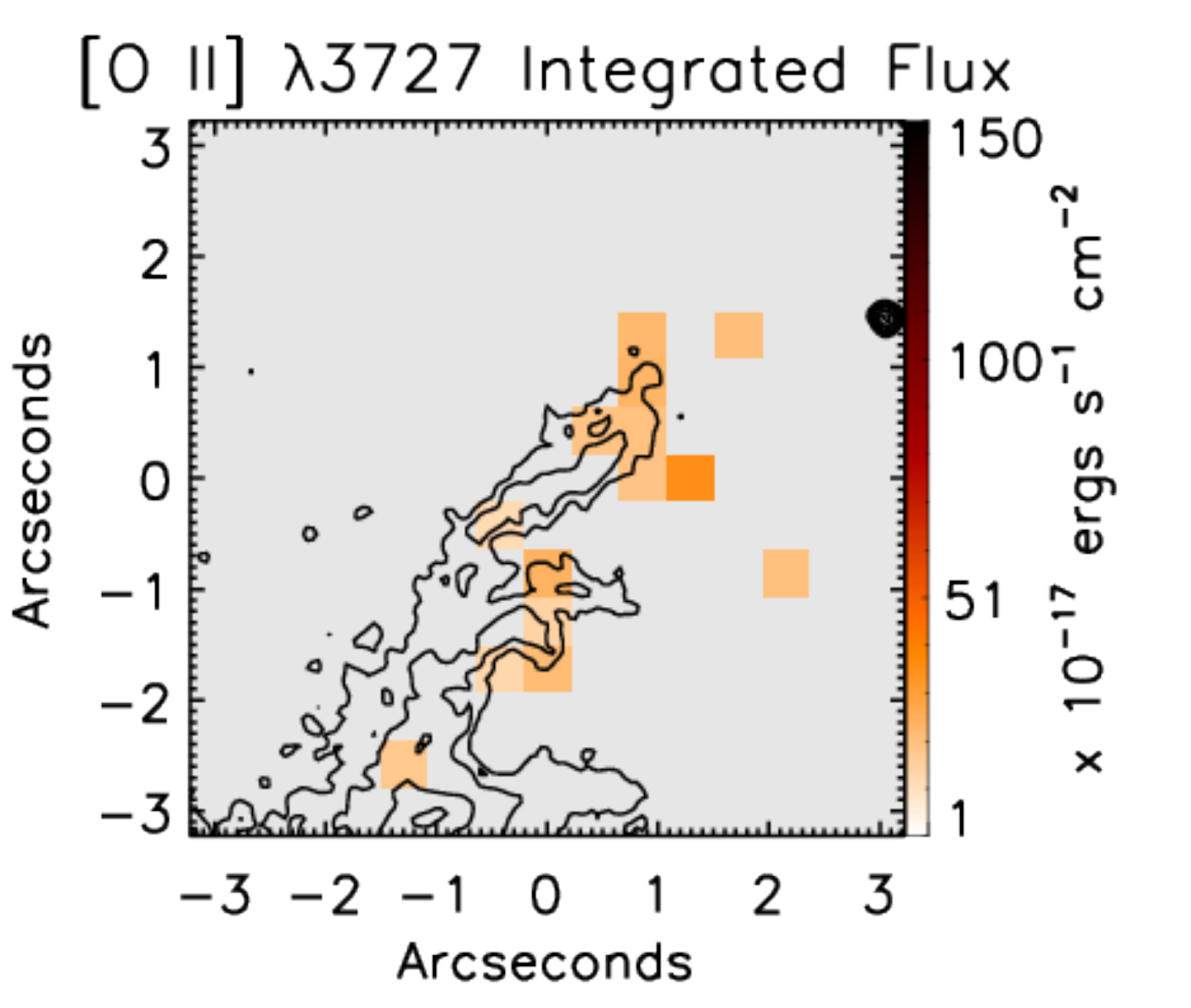}\\
\includegraphics[width=0.31\textwidth]{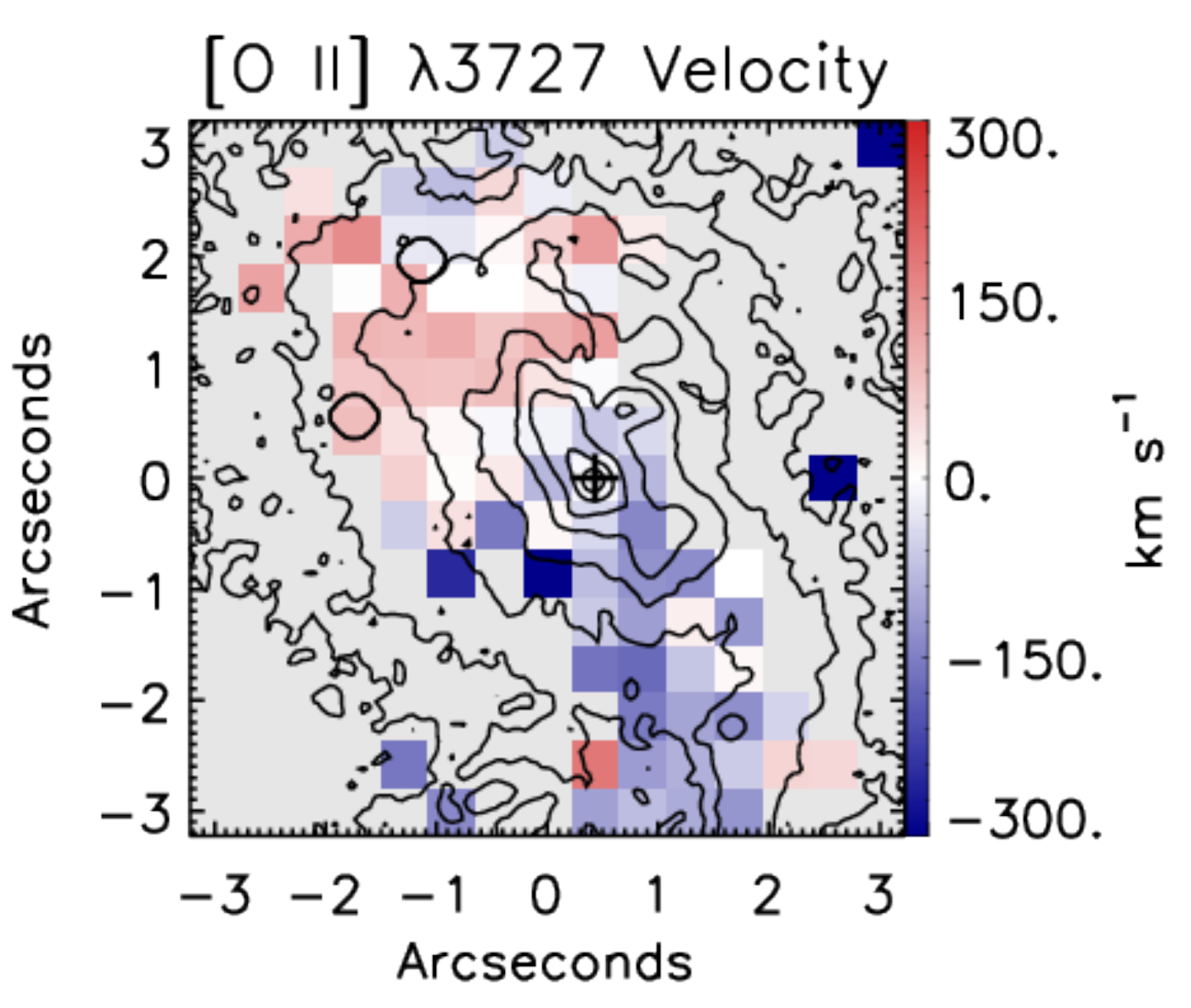}
\includegraphics[width=0.31\textwidth]{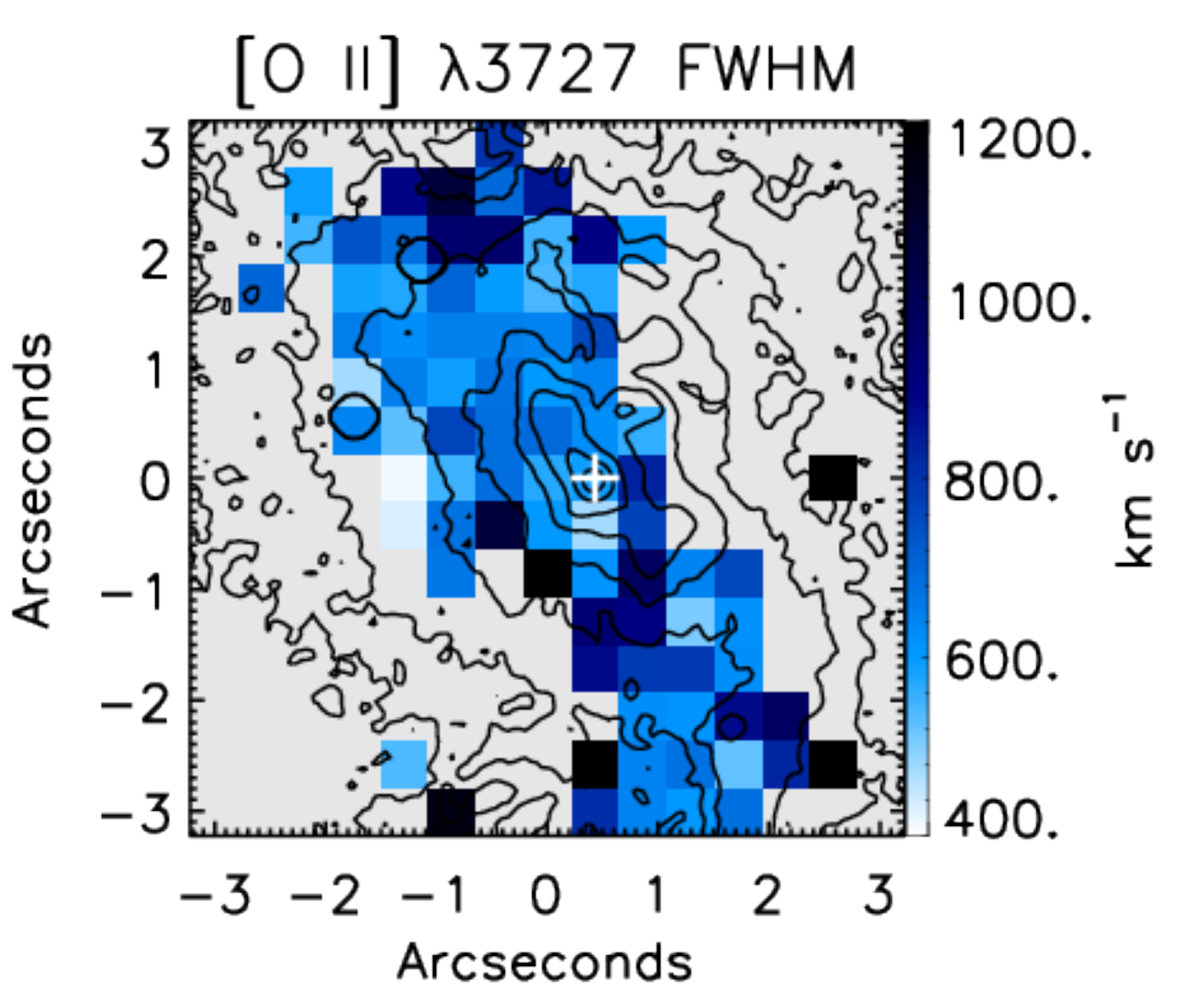}
\includegraphics[width=0.31\textwidth]{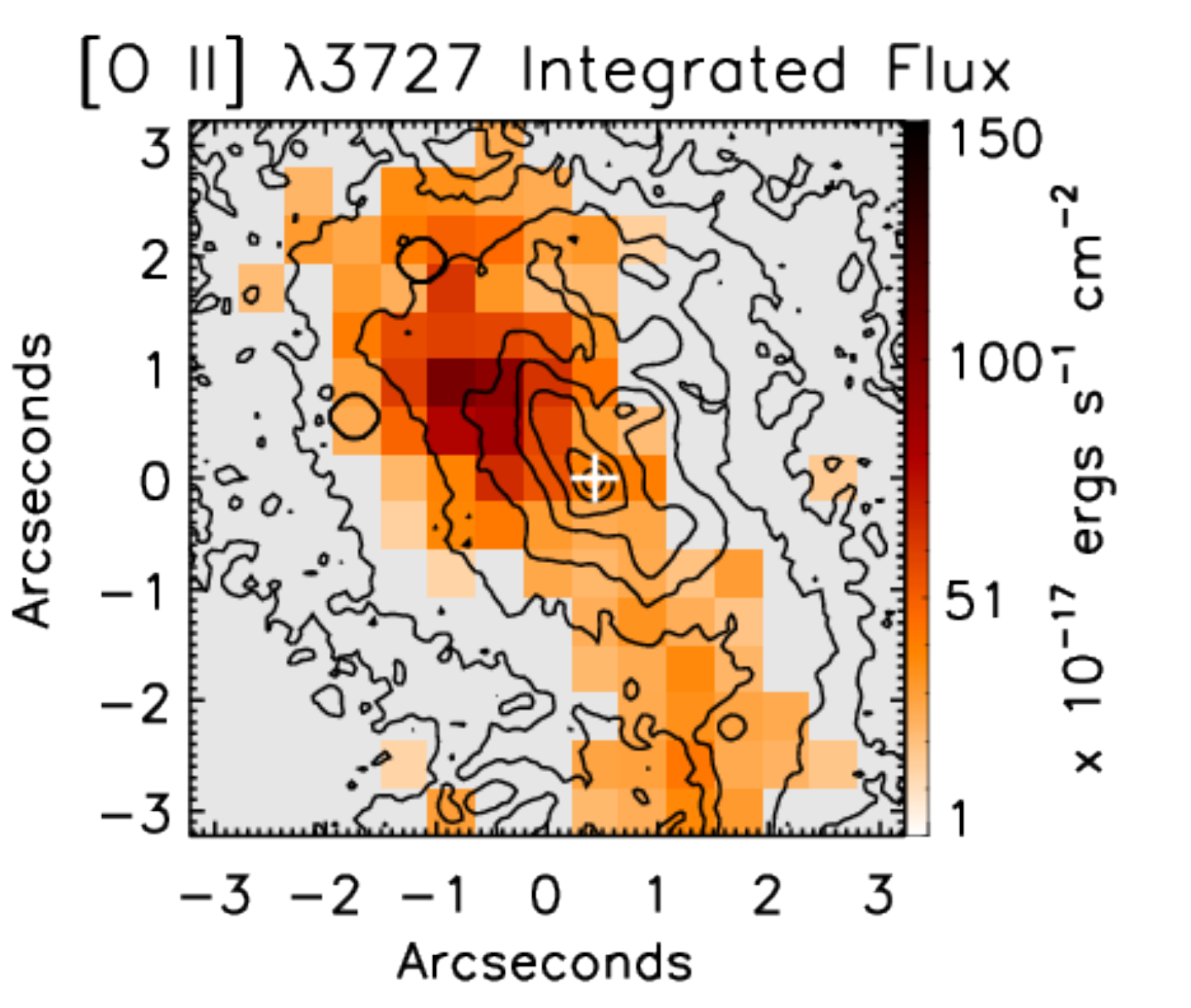}\\
\includegraphics[width=0.31\textwidth]{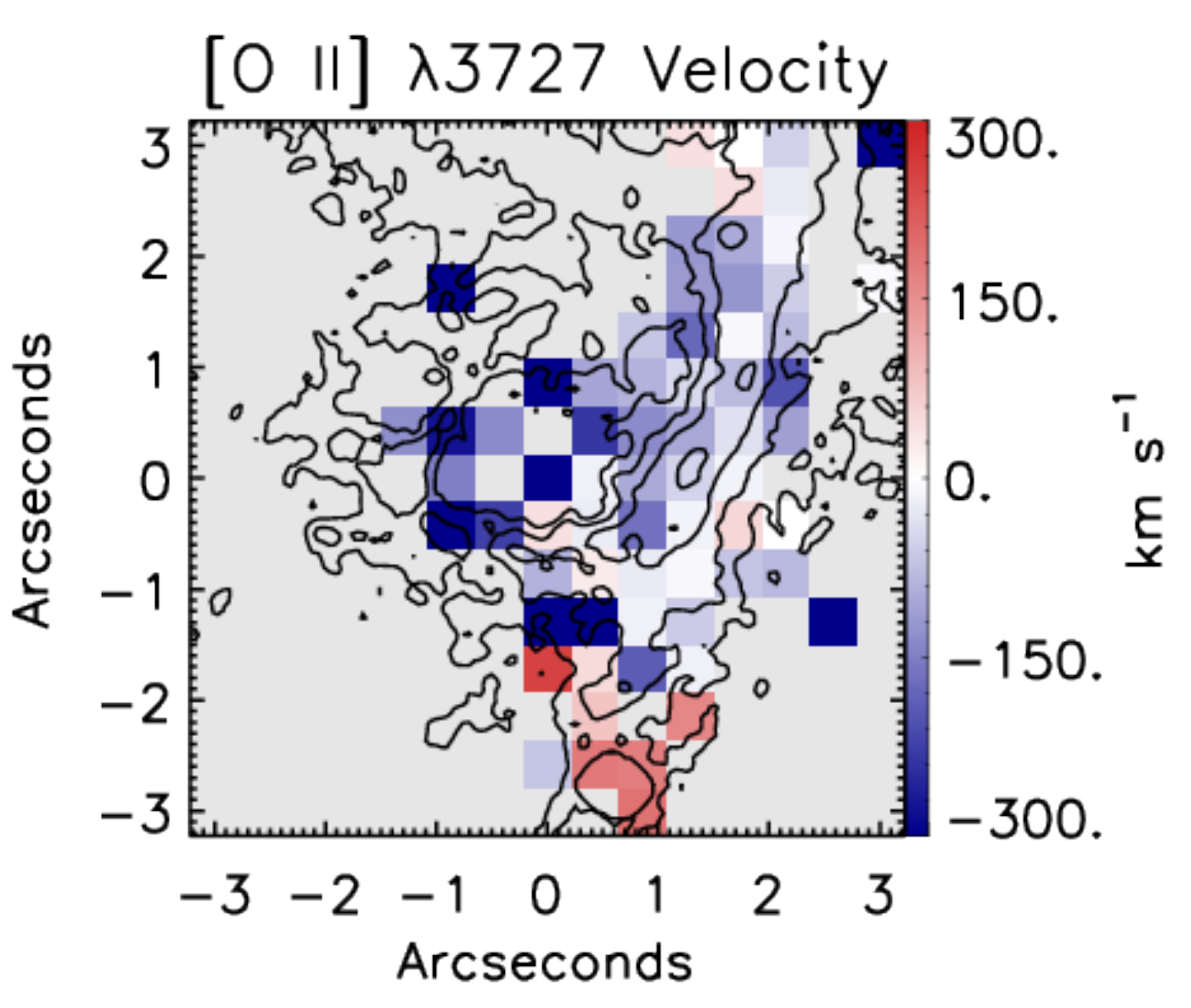}
\includegraphics[width=0.31\textwidth]{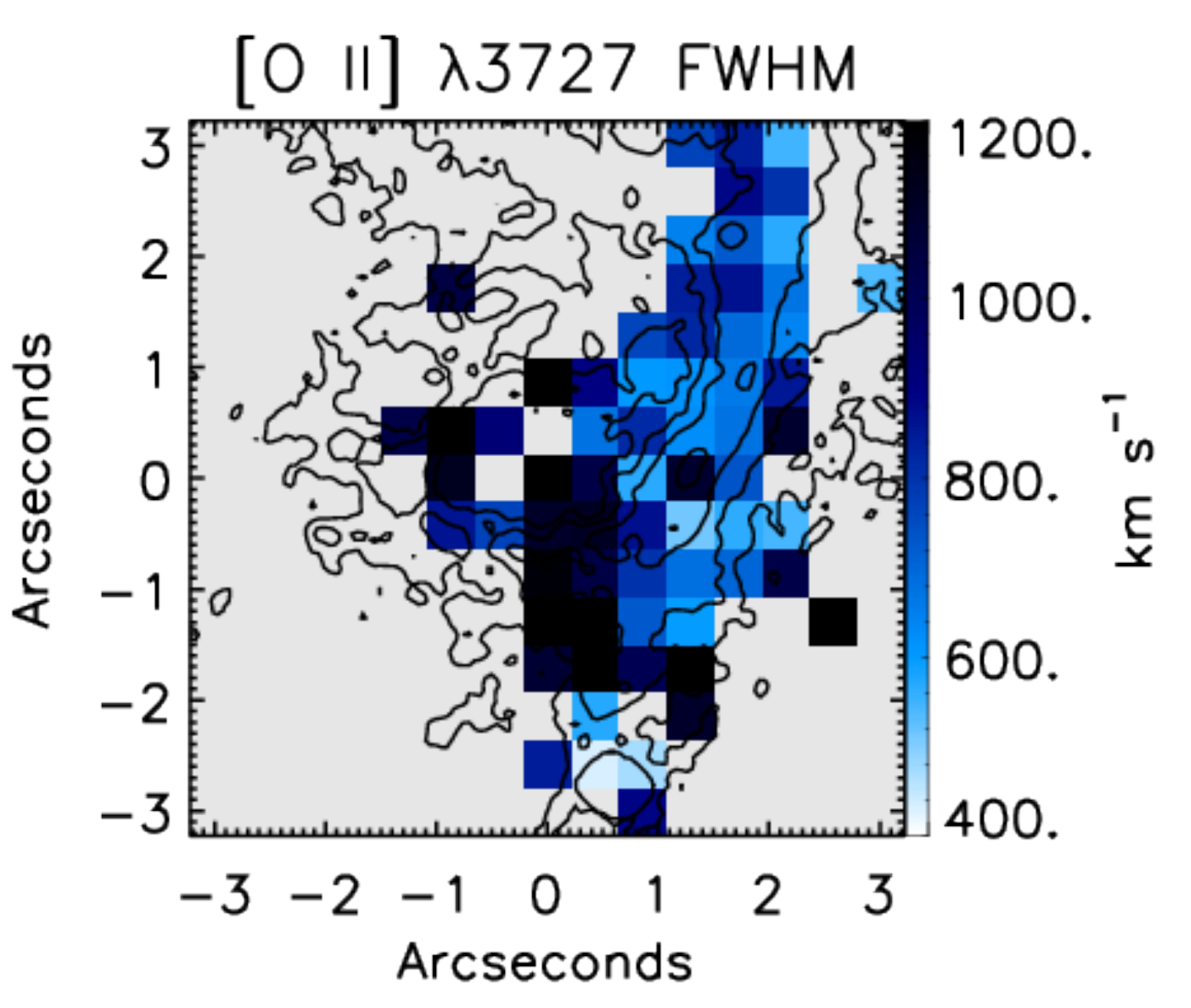}
\includegraphics[width=0.31\textwidth]{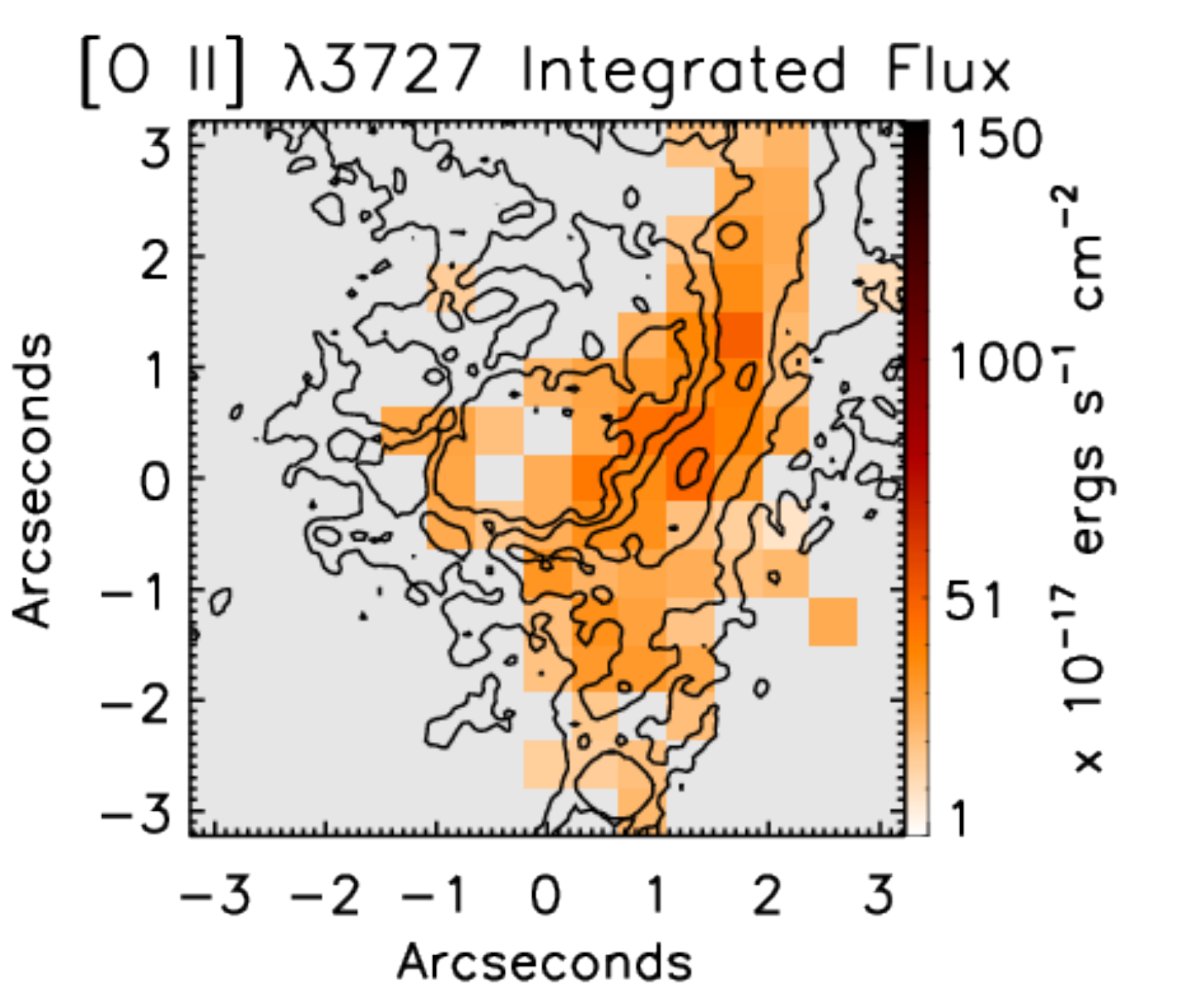}\\

\caption{[O~II] $\lambda$3727 kinematic measurements in 2MASX~J0423 from SNIFS IFU observations. First, 
second, and third columns display emission-line profile centroid velocity, FWHM, and integrated 
flux maps, respectively. First, second, and third rows display measurements for the top, 
center, and bottom fields of view, respectively. Black contours represent \emph{HST}/WFPC2 
F675W imaging. The optical continuum flux peak is depicted by a cross. One 0.43$" \times$ 0.43$"$ spaxel 
samples approximately 380\,pc $\times$ 380\,pc.}
\label{fig:oiimaps}

\end{figure*}

\begin{figure*}[!htbp]
\centering
\includegraphics[width=0.31\textwidth]{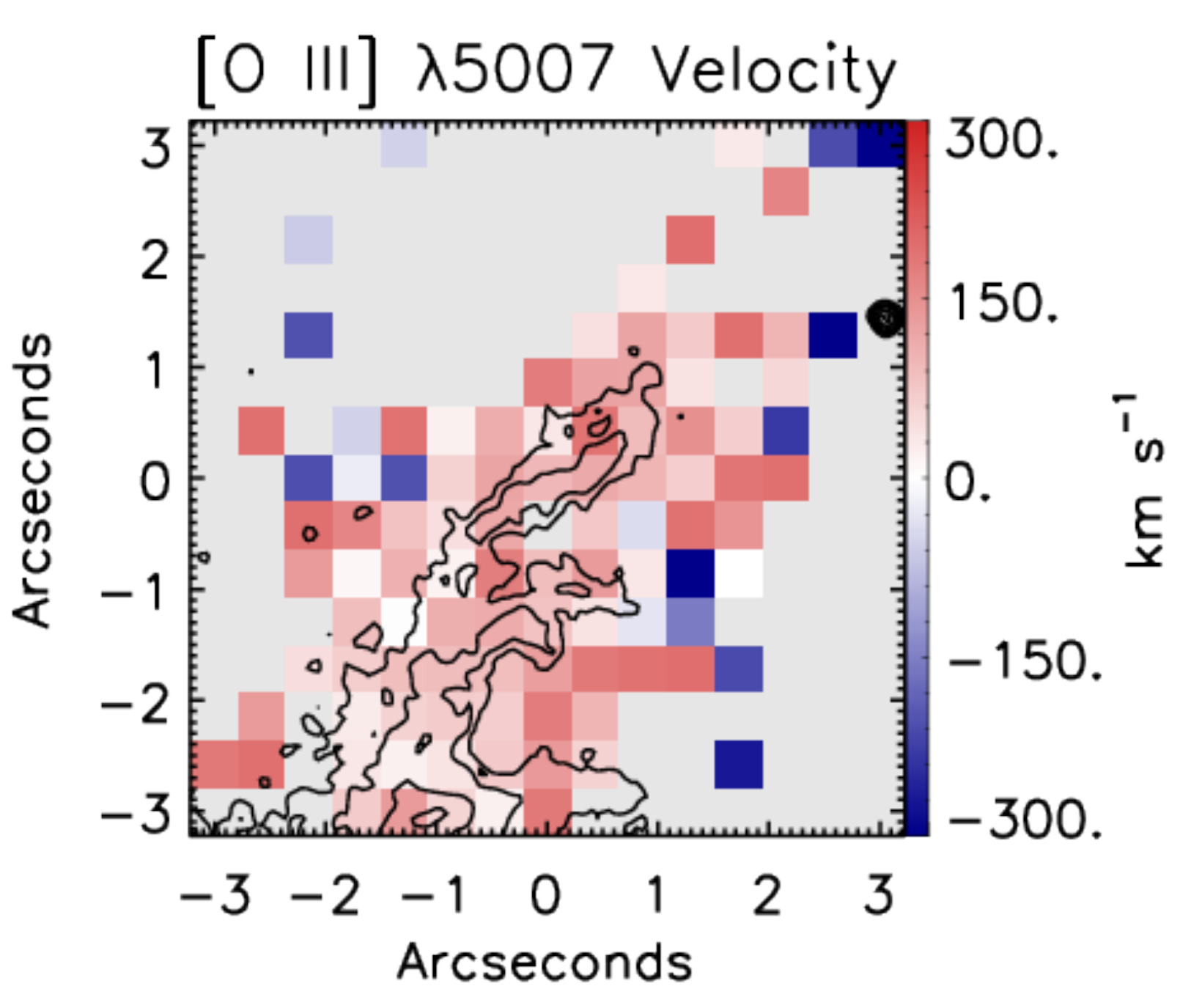}
\includegraphics[width=0.31\textwidth]{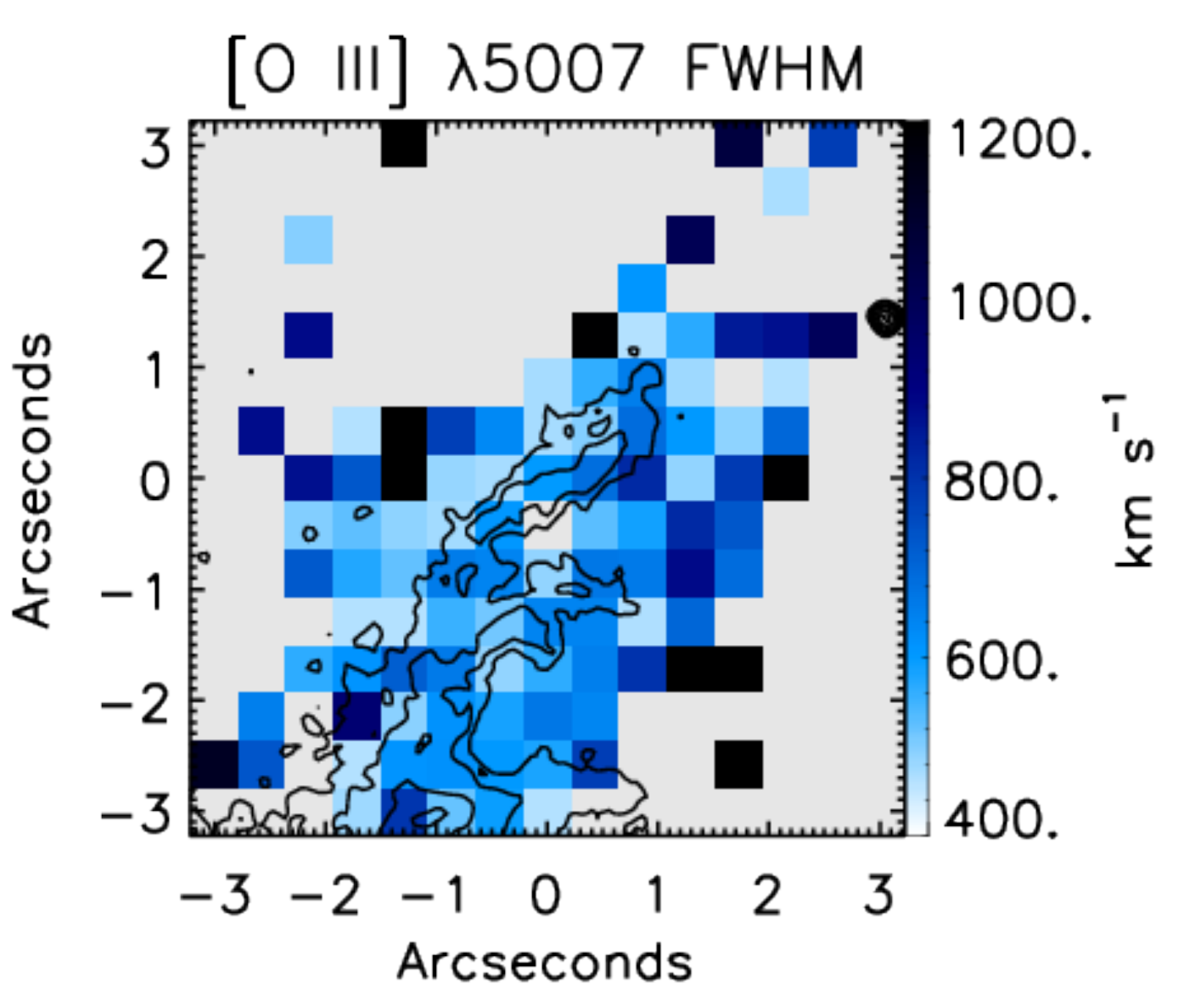}
\includegraphics[width=0.31\textwidth]{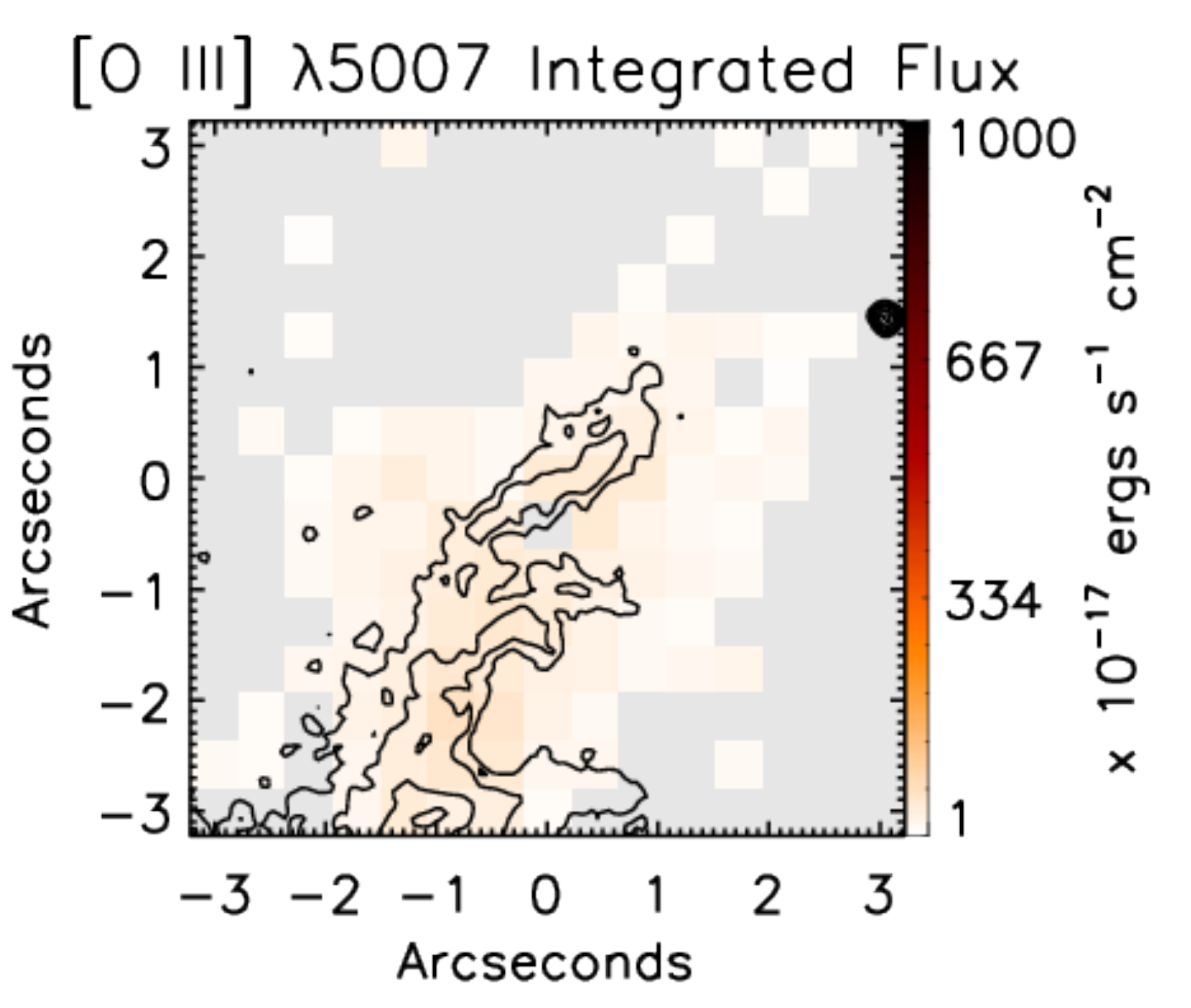}\\
\includegraphics[width=0.31\textwidth]{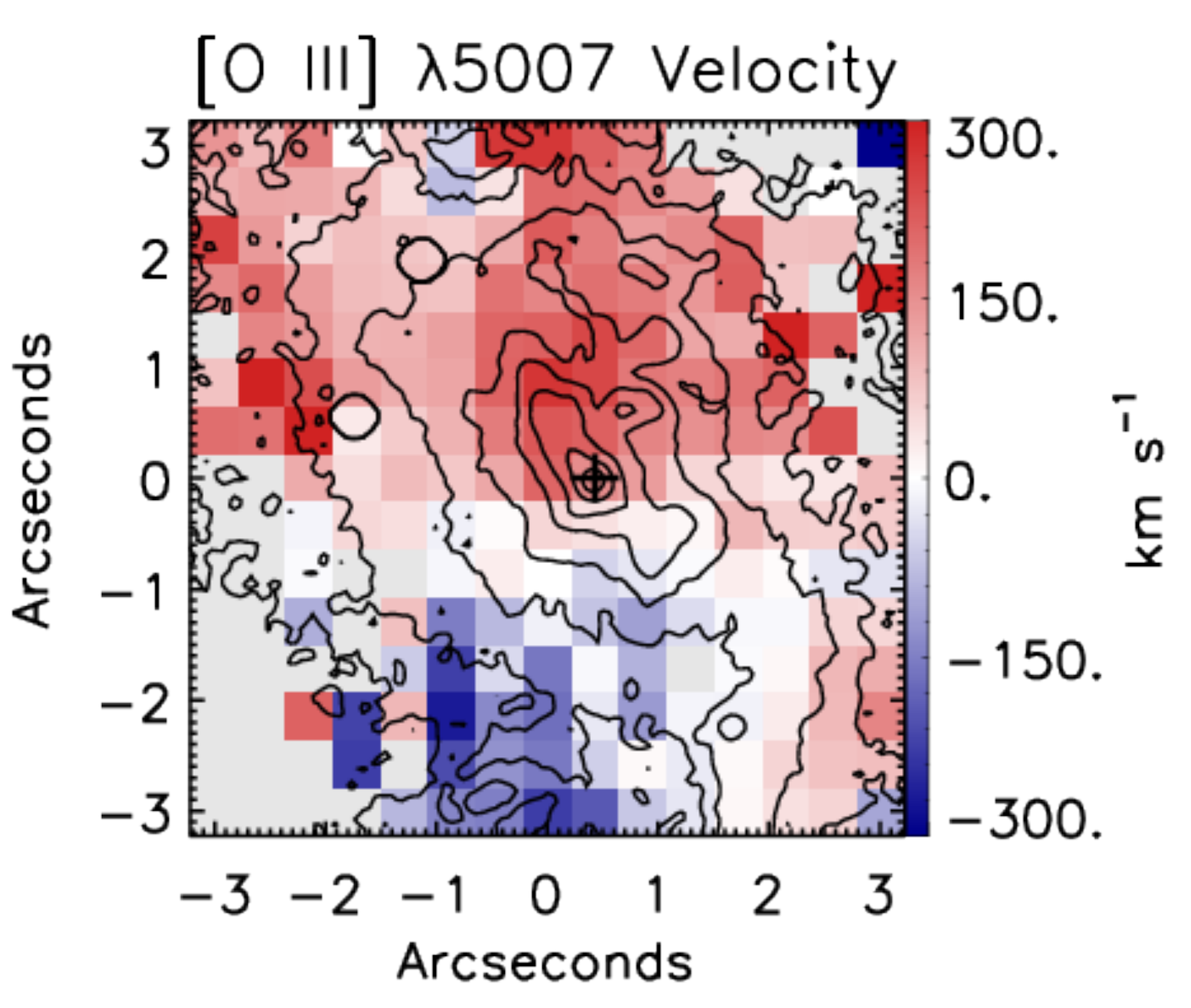}
\includegraphics[width=0.31\textwidth]{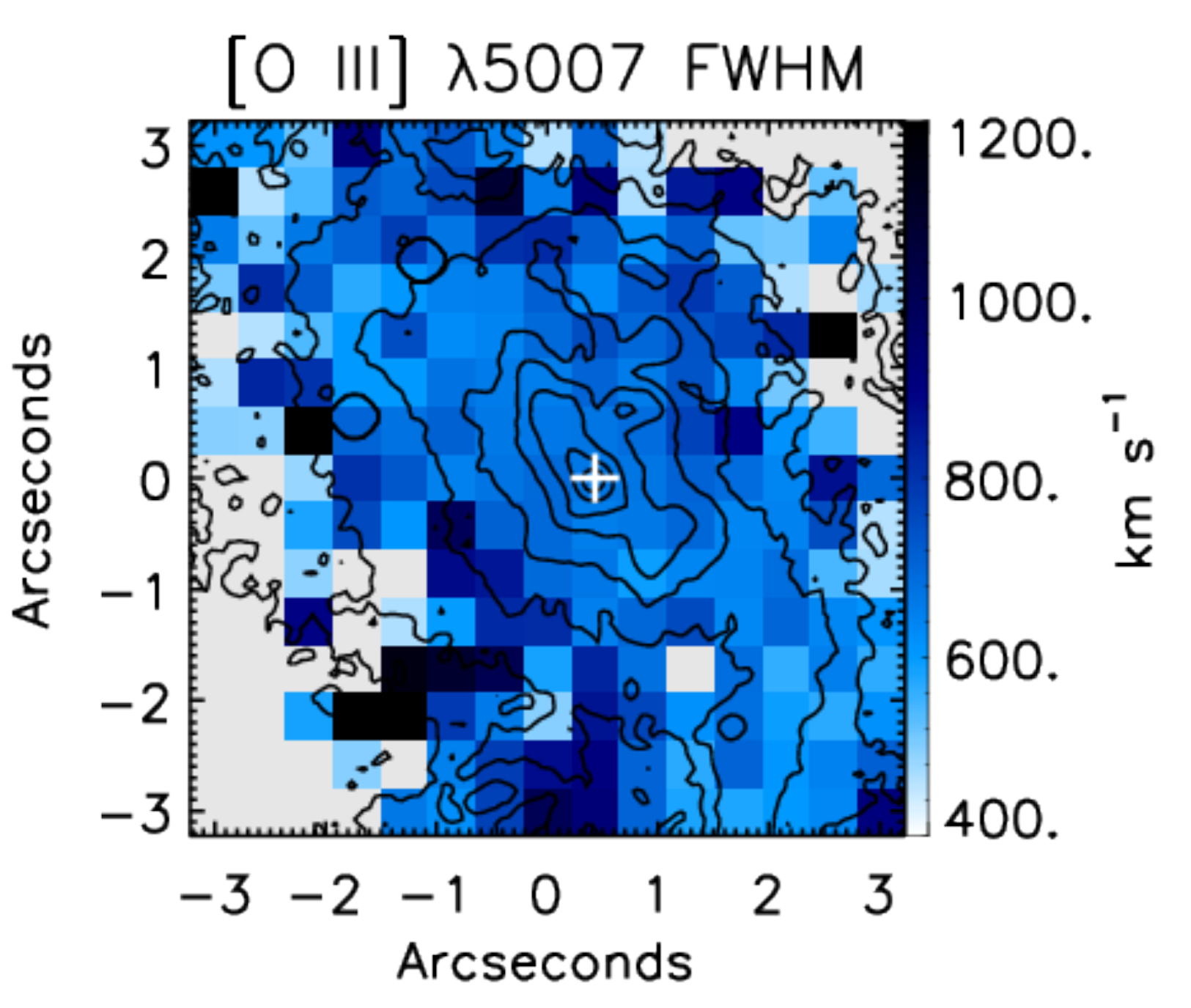}
\includegraphics[width=0.31\textwidth]{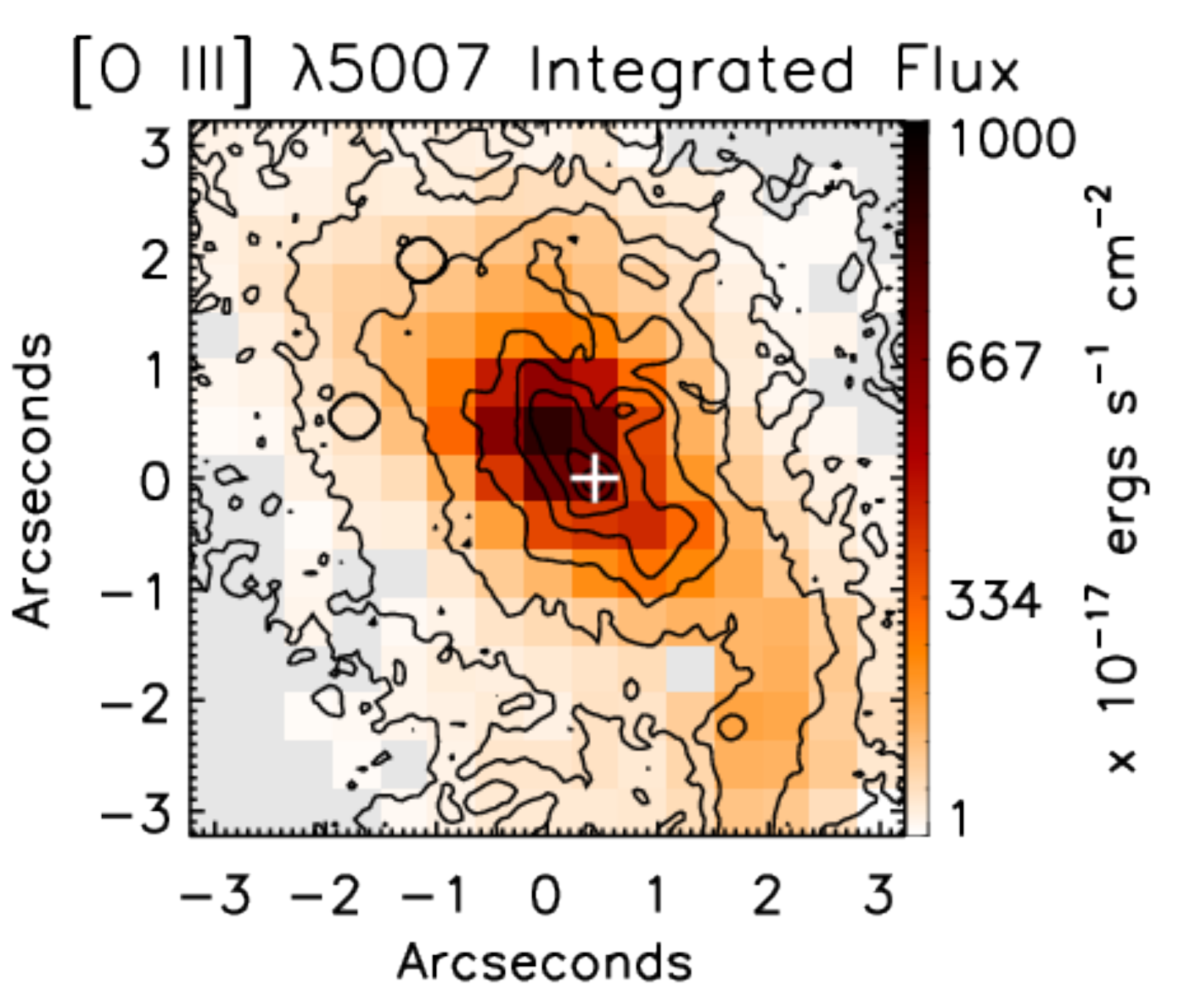}\\
\includegraphics[width=0.31\textwidth]{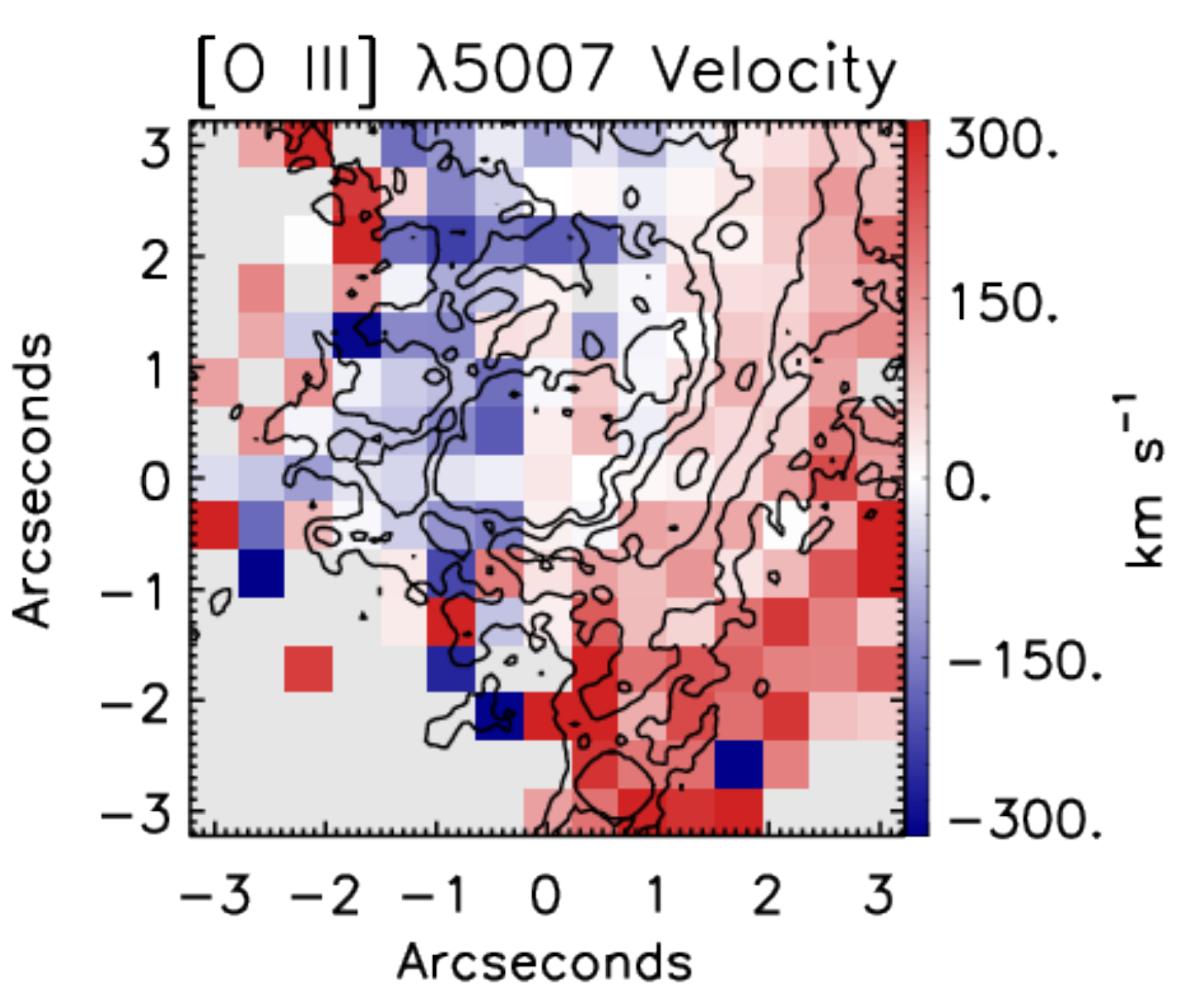}
\includegraphics[width=0.31\textwidth]{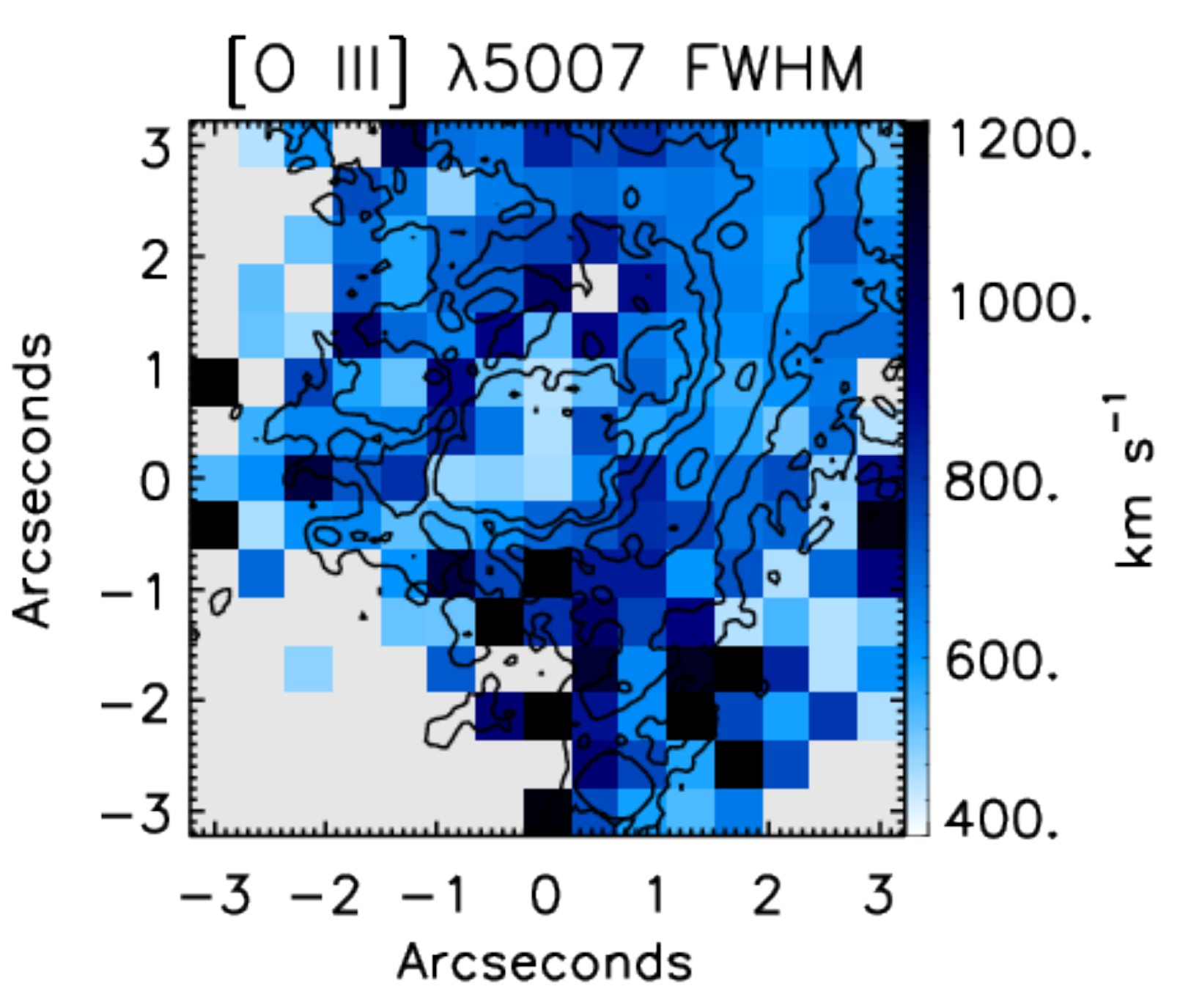}
\includegraphics[width=0.31\textwidth]{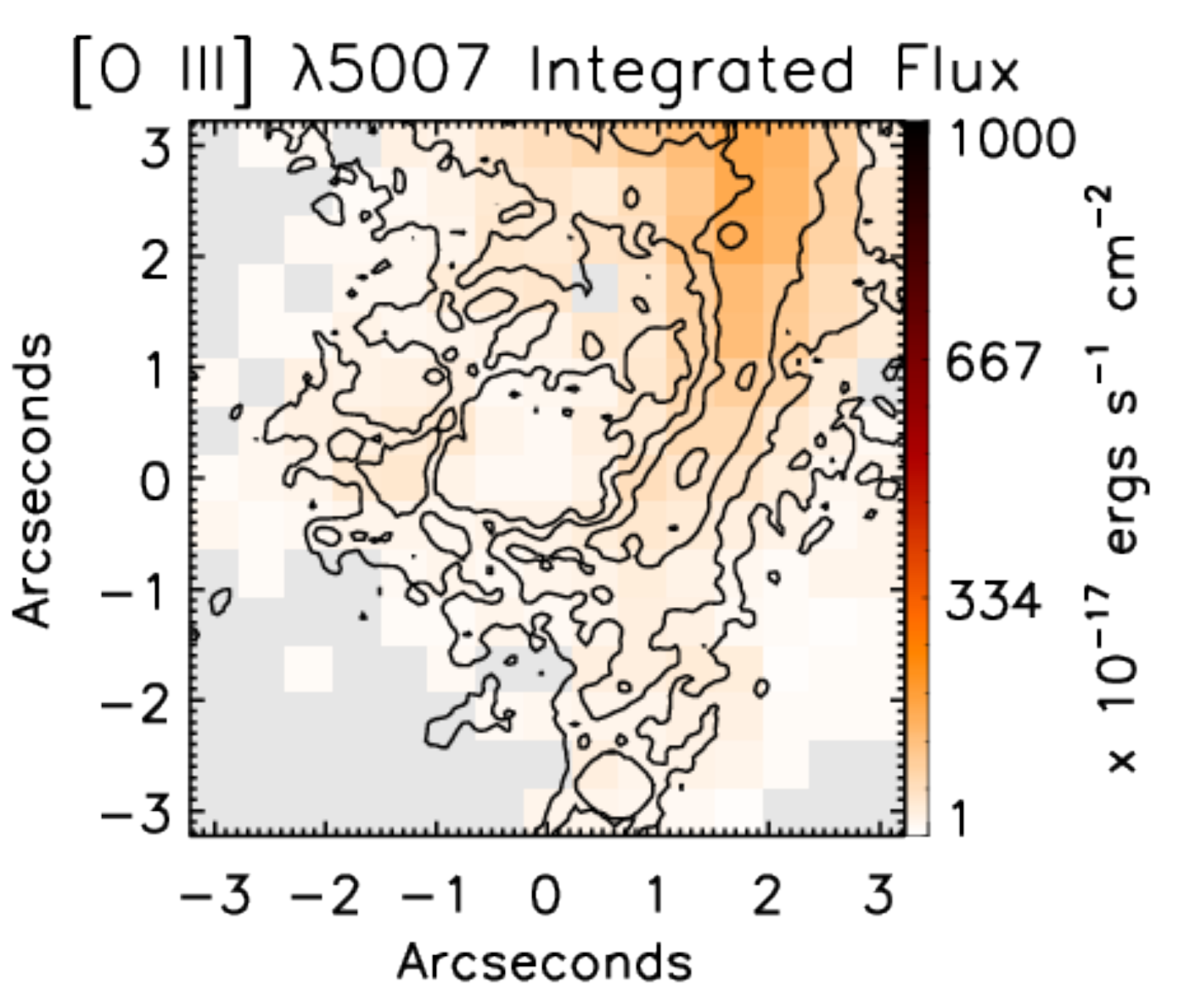}\\

\caption{[O~III] $\lambda$5007 kinematic measurements in 2MASX~J0423 from SNIFS IFU observations. First, 
second, and third columns display emission-line profile centroid velocity, FWHM, and integrated 
flux maps, respectively. First, second, and third rows display measurements for the top, 
center, and bottom fields of view, respectively. Black contours represent \emph{HST}/WFPC2 
F675W imaging. The optical continuum flux peak is depicted by a cross. One 0.43$" \times$ 0.43$"$ spaxel 
samples approximately 380\,pc $\times$ 380\,pc.}
\label{fig:oiiimaps}

\end{figure*}

\begin{figure*}[!htbp]
\centering

\includegraphics[width=0.31\textwidth]{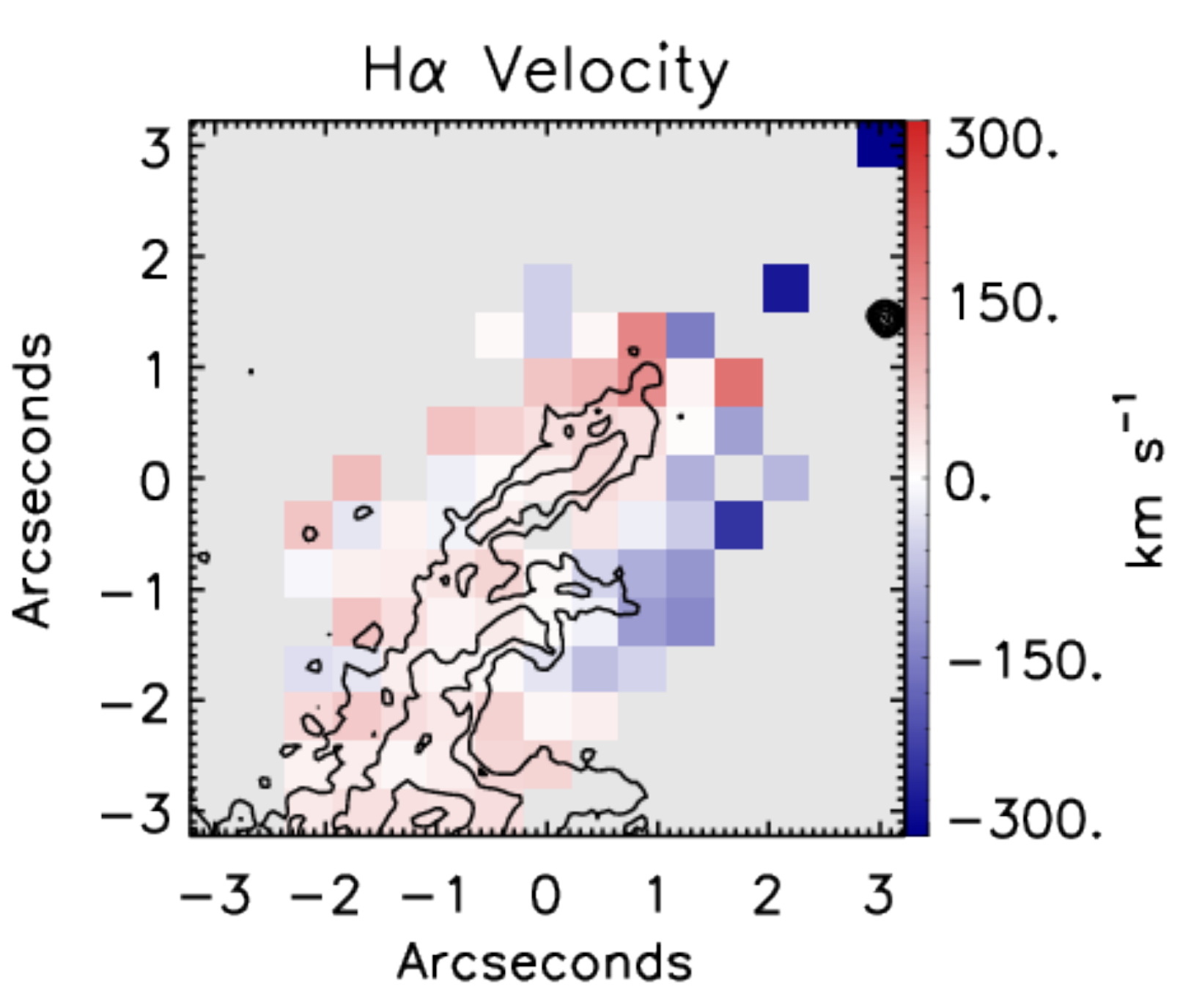}
\includegraphics[width=0.31\textwidth]{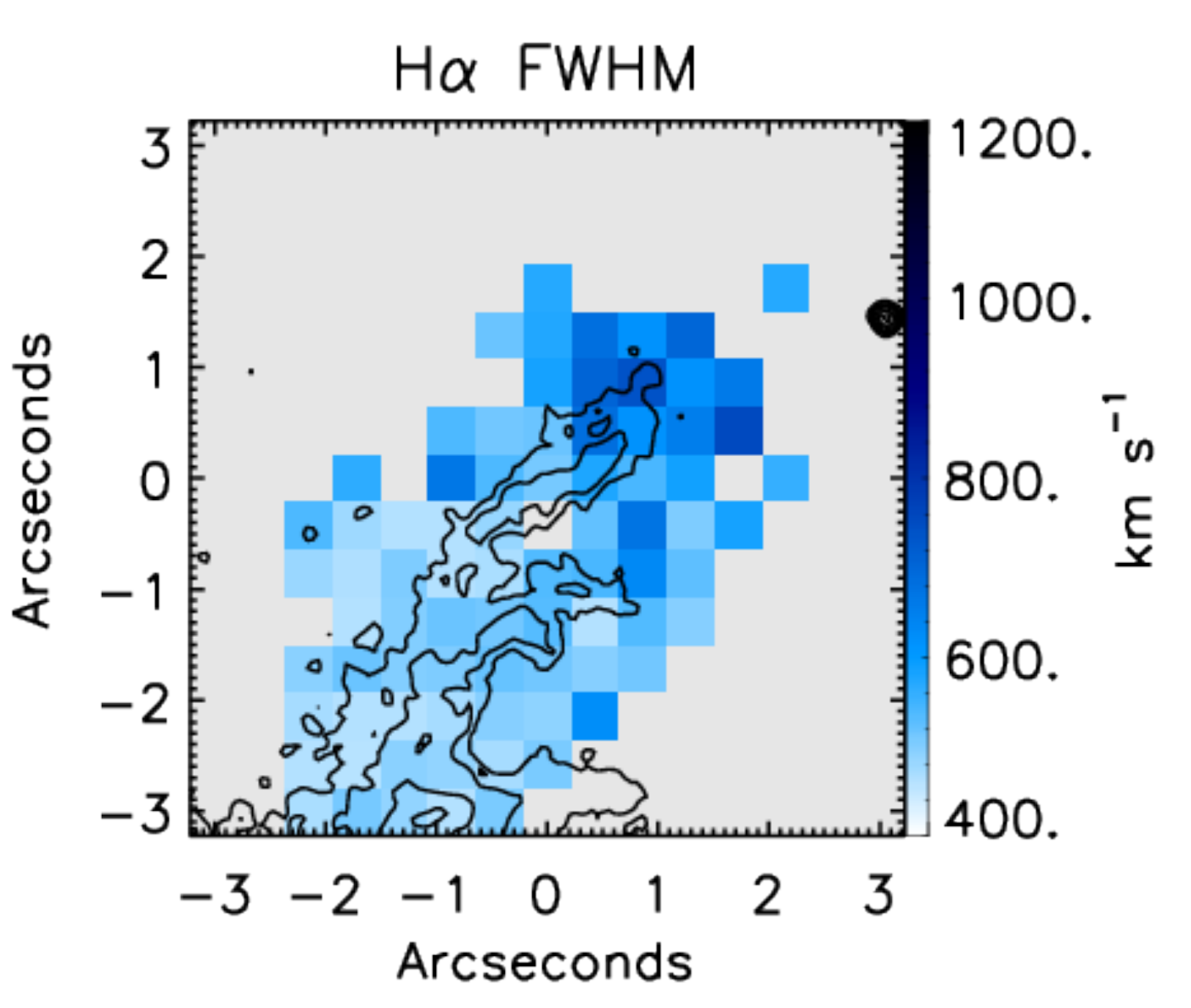}
\includegraphics[width=0.31\textwidth]{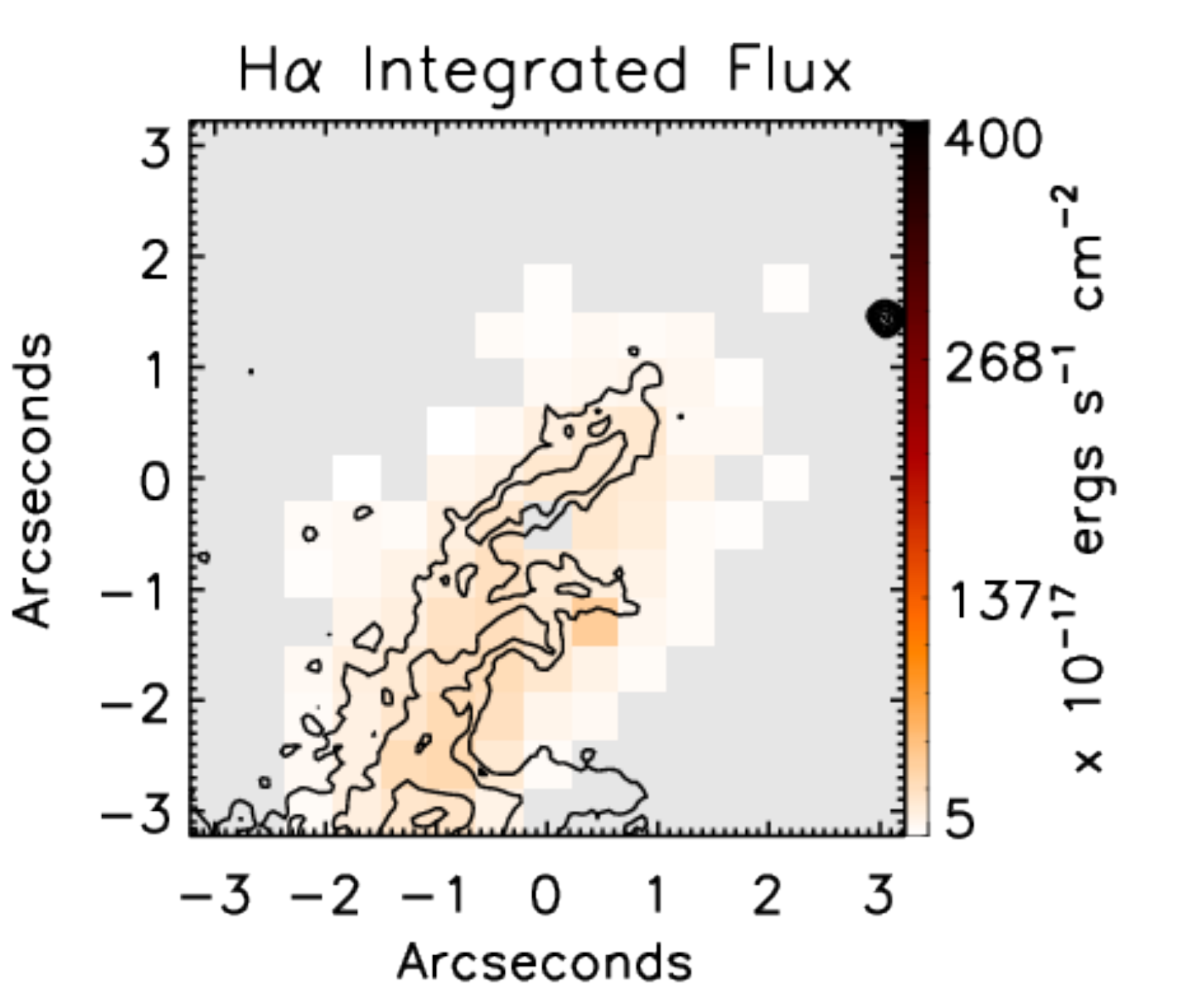}\\
\includegraphics[width=0.31\textwidth]{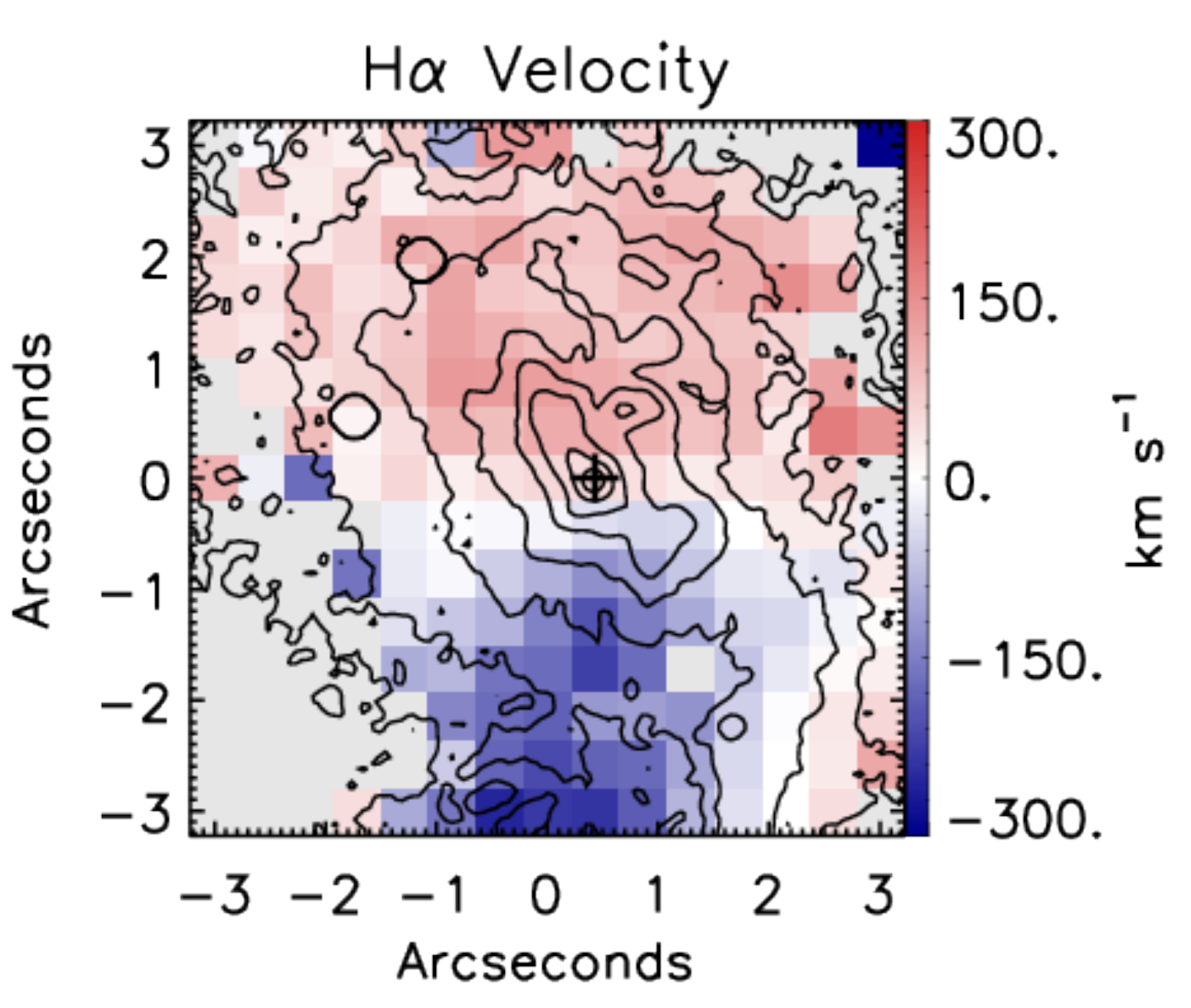}
\includegraphics[width=0.31\textwidth]{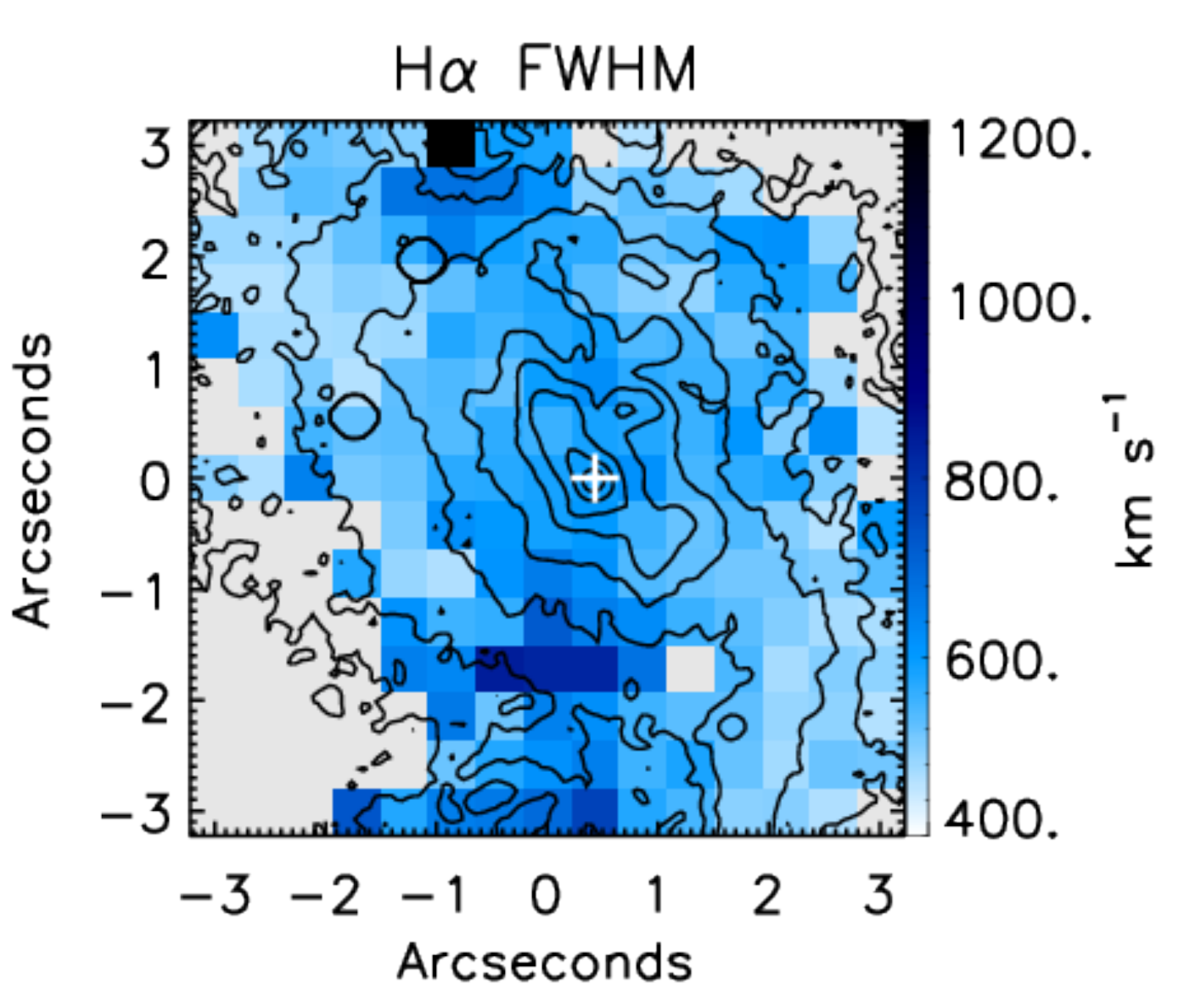}
\includegraphics[width=0.31\textwidth]{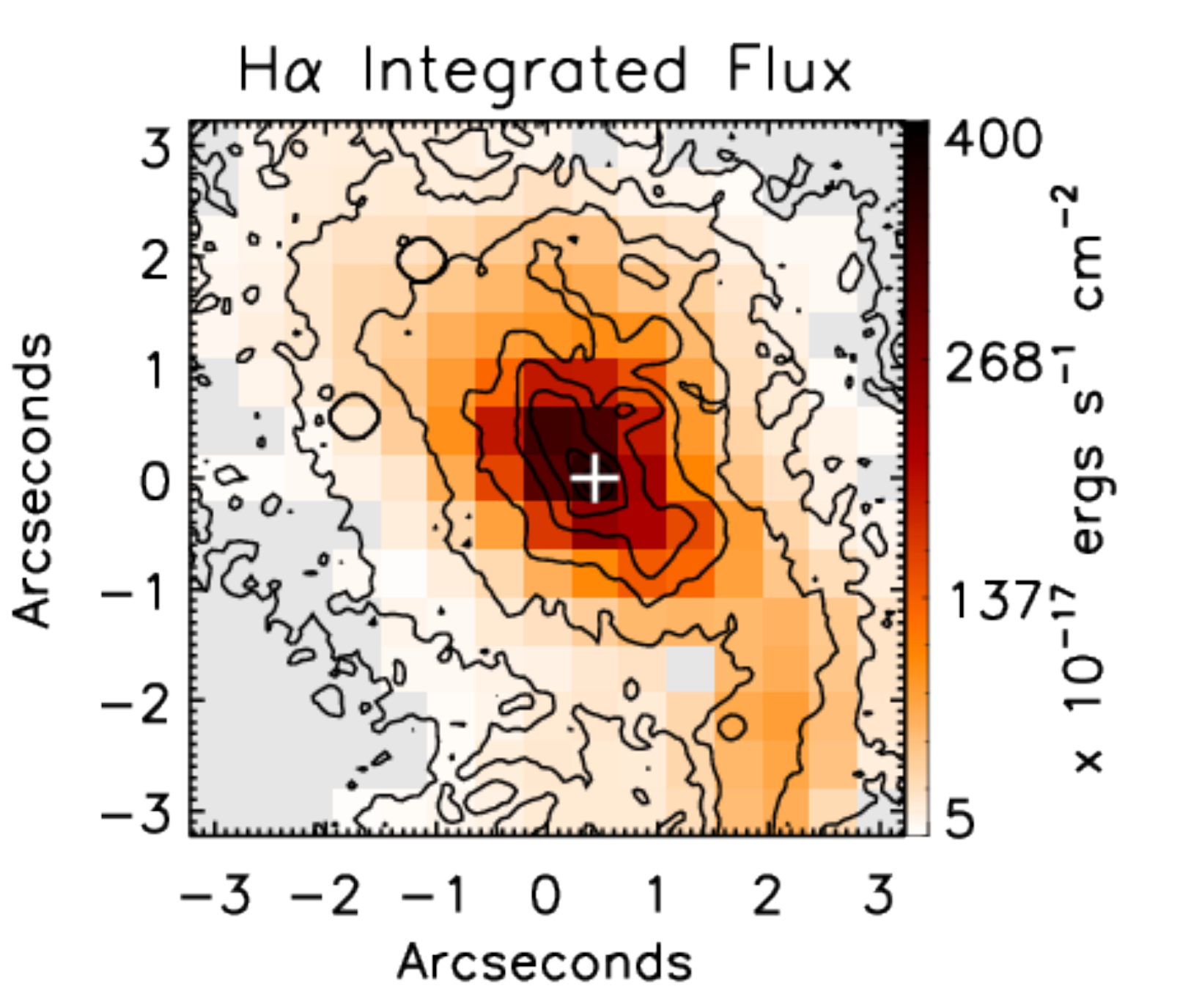}\\
\includegraphics[width=0.31\textwidth]{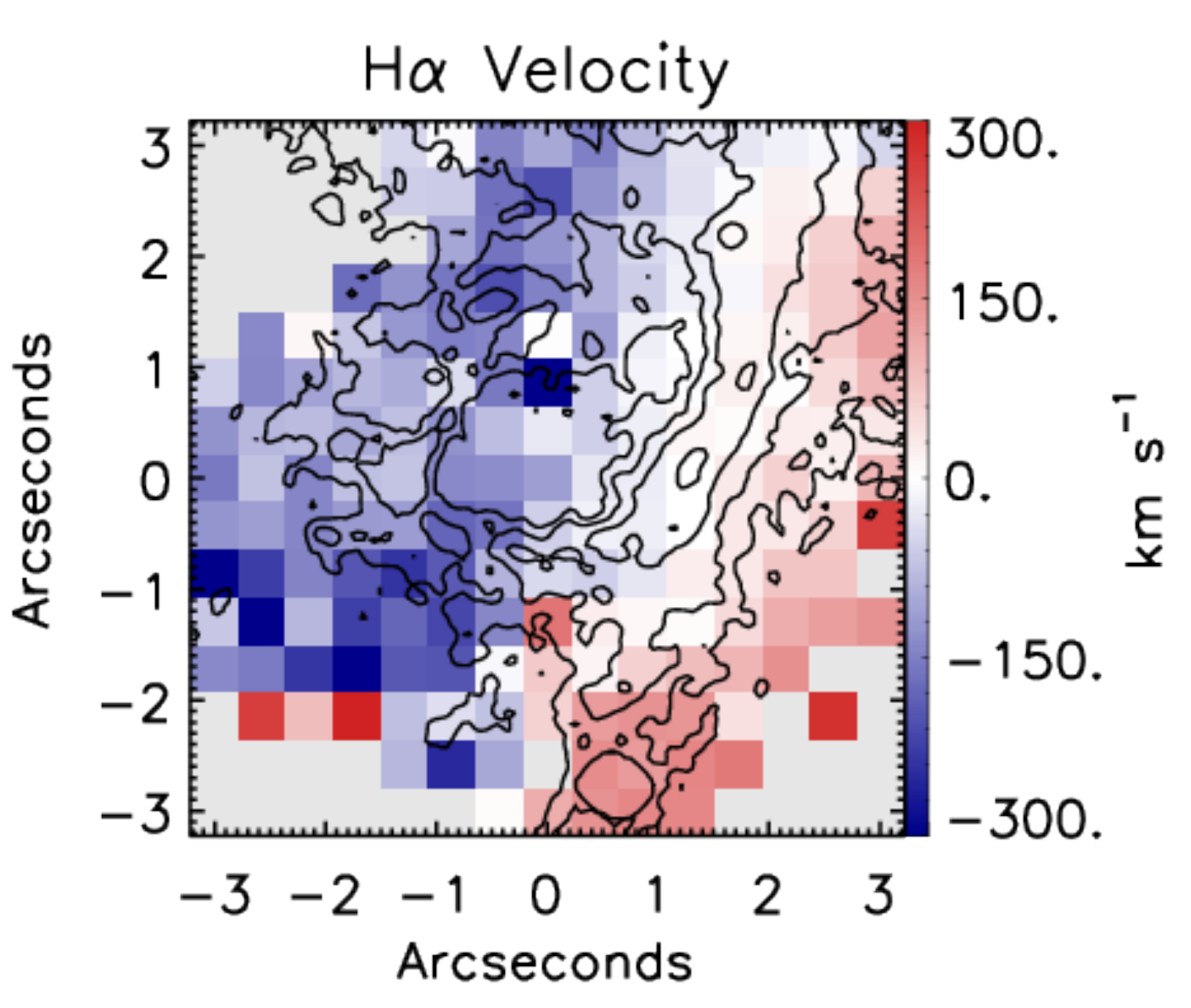}
\includegraphics[width=0.31\textwidth]{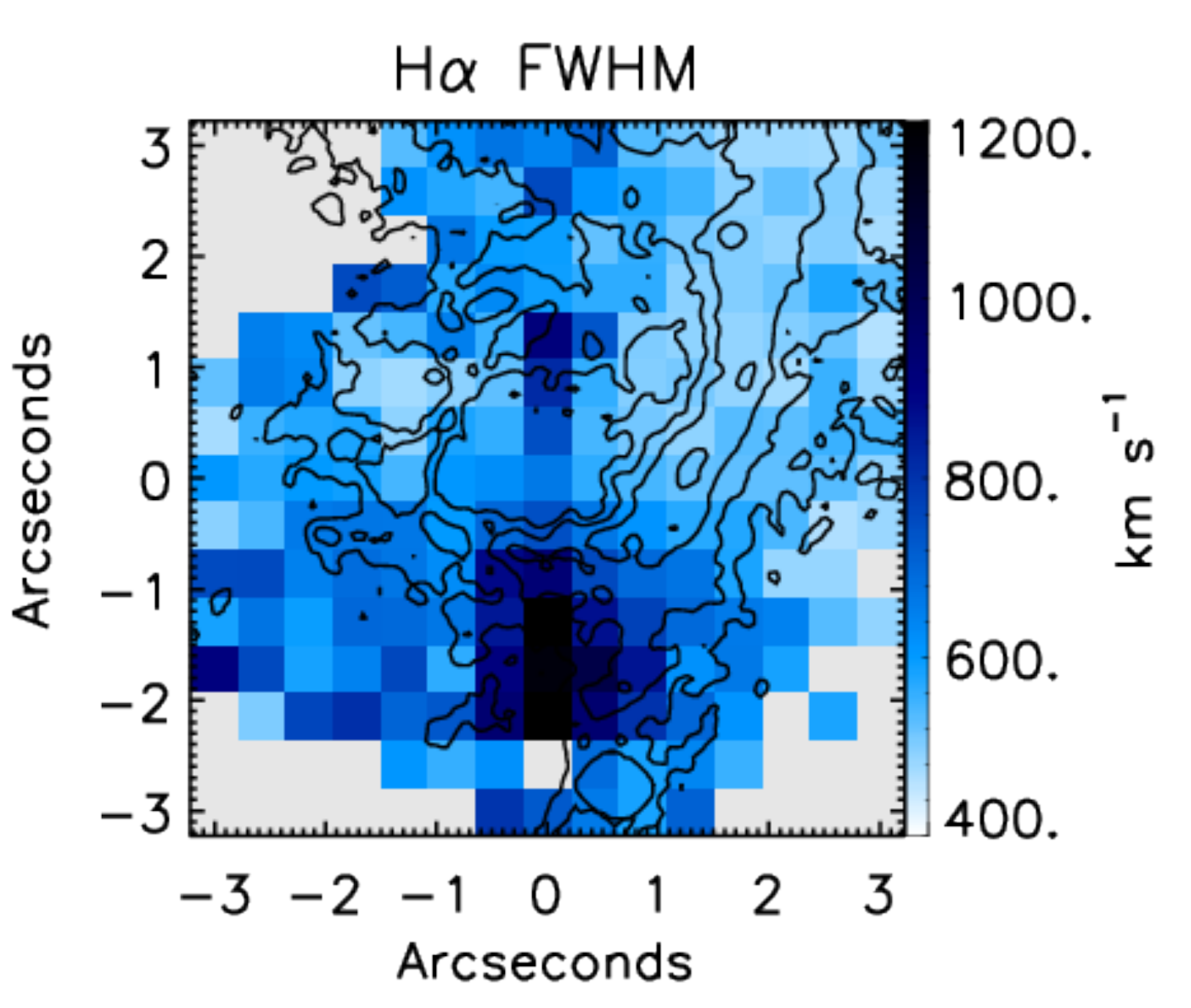}
\includegraphics[width=0.31\textwidth]{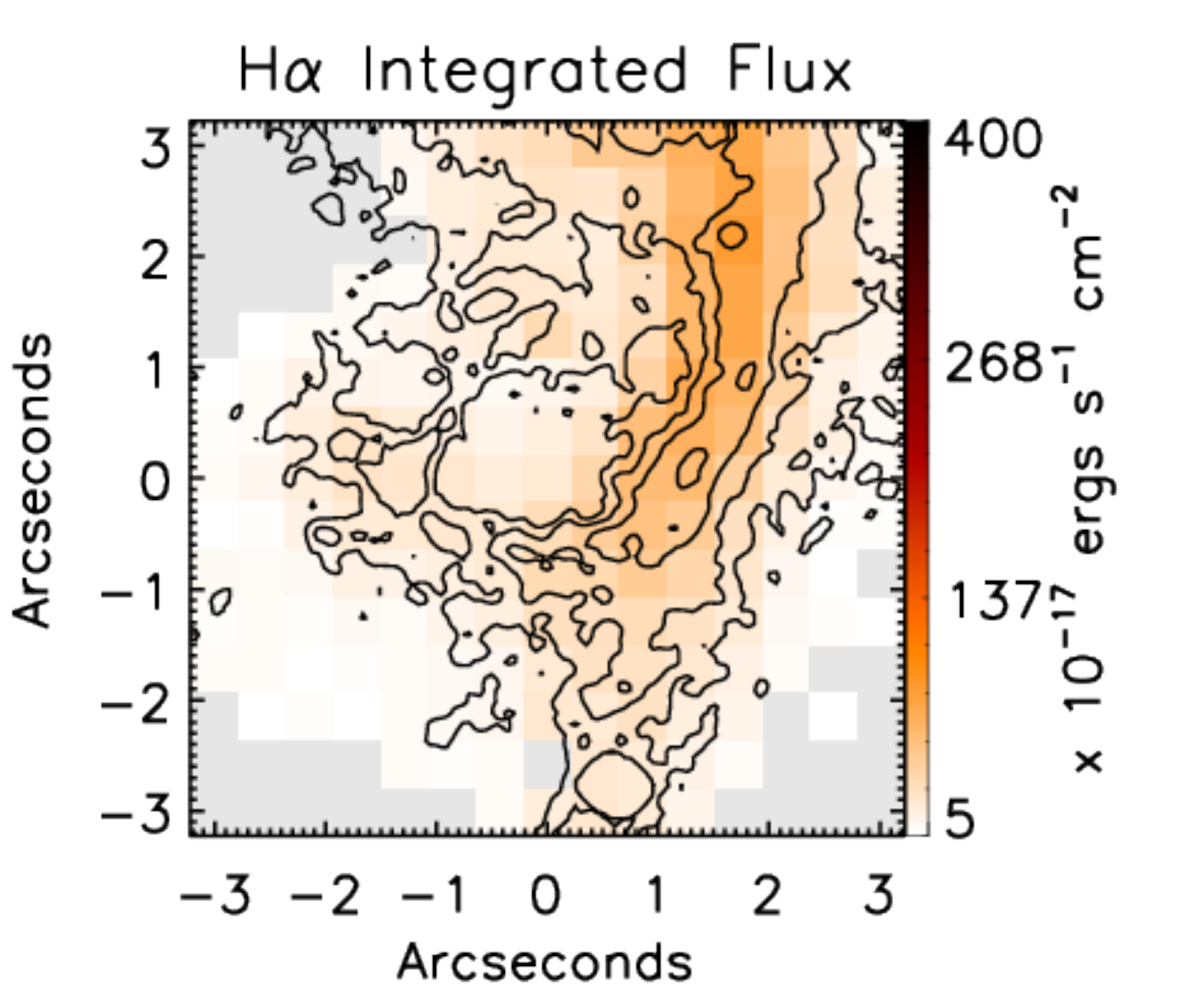}\\

\caption{H$\alpha$ kinematic measurements in 2MASX~J0423 from SNIFS IFU observations. First, 
second, and third columns display emission-line profile centroid velocity, FWHM, and integrated 
flux maps, respectively. First, second, and third rows display measurements for the top, 
center, and bottom fields of view, respectively. Black contours represent \emph{HST}/WFPC2 F675W 
imaging. The optical continuum flux peak is depicted by a cross. One 0.43$" \times$ 0.43$"$ spaxel 
samples approximately 380\,pc $\times$ 380\,pc.}
\label{fig:hamaps}

\end{figure*}
More
1–16 of 16
￼
￼

\subsection{{\it VLA} Radio Interferometry Observations}

In 2013 June 15, 2MASX~J0423 was imaged by the {\it Karl G. Jansky Very Large Array} ({\it VLA}) in C-configuration. 
Observations were obtained in the K-band, centered at 22~GHz with an 8~GHz bandwidth, resulting in 1\arcsec spatial 
resolution \citep{Smi16}.  The science integration for 2MASX~J0423 was 7 minutes in duration and was preceded and 
followed by a pointing and gain calibration scan of the reference quasar J0433+0521; the 1$\sigma$~sensitivity in the 
science image is 20\,$\mu$Jy per beam. The entire observing block was concluded with a flux and bandpass calibration scan 
of 3C~48. 

After collection, the raw data were passed through the standard {\it VLA} reduction pipeline at the National
Radio Astronomy Observatory (NRAO). We then processed the data using the Common Astronomy Software
Applications package \citep[v. 4.5, CASA; ][]{McM07}. The calibrated science observation was split 
from the parent measurement set, averaging over all 64 channels within each spectral window in order to reduce 
processing time without compromising image quality. Finally, the image was cleaned to a 0.03~mJy threshold 
using the CASA clean task with Briggs weighting. 


We also include archival images of 2MASX~J0423 in the NRAO VLA Archive Survey (NVAS)\footnote{http://archive.nrao.edu/nvas/} at 
1.45\,GHz and 4.89\,GHz, taken in A-Configuration in September 1995, which we include in our analysis that have similar 
resolutions to our 22~GHz image of 1.33\arcsec and 0.39\arcsec, respectively.

\section{Observational Analysis}
\label{sec:analysis}

\begin{figure*}[!htbp]
\centering

\includegraphics[width=\textwidth]{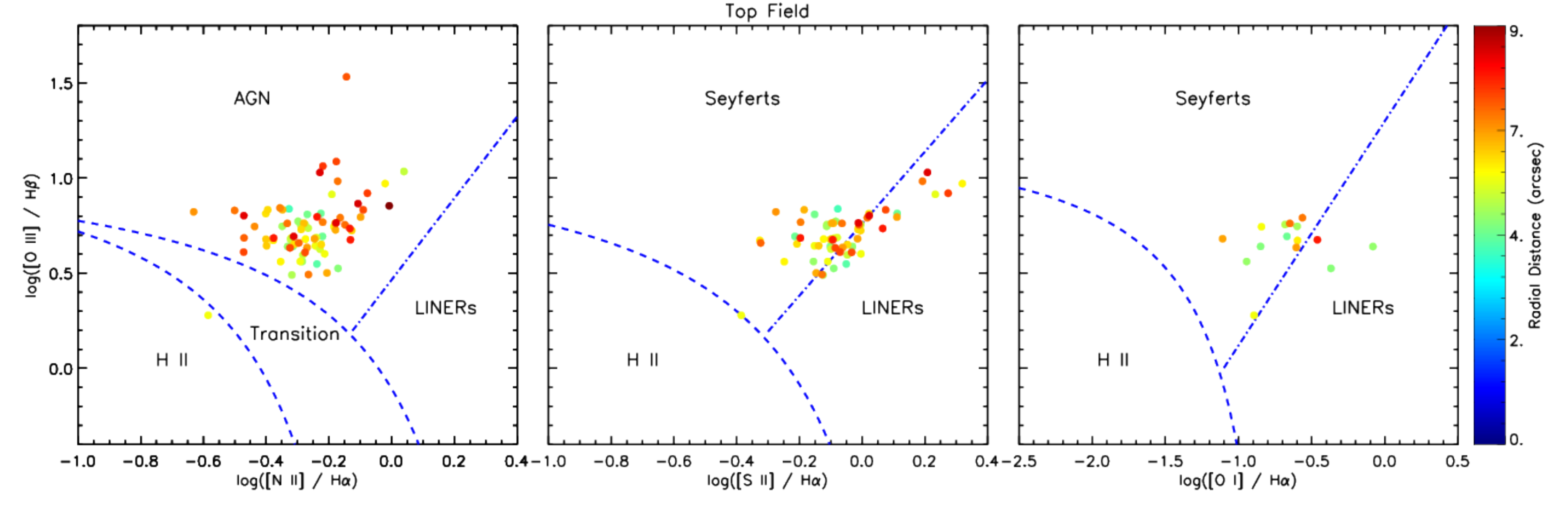}\\
\includegraphics[width=\textwidth]{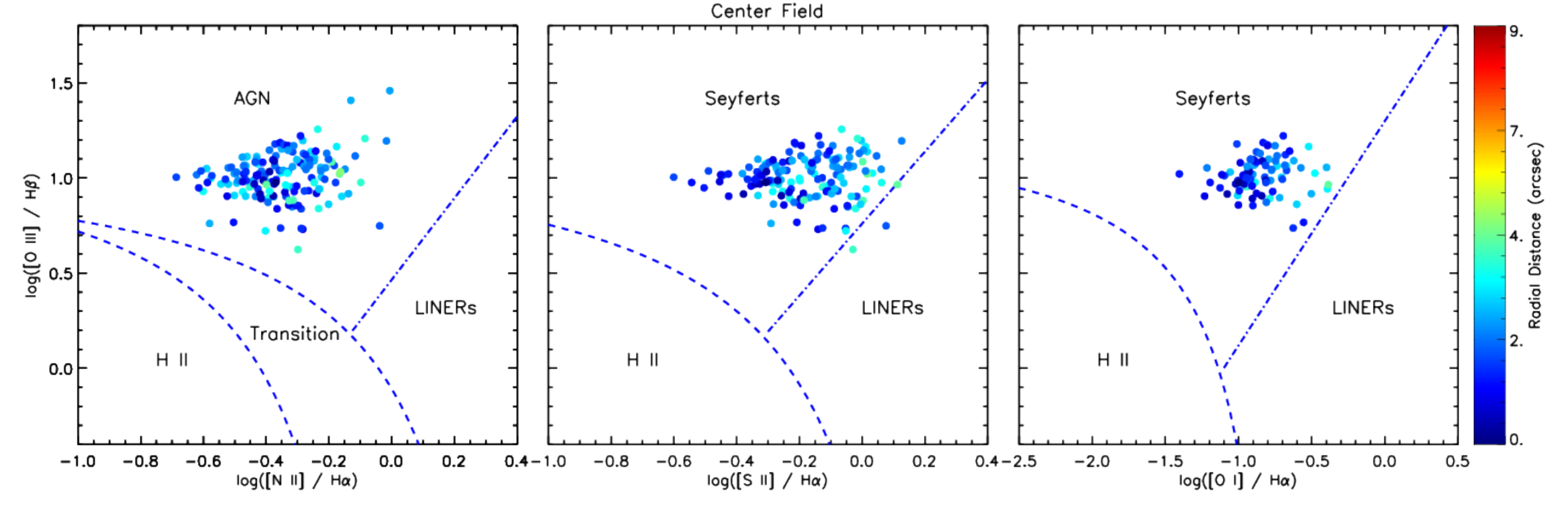}\\
\includegraphics[width=\textwidth]{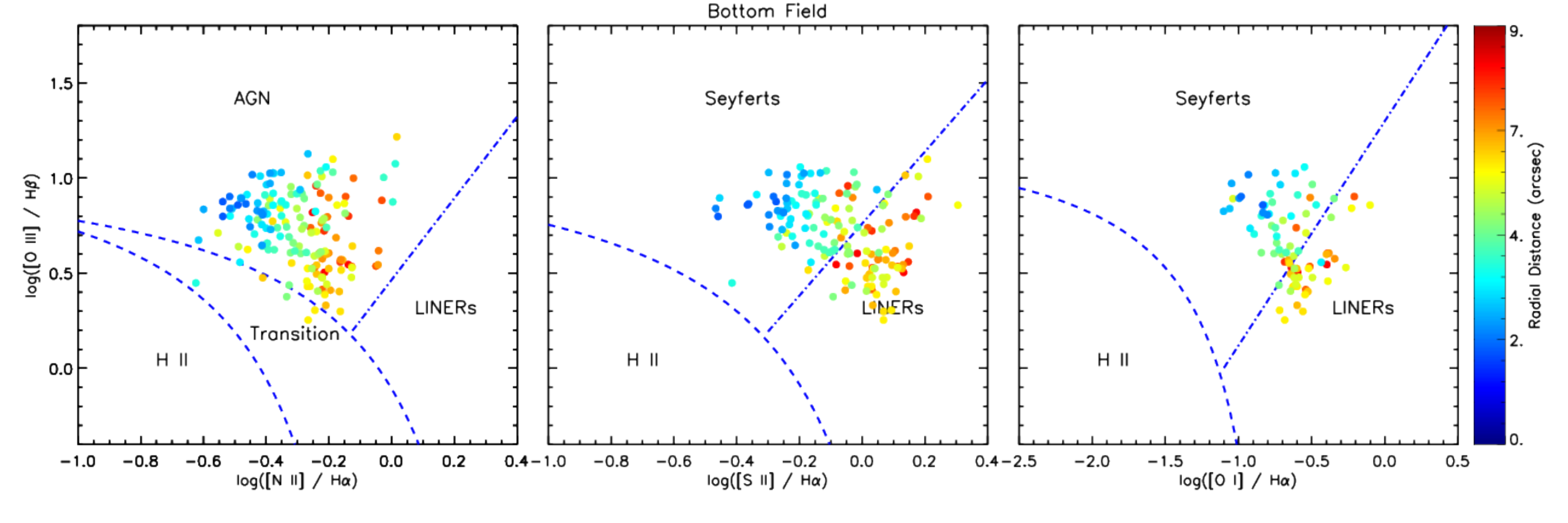}\\

\caption{Ionization source diagnostics across 2MASX~J0423 from SNIFS IFU observations. First, second, and third rows 
display diagnostics for the top, center, and bottom fields of view, respectively. Distances are projected 
and measured from the nuclear optical continuum peak in the central field of view.}
\label{fig:bpts}

\end{figure*}

\begin{figure}[!htbp]
\centering
\includegraphics[width=.48\textwidth]{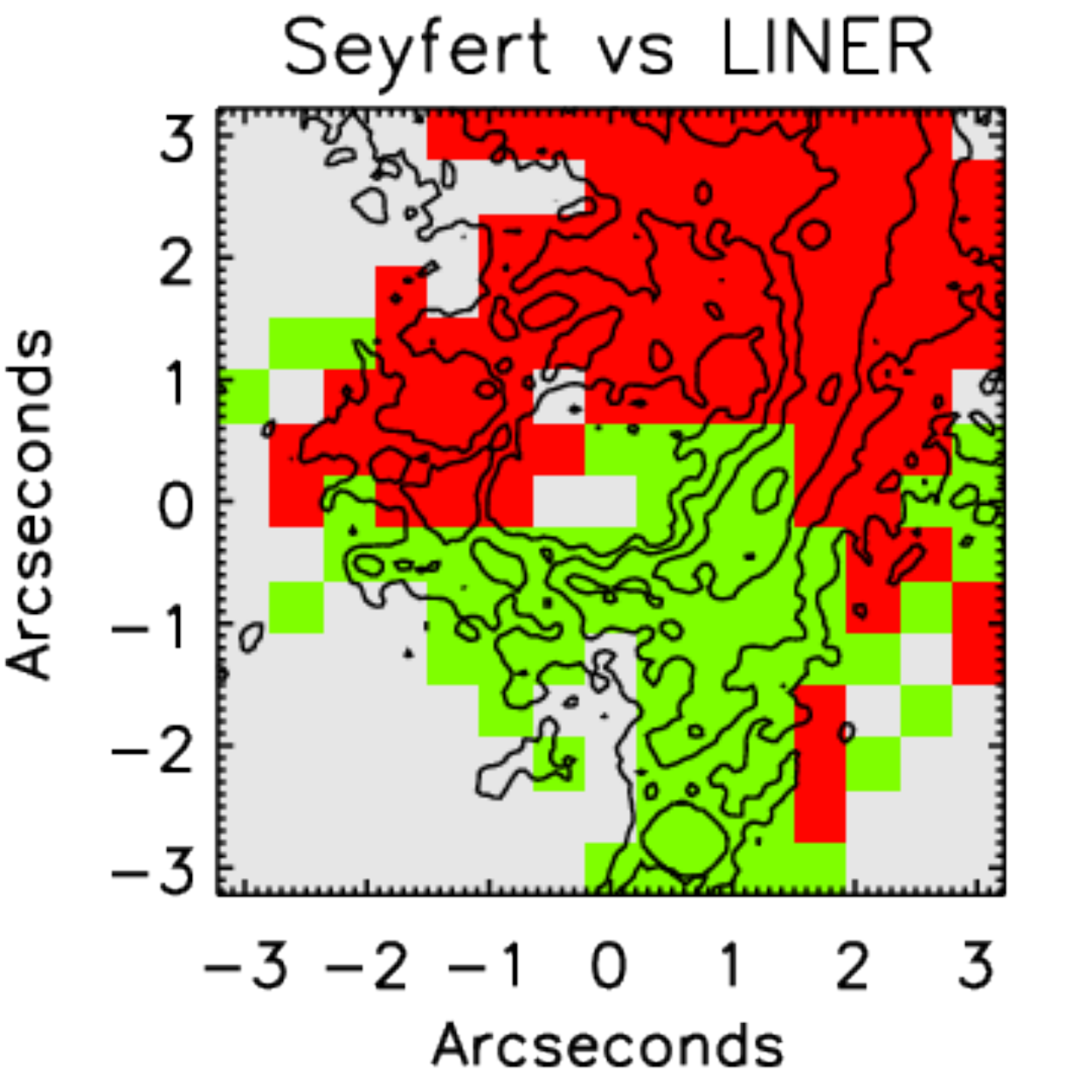}

\caption{2D distribution of the [O~III]/H$\beta$ vs [S~II]/H$\alpha$ ionization source diagnostic for 2MASX~J0423 the 
south SNIFS field of view shown in Figure \ref{fig:bpts} with \emph{HST}/WFPC2 F675W imaging contours. AGN and LINER 
dominated ionization sources are represented by red and green, respectively.}
\label{fig:bpt_map}

\end{figure}

\subsection{SNIFS Spectroscopic Fitting}

Emission-line kinematics and fluxes of [O~II]$\lambda$3727, [Ne~III]$\lambda$3869, H$\beta$, [O~III] $\lambda$5007, [O~I] $\lambda$6300, 
H$\alpha$, [N II] $\lambda\lambda$6548,6584, [S~II] $\lambda\lambda$6716,6731, and [S~III] $\lambda$9071 were measured in each 
spaxel of our SNIFS data cubes by fitting Gaussians in an automated routine, with example fits to spectra shown in
Figures \ref{fig:fig2}. Our fitting process, previously discussed in depth in \citet{Fis17}, uses the Importance Nested 
Sampling algorithm as implemented in the MultiNest 
library \citep{Fer08,Fer09,Fer13,Buc14} to compute the logarithm of the evidence, $lnZ$, for models containing a 
continuum plus zero to two Gaussian components per emission line. When comparing two models, i.e. a model with zero 
Gaussians ($M_{0}$) and a model with one Gaussian ($M_{1}$), the simpler model is chosen unless the more complex 
model, $M_{1}$, has a significantly better evidence value, $|ln(Z_{1}/Z_{0})| > 5$ (99\% more likely). 

Several models were used to fit emission lines in each spaxel. [O~III] models measured [O~III] $\lambda$5007 and 
simultaneously fit a second set of components to [O~III] $\lambda$4959 in order to properly account for possible flux 
contributions from wing emission between both lines. Gaussian wavelength centroid and width parameters of [O~III] 
$\lambda$4959 were fixed following parameters used in fitting [O~III] $\lambda$5007, with the flux of 
[O~III] $\lambda$4959 fixed to be 1/3 that of the [O~III] $\lambda$5007 flux. H$\alpha +$ [N~II] models 
fit H$\alpha$ and [N~II] $\lambda\lambda$6548,6584 simultaneously. Gaussian wavelength centroid and width 
parameters of [N~II] $\lambda\lambda$6548,6584 were fixed following parameters used in fitting H$\alpha$, 
under the assumption that the lines originate from the same emission region, with the flux of [N~II] 
$\lambda$6548 fixed to be 1/3 that of the [N~II] $\lambda$6584 flux, which was left as an open parameter. [S~II] 
models fit [S~II] $\lambda\lambda$6716,6731 simultaneously, with Gaussian width fixed to be identical between lines. 
[O~II]$\lambda$3727, [Ne~III]$\lambda$3869, H$\beta$, [O~I] $\lambda$6300, and [S~III] $\lambda$9071 lines were fit 
individually. 

Initial input parameters in our models are selected based on physical considerations. The centroid position for 
each Gaussian was limited to a 40\AA~range around the wavelength that contained the entirety of the line profiles 
throughout each data cube. Gaussian standard deviation ranged from the spectral resolution of the blue and red gratings, 
to an artificial FWHM limit of $\sim$1600 km s$^{-1}$. Gaussian height was defined to allow for an integrated flux 
that ranged from a 3$\sigma$ detection to a maximum integrated flux of 3$\sigma \times$ 10$^4$.

Fit parameters for measured lines were used to calculate their observed velocity, FWHM, and integrated flux, mapped 
for [O~II]$\lambda$3727, [O~III]$\lambda$5007, and H$\alpha$ in Figures 
\ref{fig:oiimaps} - \ref{fig:hamaps}, with additional measurements for [O~I]$\lambda$6300, [S~II]$\lambda$6716, and
[Ne~III]$\lambda$3869, H$\beta$, and [S~III]$\lambda$9071 lines for the nuclear field of view in the Appendix. 
The Doppler shifted velocity for each emission-line component is given in the rest frame of the galaxy using air rest 
wavelengths of each line. We found emission lines present in most spaxels to be best fit with a single Gaussian. 

Individual spaxel fits for H$\beta$ were largely unsuccessful as the emission line was located near the edge of 
the detector in the blue data cubes. Therefore, we instead estimate observed H$\beta$ fluxes from H$\alpha$ 
observations by measuring the intrinsic reddening in each SNIFS FOV. We bin the blue and red cubes from each 
FOV into single spectra and compare the resultant H$\beta$ and H$\alpha$ flux peaks, assuming that the FWHM 
across both lines are identical, to measure H$\alpha$/H$\beta$ ratios for the north, central, and south fields 
of 3.18, 4.02, and 4.64, respectively. We used these ratios along with the observed H$\alpha$ flux maps to estimate 
H$\beta$ fluxes across the system.

\subsection{Optical Morphology}

From the {\it HST} F547M/F675W/F814W imaging shown in Figure 1, the optical morphology of 2MASX~J0423 is 
largely confined to an S-shape provided by two gas lanes north and south of the nucleus. The brightest structure is 
contained inside a radial distance of 1.2$''$ ($\sim$ 1 kpc) from the nucleus, observed in all three bands extending 
along the inner portion of the gas lanes. A background galaxy resides 4$''$ ($\sim$ 3.6 kpc) from the nucleus along 
the northern lane gas lane, visible in F675W/F814W imaging and absent in F547M imaging. Gas lanes extend north and 
south of the nucleus to approximately 7.75$''$ ($\sim$ 7 kpc) and the merger companion is located at a projected 
distance of approximately 10.65$''$ ($\sim$ 9.5 kpc) from the nucleus of 2MASX~J0423.

It is unclear whether the observed lanes are loose spiral arms in the galaxy, as they are largely absent in the near-IR F814W
imaging, suggesting little to no contribution from stellar emission \citet{Hill1988}, or filaments carved from the host disk by 
some form of AGN feedback. The host may also be morphologically disturbed due to the interaction with the nearby companion 
galaxy to the northeast as shown in Figure \ref{fig:fig1}. The F814W imaging also suggests a stellar stream connects the two 
galaxies, implying that at least one merger pass has occurred. 

\subsection{Ionization Source Diagnostics}

We compare measured line flux ratios via three ionization diagnostic diagrams (i.e. BPT diagrams; 
\citealt{Bal81}) to spatially resolve the source of ionization throughout the system. Figure \ref{fig:bpts} shows 
[O III]/H$\beta$ vs [N II]/H$\alpha$, [S II]/H$\alpha$, and [O I]/H$\alpha$ as a function 
of distance from the nuclear optical continuum peak for each of the three fields of view. We find most of the gas to be 
AGN ionized across all three fields of view, with [S~II] and [O~I] diagnostics suggesting a large portion of the southern 
most regions and a few regions in the north lane exhibit LINER-like ionization. We can see that the 2D distribution of 
ionization sources, as shown in Figure \ref{fig:bpt_map}, shows LINER-like emission in the south to be distinctly located 
near the end of the southern gas lane below x $\approx$ -0.2$''$. 

\subsection{Optical Ionized Gas Kinematics}

We overlay the {\it HST} F675W flux map onto the independently measured SNIFS kinematics of each emission line, by aligning 
the flux peak in the central cube with that observed in imaging, to provide high resolution structural reference to our 
kinematics. 

The central field of view is largely redshifted north of the nucleus, with bright flux profiles corresponding to the 
bright structures observed in imaging likely due to adjacent molecular gas lanes being ionized by the AGN \citep{Fis17}. 
South of the nucleus is a low-flux region that exhibits blueshifted velocities, adjacent to a bright gas lane which 
exhibits generally systemic velocities over the brightest regions, blueshifts on the concave side, and 
redshifts on the convex side. Velocities for more ionized line species are generally redshifted from lower ionization potential 
emission lines across the field. FWHM measurements are lower along the bright structure in the field, and are larger in 
fainter regions directly north and south of the nucleus.

The bottom field of view follows the remainder of the south gas lane. Kinematics near the gas lane 
continue to exhibit blueshifts on the concave side and redshifts on the convex side, except for [O~II] kinematics
which are largely blueshifted along the lane. Again, velocities of more ionized line species are generally 
redshifted in comparison to lower ionization potential emission lines across the field.
The largest velocity amplitudes across all emission lines are observed at the end of the lane, with peak blueshifts and redshifts 
corresponding to left and right splits of the gas lane morphology at the bottom of the field. The largest FWHM measurements 
in the system are also located at the the splitting point in the south gas lane, reaching values greater than 1000 km s$^{-1}$ 
with velocity offsets near systemic. Both the peaks in velocity offset and FWHM reside the region predominantly 
exhibiting LINER-like emission in the bottom half of the southern field. 

The top field of view traces the north gas lane. Flux distributions, offset velocities, and FWHMs in this field are similar 
but generally of smaller amplitude than what is observed in the southern lane. Kinematics of the lane are generally redshifted, 
with evidence of blue shifts on the concave side of the lane in lower ionization states. The tip of the lane also exhibits an 
increase in FWHM, similar to the southern lane.  This field also partly contains the background field galaxy along the north 
lane observed in Figure \ref{fig:fig1}, however no corresponding spectral signatures are detected.

\subsection{X-ray Analysis}

\begin{figure*}[!htbp]
\centering

\includegraphics[width=0.98\textwidth]{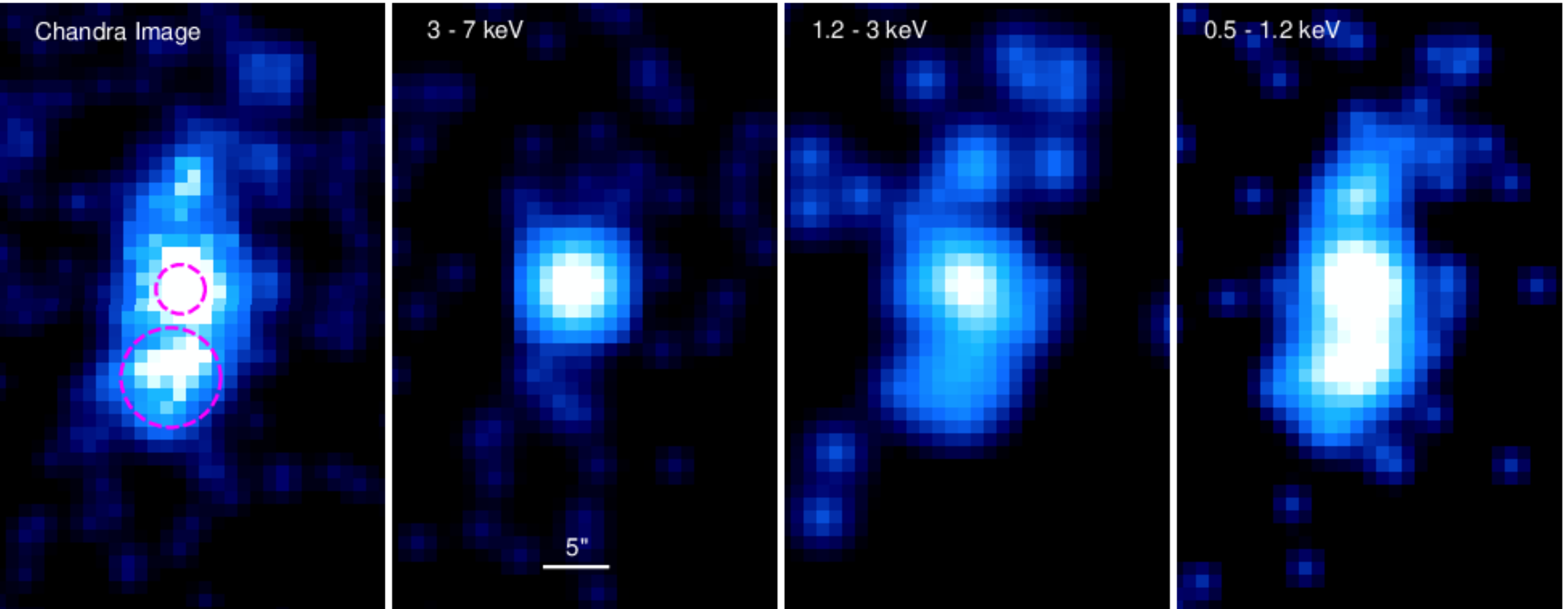}

\caption{\emph{Chandra} images of 2MASX~J0423 in the full bandpass (left) and in the decomposed hard (left-center), 
medium (right-center), and soft (right) bandpasses. A 5\arcsec scale bar is shown in the left-center panel; all images 
are on the same physical scale. The magenta dashed circles in the left panel denote the regions from which the spectra 
shown in Figure~\ref{fig:chandra_spectra} were extracted. 5$''$ corresponds to a spatial scale of $\sim$4.5\,kpc.}
\label{fig:chandra_images}
\end{figure*}

\begin{figure*}[!htbp]
\centering

\includegraphics[width=0.46\textwidth]{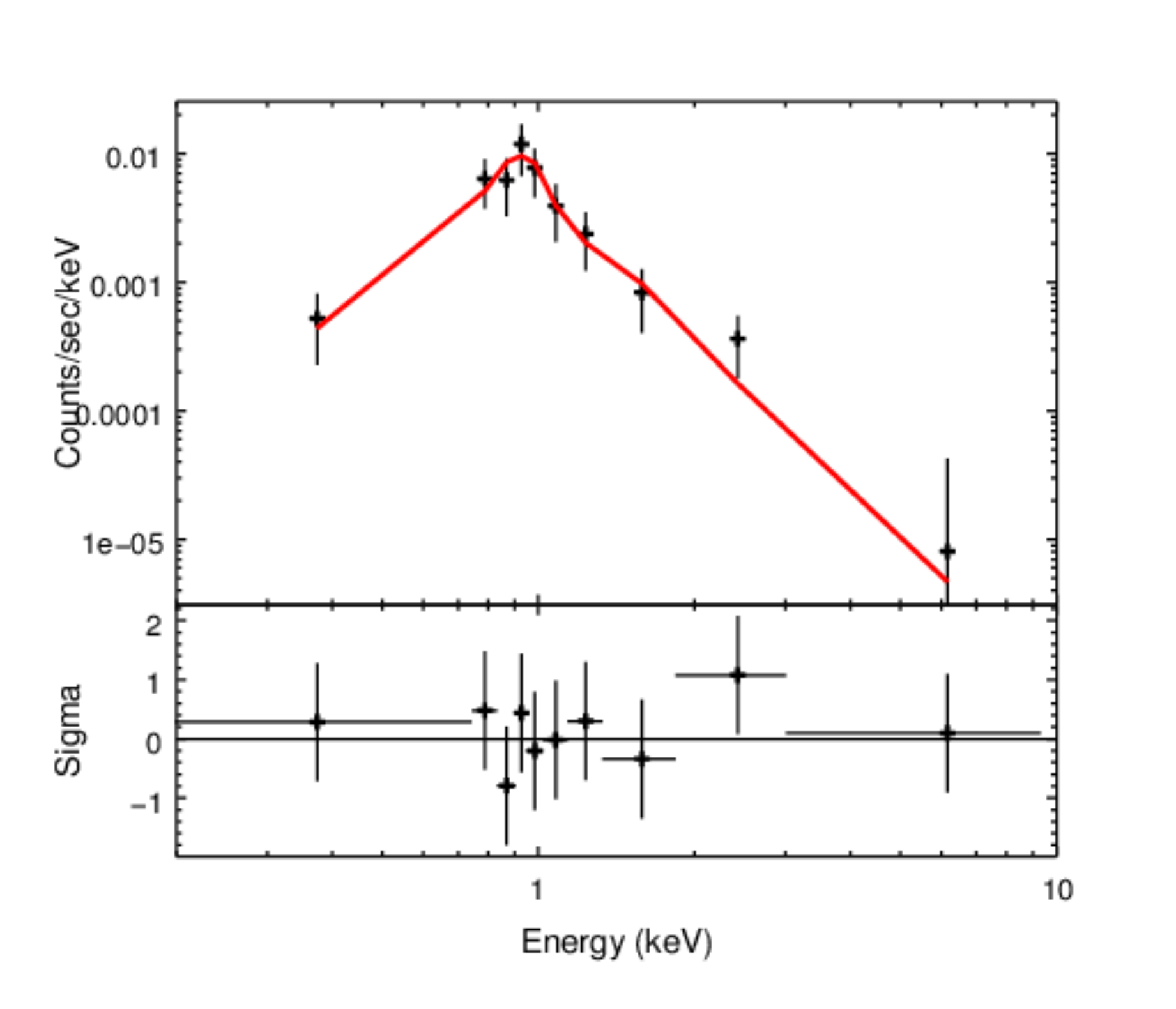}
\includegraphics[width=0.47\textwidth]{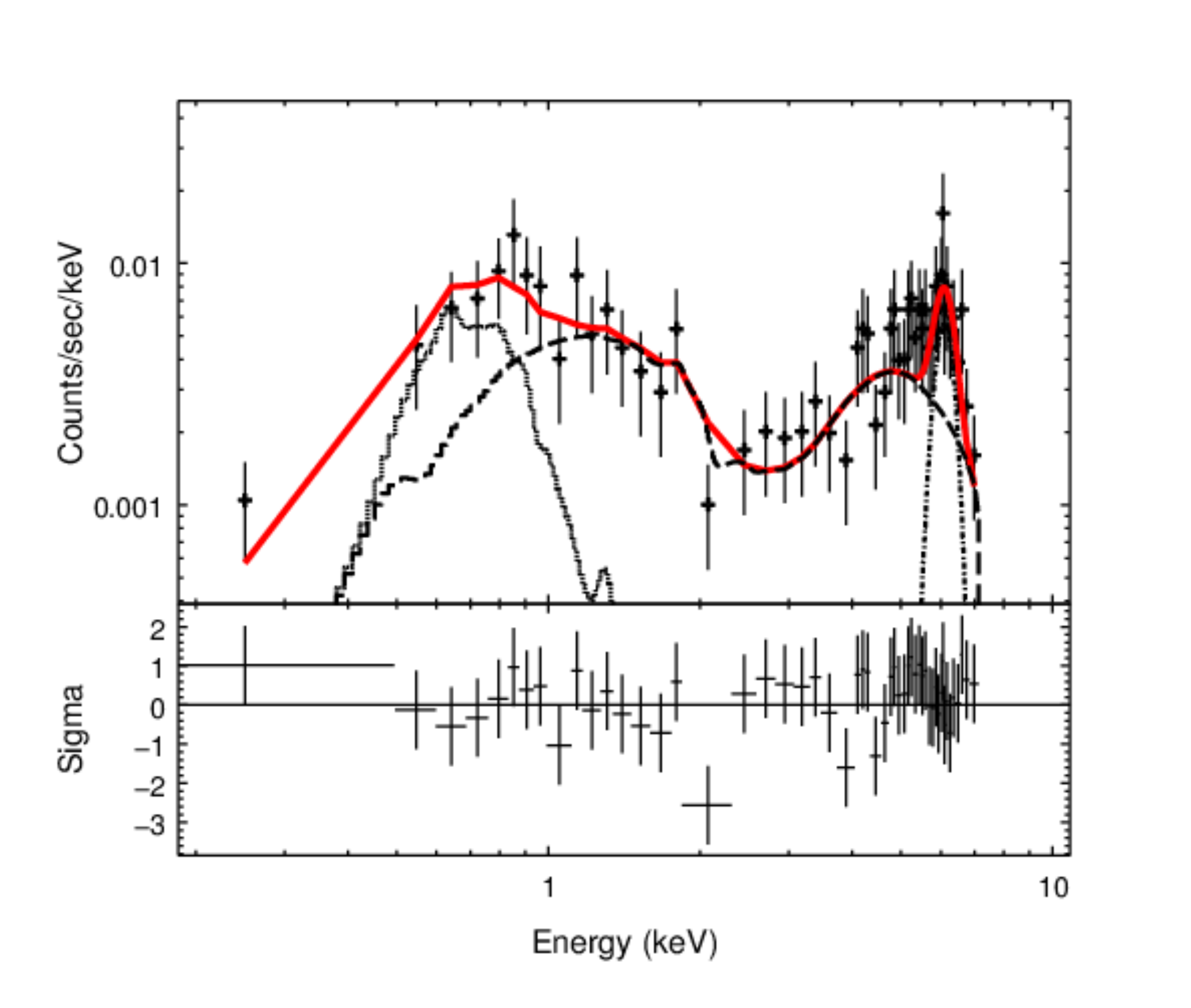}\\

\caption{\emph{Chandra} X-ray spectra and best-fitting models with residuals of the southern X-ray lobe (left) and the nucleus (right), 
extracted from the regions shown in Figure~\ref{fig:chandra_images}. The southern lobe model consists of only a thermal plasma model 
affected by Galactic absorption. The nucleus is best-fit by a model consisting of a thermal plasma (dotted), an absorbed power law 
(dashed), and a Gaussian emission Fe~K$\alpha$~line (dot-dashed).}
\label{fig:chandra_spectra}
\end{figure*}

From the available {\it Chandra} observations, we produced images in three separate bands: soft (0.5-1.2~keV), medium (1.2-3~keV) 
and hard (3-7~keV), shown in Figure~\ref{fig:chandra_images}. We find the hard X-ray (3.0 - 7.0 keV) emission 
to be largely concentrated in the nucleus, while soft X-ray (0.5 - 3.0 keV) is extended, with 3$\sigma$ detections largely 
extending to $\sim$12.6$"$ (11.3 kpc) and one extended cloud to the northwest at $\sim$16.7$"$ (15.0 kpc). 

From the full-spectrum event file we extracted spectra using the task \texttt{specextract}. Nuclear and off-nuclear 
X-ray spectral fitting was performed of the two regions shown in Figure~\ref{fig:chandra_images},
with extracted spectra and the resultant fits with residuals shown in Figure~\ref{fig:chandra_spectra}.
The nuclear spectrum is consistent with most Type 2 AGN \citep{Tur97}. It is well-fit by a soft 
thermal component at kT $= 0.26\pm 0.04$~keV with 
an abundance fixed at solar, an absorbed power law with a high column density (4.9$\times$10$^{23}$cm$^{-2}$) and 
covering fraction (0.98), and a significant Fe~K$\alpha$~line at 6.4~keV in the rest frame. Combining the {\it Chandra} 
and {\it NuSTAR} spectra, we derive a power law spectral index of $\Gamma = 1.45^{0.15}_{-0.16}$. 

The X-ray spectrum of the southern X-ray lobe is best fit by a soft thermal plasma model with kT $= 0.97^{+0.106}_{-0.133}$~keV 
with an abundance fixed at solar. However, the signal-to-noise is not sufficient to differentiate significantly between photoionized, 
shocked, or collisional equilibrium plasmas. Similar analysis of the norther lobe is not possible due to its relative faintness, 
as the spectrum from the northern lobe has too few data points to constrain the fitting parameters. However, the observed 
emission is consistent with the thermal properties of the southern lobe. 

\begin{figure}[!htbp]
\centering

\includegraphics[width=0.45\textwidth]{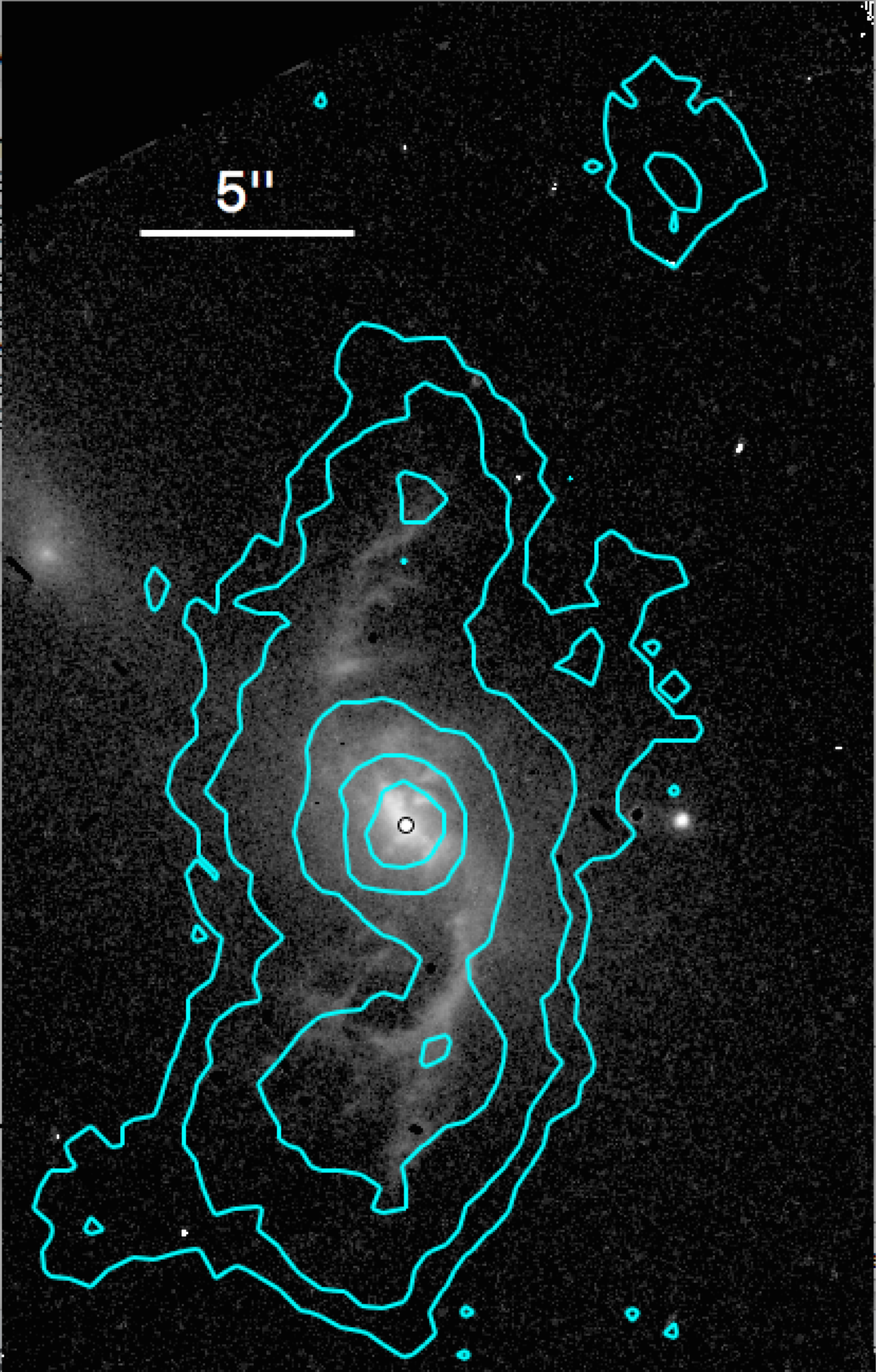}

\caption{{\it HST} F675W imaging and full bandpass {\it Chandra} 
X-ray imaging (cyan contours) in 2MASX~J0423. A black circle represents the optical emission peak 
and assumed nucleus position. 5$''$ corresponds to a spatial scale of $\sim$4.5\,kpc.}
\label{fig:x_vs_opt}
\end{figure}

Comparing the X-ray emission with {\it HST} imaging, as shown in Figure \ref{fig:x_vs_opt}, the X-ray morphology follows a similar 
extended morphology to what is observed in the optical. X-ray flux peaks outside the nuclear region are located near the ends 
of the observed optical gas lanes and coincident with the pockets of high FWHM emission exhibited in our SNIFS IFU observations. 
Neither the merger companion or background galaxy in this system appear to have corresponding X-ray signatures.

\subsection{Radio Analysis}

\begin{figure*}[!htbp]
\centering

\includegraphics[width=0.98\textwidth]{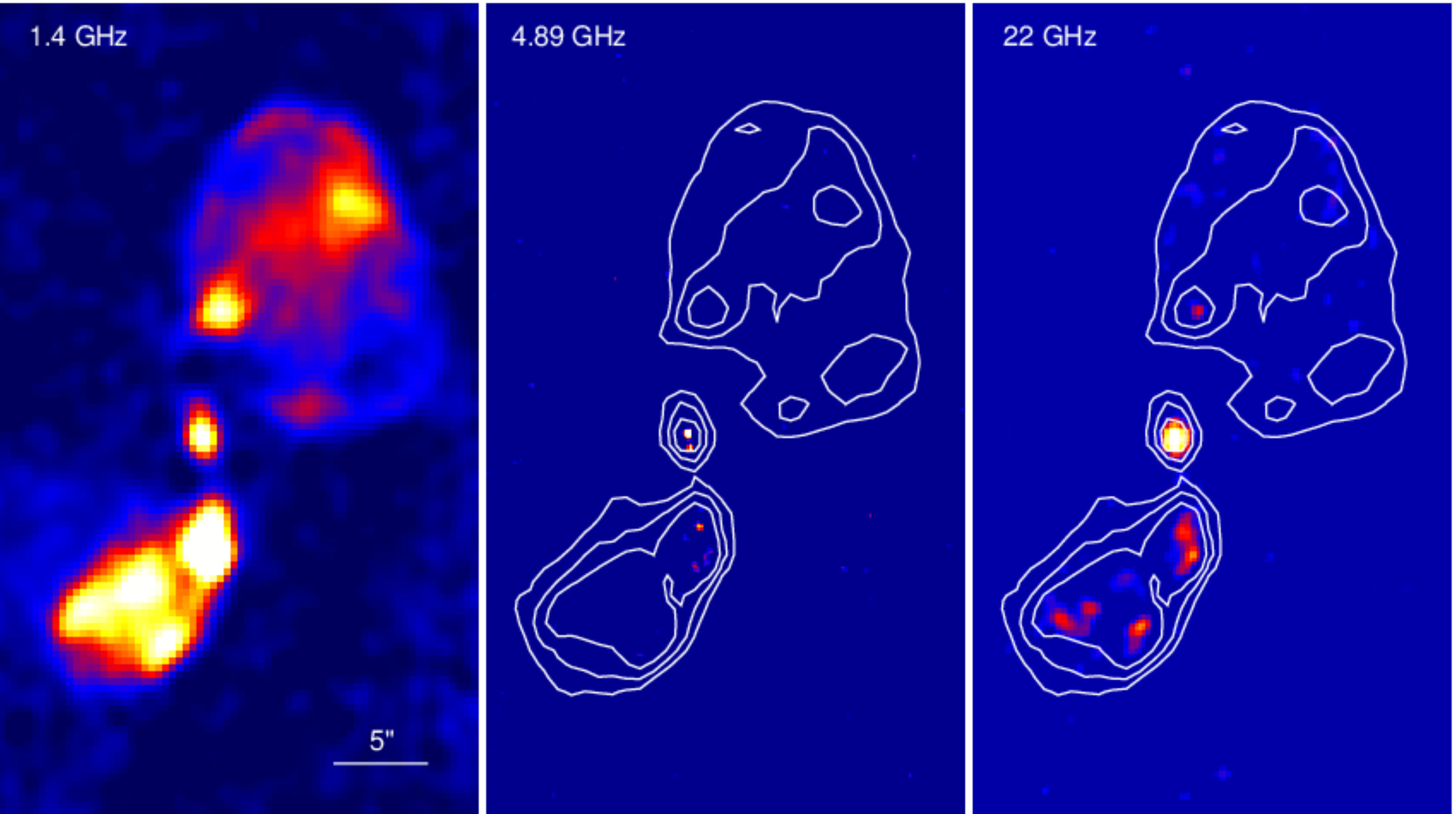}

\caption{Left to right: {\it VLA} imaging of 1.45\,GHz, 4.89\,GHz, and 22\,GHz emission in 2MASX~J0423. 
Contours of 1.45\,GHz are overlaid in the other bands for registration. North is up and east is to the left.
5$''$ corresponds to a spatial scale of $\sim$4.5\,kpc.}
\label{fig:contours}

\end{figure*}

Radio morphologies via VLA observations across 1.45\,GHz, 4.89\,GHz, and 22\,GHz bands are shown in Figure \ref{fig:contours}, and 
are largely divided into three regions; a nuclear component and two extended lobe-like components north and south of the nucleus. 
The nuclear component can be resolved into two knots of emission in the 4.89\,GHz observations, and is otherwise a single elliptical knot. 
The southern lobe is brightest directly south of the nuclear component and extends to the southeast in a fan shape primarily consisting 
of two elongated knots forking from the brightest knot. 4.89\,GHz and 22\,GHz observations show that the brightest knot in the southern 
lobe is also comprised of two elongated knots parallel with the knots at larger radii. The northern lobe is only observed in 1.4\,GHz 
and 22\,GHz observations, is brightest directly north of the nuclear component and extends northeast. The two bright knots in the 
northern lobe are connected by fainter emission and surrounded by a bubble with enhanced structure around the edges, particularly to 
the north of the brightest knots.

The spectral index $\alpha$, where the flux density $S_\nu \propto \nu^{\alpha}$, of a radio source is often a useful diagnostic 
of the nature of the emission. Flat spectral indices ($-0.5 \leq \alpha \leq 0$) typically indicate extended thermal sources, while 
steep indices ($\alpha \leq -0.7$) indicate non-thermal synchrotron radiation; a compact non-thermal source that is optically thick 
will also result in a flat spectral index \citep{Pet97}. 

\begin{figure}[!htbp]
\centering

\includegraphics[width=0.45\textwidth]{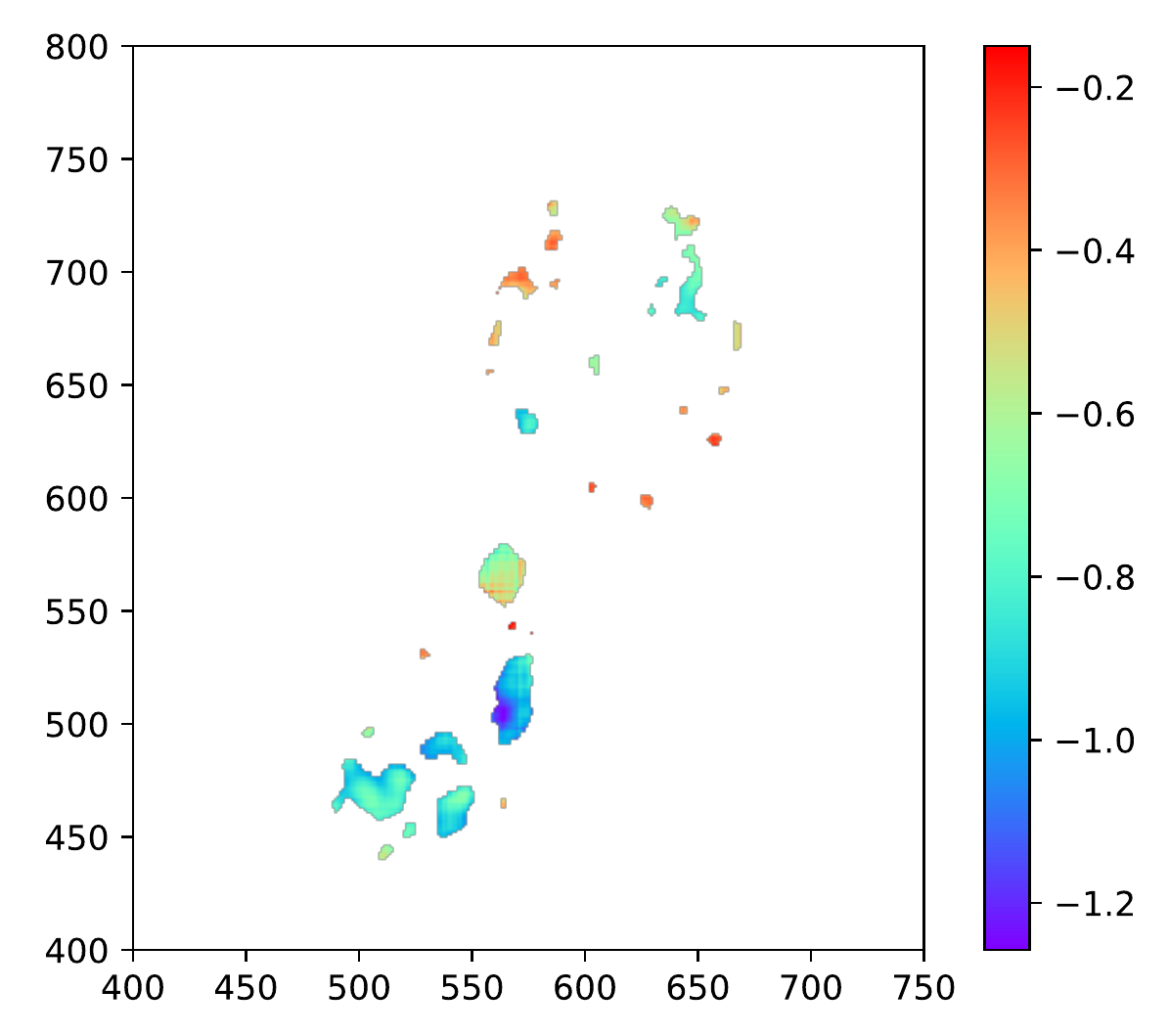}

\caption{Spectral index map based on 3$\sigma$ measurements from {\it VLA} 1.45\,GHz and 22\,GHz observations.}
\label{fig:indices}

\end{figure}

In order to create a spatially resolved spectral index map, we have reprojected the archival 1.45\,GHz image and our 22\,GHz image 
using the Astropy package \texttt{reproject}. After measuring the mean rms noise in the background of both the 1.45 and 22\,GHz 
images, we calculate the slope $\alpha$ between the 1.45~GHz and 22~GHz flux densities at every pixel where both images have 
significant emission detected at $\geq 3\sigma$. The resulting spectral index map is shown in Figure~\ref{fig:indices}, indicating 
that bright, extended regions are due to non-thermal synchotron radiation, discussed further in Section \ref{sec:dis},
with emission in the surrounding edges of the northern lobe originating from thermal sources such as shock heating. 
The nuclear knot exhibits a gradient in indices, with the southern portion that is closest to the 
optical and X-ray nucleus exhibiting a relatively flat index and the northern portion exhibiting a steep index. Again, compact 
emission near the nucleus can have a flat index because it is optically thick, or because multiple synchotron emitting regions 
are contained within the source, which we suggest is likely occurring here\footnote{Images are disparate in time by about 20 years, 
although on this timescale only the unresolved radio core emission would vary.}.

\begin{figure}[!htbp]
\centering

\includegraphics[width=0.45\textwidth]{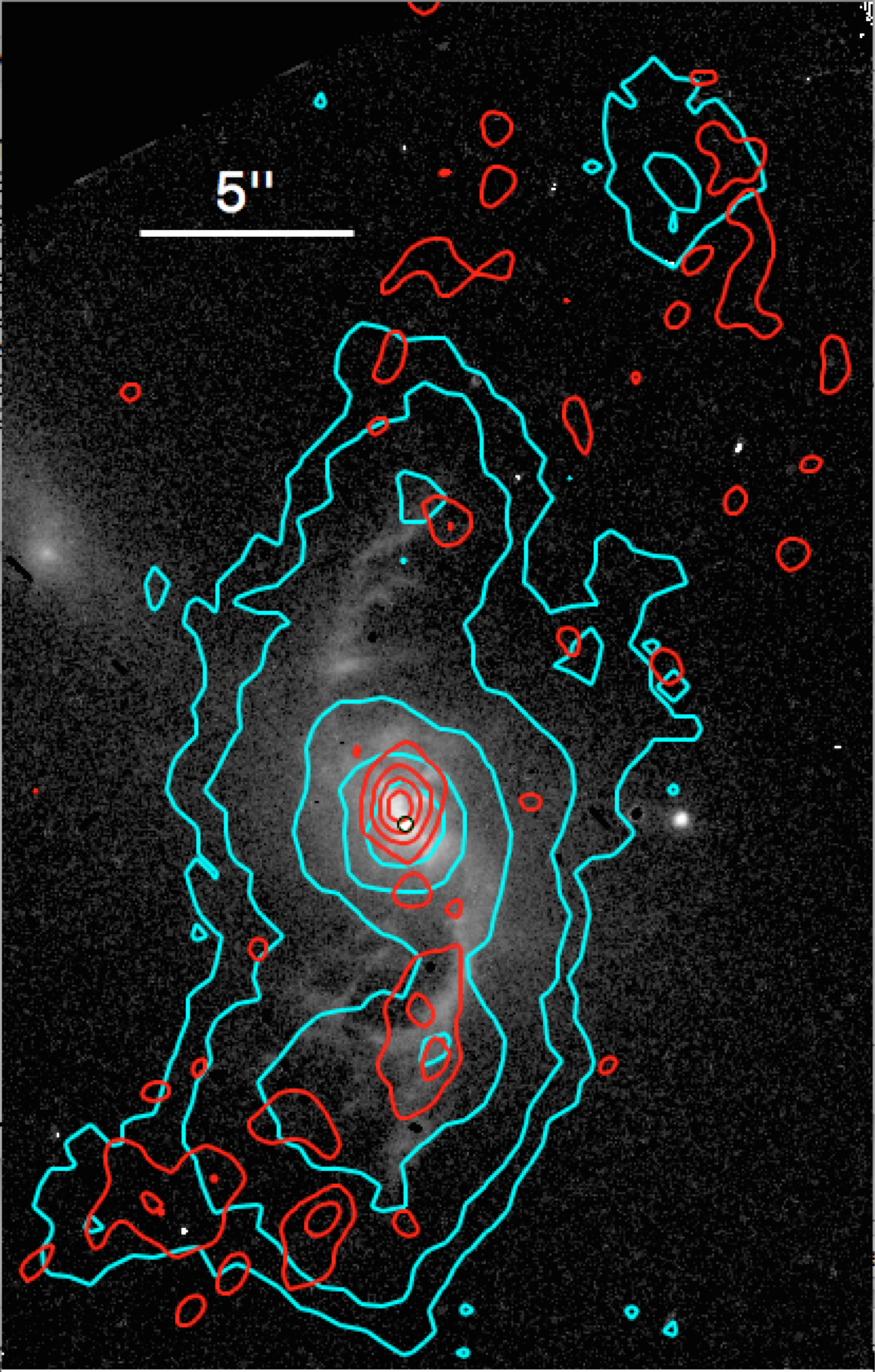}

\caption{{\it HST} F675W imaging and full bandpass {\it Chandra} 
X-ray imaging and {\it VLA} 22\,GHz emission as cyan and red contours, respectively, 
in 2MASX~J0423. A black circle represents the optical emission peak and assumed nucleus position. 
5$''$ corresponds to a spatial scale of $\sim$4.5\,kpc.}
\label{fig:optxrad}

\end{figure}

We compare the morphologies and locations of the optical, X-ray, and radio emission in Figure \ref{fig:optxrad}. Immediately, 
we observe that the nuclear radio source is not centered on the optical and X-ray nucleus. The emission is instead located 
to the north in the region where we see bright optical structures and outflows adjacent to the nucleus. South of the nucleus, 
radio emission in the brightest knot is split into two elongated regions which straddle the optical gas lane before extending 
along the path of the gas lane arc to the southeast. X-ray and radio emission also follow the optical structure to the north, 
with X-ray emission located radially interior to adjacent radio structures. Assuming that the nuclear and southern straddled 
knots are adjacent to nuclear X-ray emission, X-ray emission being radially interior to adjacent radio structures is then 
prevalent throughout the system, likely structured by the location of molecular gas lanes highlighted by the observed optical 
emission.

\section{Interpretive Analysis : Optical Kinematics}

\subsection{Rotation Kinematics}

Observations which may provide spatially resolved stellar kinematics, and thus 
determine where the observed ionized gas kinematics are affiliated 
with rotation, are unavailable. As such, we use the observed [O~II] kinematics 
in the central SNIFS field as a proxy, as large-scale [O~II] emitting gas is less likely 
to be directly interacting with ionizing radiation from the AGN. However, [O~II] in the 
north and south SNIFS fields are likely enhanced from other processes described below 
and are not included in this kinematic analysis. [O~II] structure in the center SNIFS field 
is aligned with the gas lane structure observed in optical imaging and the observed kinematics 
suggest that the north gas lane is redshifted, rotating into the plane of the sky, and the south 
gas lane is blueshifted, rotating out of the plane of the sky. 

\begin{figure}[!htbp]
\centering
\includegraphics[width=.48\textwidth]{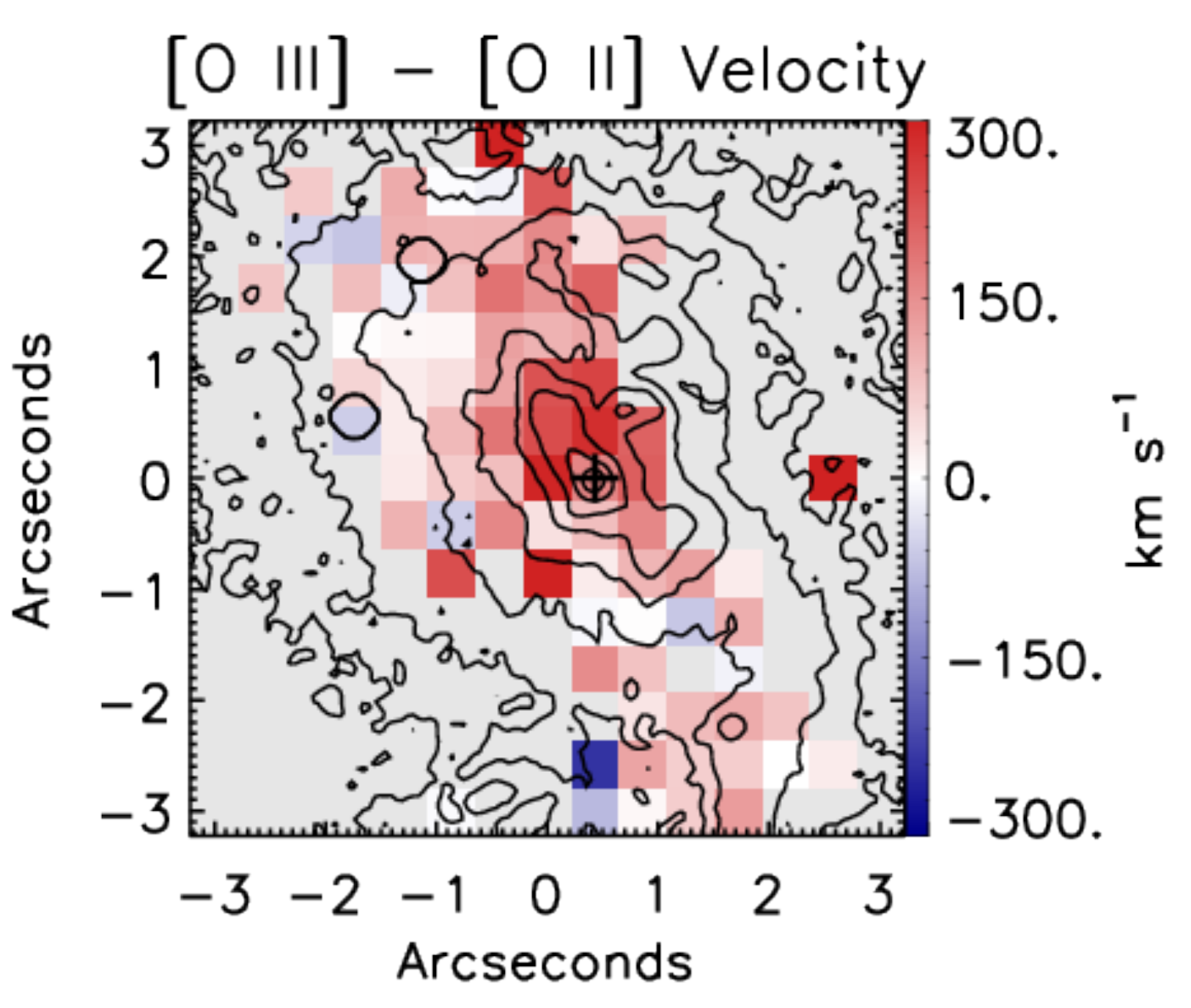}
\includegraphics[width=.48\textwidth]{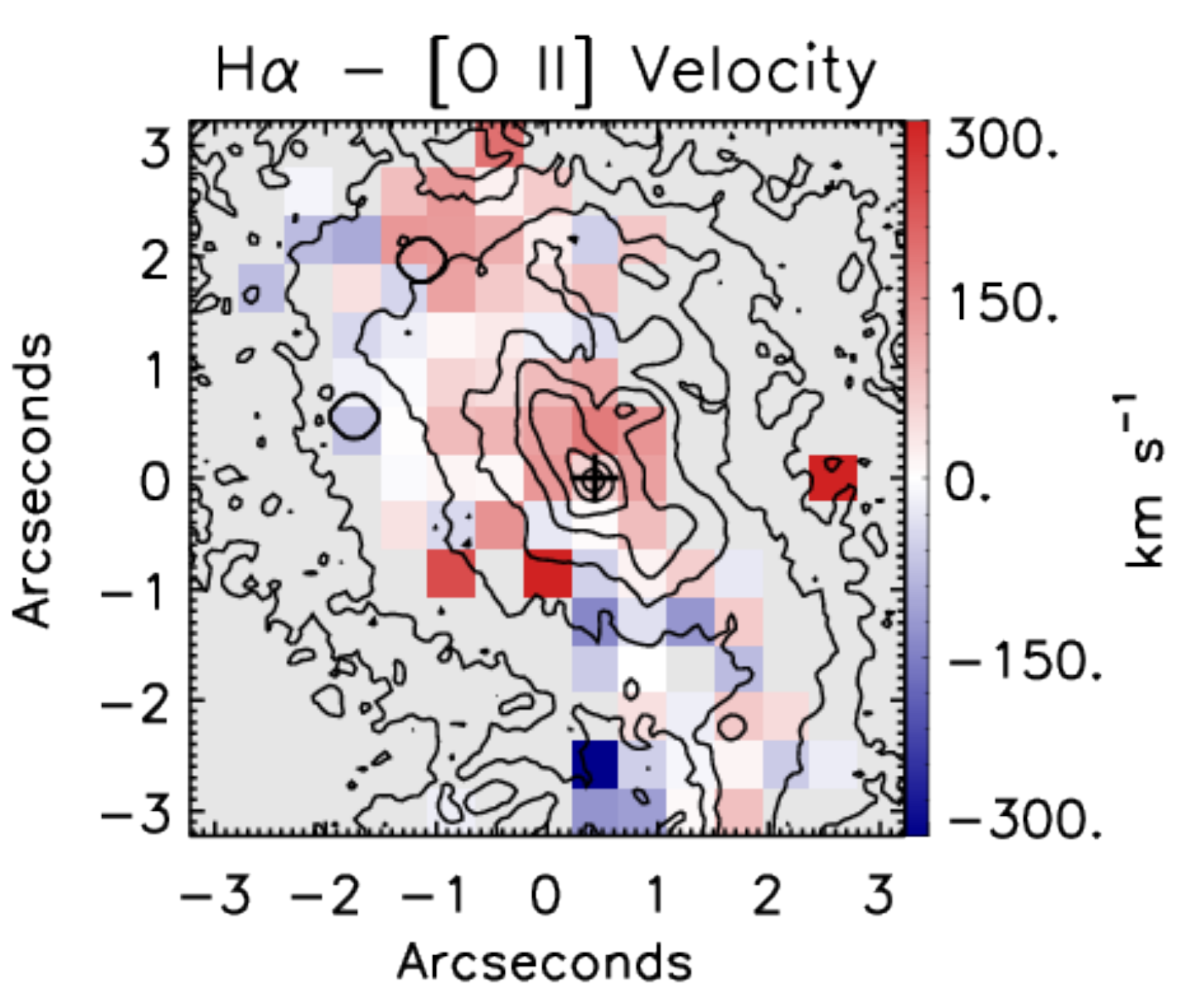}

\caption{Velocity offsets between outflowing [O~III] (top) and H$\alpha$ (bottom) and rotating [O~II] 
in the central SNIFS field of view. Velocity offsets are largest directly north of the nucleus. 
Black contours represent \emph{HST}/WFPC2 F675W imaging.}
\label{fig:vel_diff}

\end{figure}

\subsection{Radiatively Driven Kinematics}

Comparing rotation kinematics to the velocities observed in [O~III] and H$\alpha$ emitting gas, 
as shown in Figure \ref{fig:vel_diff}, we find a clear offset between the likely rotating gas 
and gas influenced by the AGN. Particularly, large offsets in [O~III] and H$\alpha$ velocities are most 
visible immediately north of the nucleus. In order to determine whether radiative driving can produce 
the observed optical ionized gas velocity offsets, we compare the gravitational deceleration to the 
radiative acceleration experienced by the gas within the inner 10 kpc of the system.

\begin{figure*}[!htbp]
\centering

\includegraphics[width=0.95\textwidth]{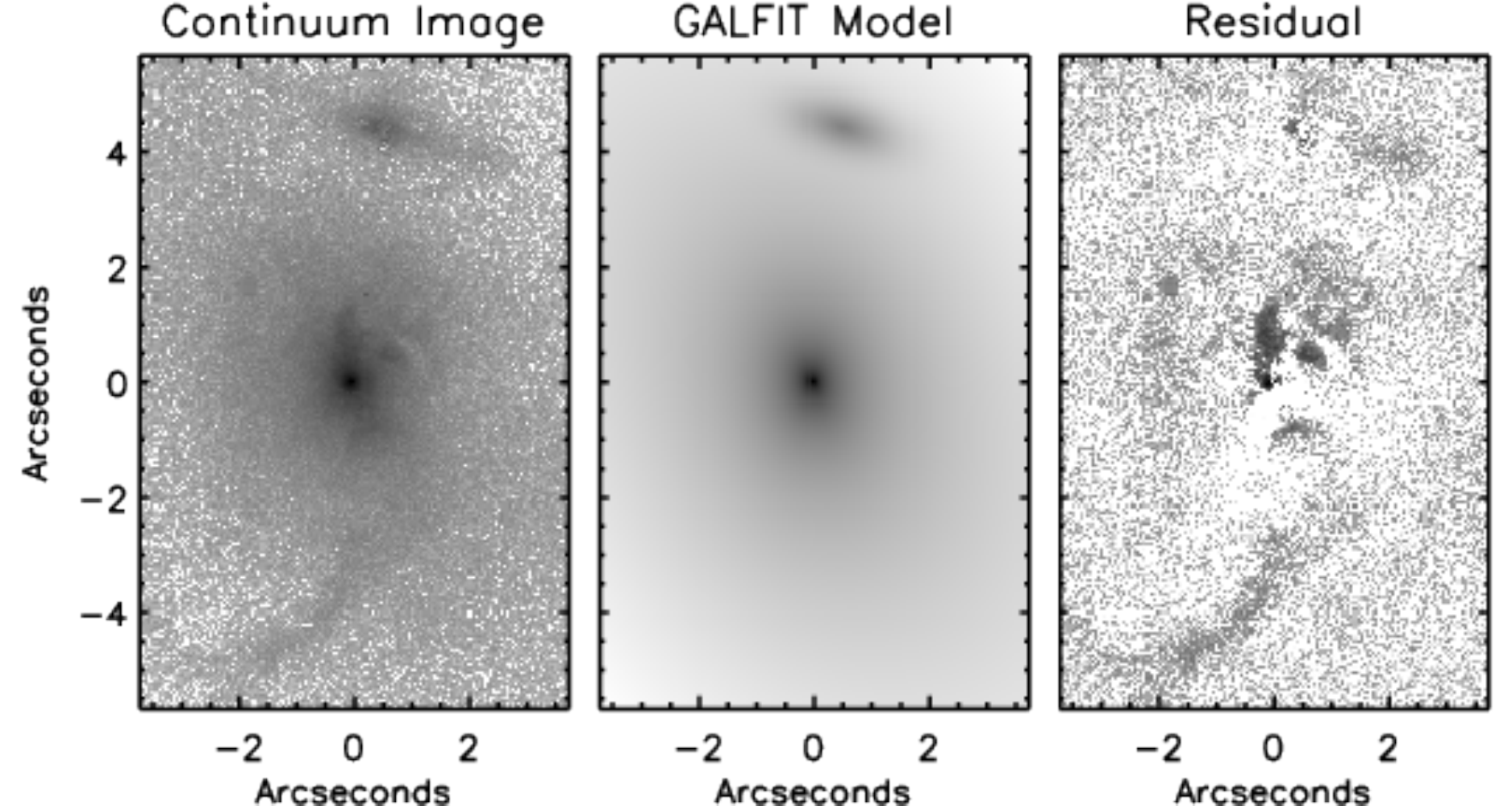}

\caption{Left: {\it HST} WFPC2/PC F814W continuum image of 2MASX~J0423. Center: best fit galaxy decomposition model (2 components + background galaxy) for 2MASX~J0423. Right: residuals
between image and model.}
\label{fig:galfit}

\end{figure*}

Using a method similar to the one used by \citet{Fis17} for Mrk 573, {\it HST} WFPC2/PC F814W imaging 
of 2MASX~J0423 was decomposed using GALFIT version 3.0.5 \citep{Pen02,Pen10} in order to measure 
enclosed mass and calculate gravitational deceleration in the galaxy as function of radius. We find 
that the best fitting model is composed of three S\'ersic components with parameters described in Table 
\ref{tab:galfit}. The original image, GALFIT model, and resulting residual map, are presented in Figure 
\ref{fig:galfit}. Component 1 can be identified as a bulge which is aligned with the spiral arm structure. 
Component 2 is disk-like and elongated in the direction of the merger companion, possibly due to merger processes. 
Component 3 is the background galaxy north of the nucleus visible in Figure \ref{fig:fig1}.

\begin{table}[h!]
  \centering
  \ra{1.3}
  \caption{GALFIT Model Results}
  \label{tab:galfit}
  \begin{tabular}{ccrrlc}
    \toprule	
    Component	& $I$		    & r$_{e}$   & $n$		    & $b/a$			& PA         \\
  				& (mag) 		& (kpc)      &			    &	            & (deg) 	 \\                                                                      
  	1       & 15.78         & 3.34     & 4.44          & 0.71          & 36.73      \\
  	2       & 15.79         & 9.48     & 0.75          & 0.48          & 67.27      \\
  	3       & 19.78         & 0.57     & 0.83          & 0.48          & 100.63     \\

    \bottomrule
  \end{tabular}
  
  \vspace{.5cm}
  \raggedright Col. (1) indicates the S\'ersic component; Col. (2) integrated $I$ band magnitude; 
  Col. (3) effective radius; Col. (4) S\'ersic index; Col (5) axial ratio ; Col. (6) 
  position angle east of north; Col. (7) fraction of the integrated flux from each component.

\end{table}

\begin{figure}[!htbp]
\centering

\includegraphics[width=0.45\textwidth]{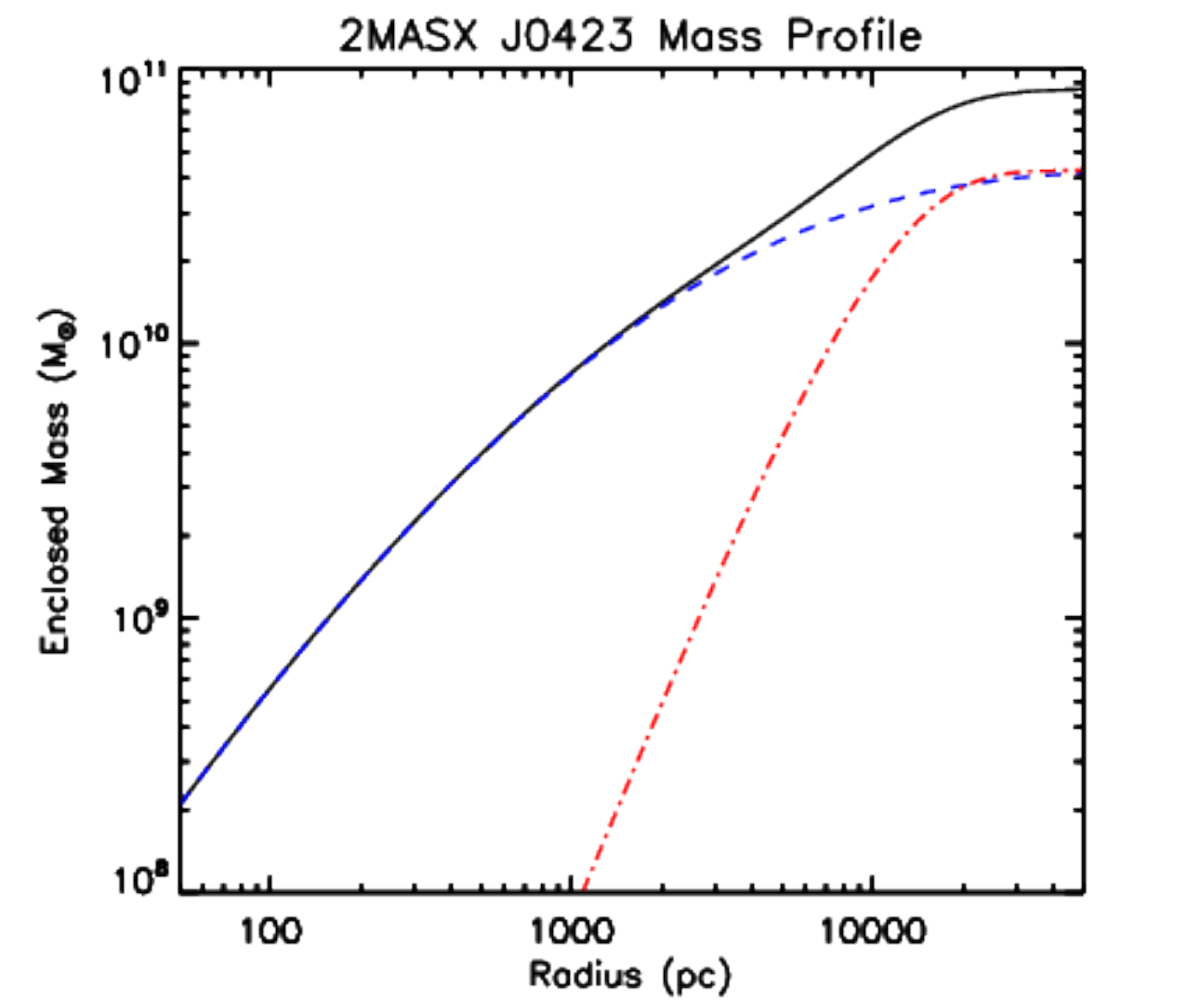}

\caption{Mass distribution profiles for each component of 2MASX~J0423 in our model. Blue dashed,
red dotted-dashed, and black solid lines represent the inner and outer components, 
and the sum of the two components, respectively. The grey line represents the total 
mass distribution in Mrk 573 from \citet{Fis17}. Our radial
mass distribution is calculated using the expressions from \citet{Ter05} 
assuming a mass-to-light ratio of 5.}
\label{fig:mass_profile}

\end{figure}

The radial mass distribution of these three components, shown in Figure \ref{fig:mass_profile}, was 
calculated using the expressions from \citet{Ter05}. The S\'ersic profile is given by the following 
expression \citep{Pen10}:

\begin{equation}
\Sigma(r) = \Sigma_e exp [ -\kappa ((\frac{r}{r_e})^{1/n} - 1)]
\end{equation}

where $\Sigma(r)$ is the surface brightness, $\Sigma_e$ is the surface brightness at the effective radius, 
$\kappa$ is a constant that depends on $n$, the index of the S\'ersic profile, and $r_e$ is the effective radius.

The value $\Sigma_e$ is calculated using the equation:
\begin{equation}
F_{tot} = 2 \pi  r_e^2  \Sigma_e e^\kappa n \kappa^{-2n} \Gamma(2n) q/R(C_0,m)
\end{equation}

where $\Gamma$ is the gamma function, $q=b/a$ is the axial ratio of the S\'ersic component 
and $R(C_0,m)$ represents deviations from a perfect ellipse \citep{Pen10}. This term has a 
value of the order of unity and is disregarded in our calculations.

Following Equation~4 in \citet{Ter05} we have that the mass density of a S\'ersic component is given by the following expressions:

\begin{equation}
\rho(r) = \rho_0 (\frac{r}{r_e})^{-p} e^{\kappa} e^{(-\kappa (\frac{r}{r_e})^{1/n})}
\end{equation}

\begin{equation}
p = 1 - \frac{0.6097}{n} + \frac{0.05563}{n^2}
\end{equation}

\begin{equation}
\rho_0 = \frac{M}{L} \Sigma_e \kappa^{n(1-p)}  \frac{\Gamma(2n)}{(2 r_e \Gamma(n(3-p))} 
\end{equation}

where $\frac{M}{L}$ is the mass to light ratio, assumed to be 5, per \citet{Ter05}. Notice that due 
to a difference in notation between \citet{Pen10} and \citet{Ter05}, the expression for $\rho(r)$ has 
an addition $e^{\kappa}$ term. Also, the expression for $p$ corresponds to $n$ values in the 
range $0.6<n<10$. 

Finally, we have from Equation~A2 in \citet{Ter05} that the mass profile is given by:

\begin{equation}
M(r) = 4 \pi \rho_0 r_e^3 n \kappa^{n(p-3)} \gamma(n(3-p),Z)
\end{equation}

where $\gamma(n(3-p),Z)$ is the incomplete gamma function and $Z$ is given by $Z = \kappa (\frac{r}{r_e})^{1/n}$. 
Using Equation 10 of \citet{Ter05}, we can calculate the enclosed mass at a given radius, as shown in Figure 
\ref{fig:mass_profile} and thus determine the gravitational deceleration at said radius. 

In order to determine whether the emission-line gas can be radiatively
accelerated in situ, we used the radiation-gravity formalism detailed in
\citet{Das07} and \citet{Fis17}. Assuming an azimuthally symmetric distribution, 
velocity as a function of radial distance, $v(r)$, in units of km s$^{-1}$ and pc, 
can be written as:

\begin{equation}
v(r) = \sqrt{\int_{r_{1}}^{r}\big[6840L_{44}\frac{\mathcal{M}}{r^{2}} - 
8.6\times10^{-3} \frac{M(r)}{r^{2}}\big]dr},
\end{equation}

where $L_{44}$ is the bolometric luminosity, $L_{bol}$ in units of 10$^{44}$ ergs s$^{-1}$,
$\mathcal{M}$ is the Force Multiplier, or ratio of the total photo-absorption
cross-section to the Thomson cross-section, $M(r)$ is the enclosed mass at the
distance $r$, determined from the radial mass distribution described above,
and $r_{1}$ is the launch radius of the gas. Calculating $M(r)$ required solving an 
incomplete Gamma function, hence we determined it at 10 pc intervals over a
range of 10 pc $<$ r $<$ 10 kpc. We derived an expression for the
enclosed mass as a function of r, in each 10 pc interval from r$_1$
to r$_2$, using a power law of the form $M(r) = M(r_{1})\times (\frac{r_{1}}{r_{2}})^{\beta}$.
We were then able to solve for $v(r)$ analytically, by integrating
within each interval, using an average $\beta$ of 0.8.

We estimate the bolometric luminosity of the AGN in 2MASX~J0423 using the intrinsic hard X-ray luminosity reported 
in \citet{Ric17} of L$_{2-10\,keV}$ = 6.6$\times$10$^{43}$ erg s$^{-1}$ and a bolometric correction of
30 \citep{Awa01}, for a bolometric luminosity of L$_{bol}$ = 2.0$\times$10$^{45}$ erg s$^{-1}$. Additionally, we can 
estimate the bolometric luminosity from the integrated [O~III] luminosity across the central 
SNIFS field, L$_{[O~III]}$= 1.04$\times$10$^{42}$ erg s$^{-1}$ and using a bolometric correction of 3500 \citep{Hec04}, 
for a similar bolometric luminosity of L$_{bol}$ = 3.6$\times$10$^{45}$ erg s$^{-1}$. Assuming that the AGN 
is radiating at Eddington, (i.e. L$_{bol} = $L$_{Edd}$), we can use the calculated bolometric luminosities 
for 2MASX~J0423 to estimate a range in minimum SMBH masses of M$_{BH} \geq$ 1.6 - 2.9 $\times$10$^{7}$ M$_{\astrosun}$. 

Photoionization models with log$U$ $\approx$ -2 - -2.5 predict that O$^{+2}$ (i.e. [O~III]) is the peak ionization 
state for oxygen \citep{Fer83}, and therefore can be used to constrain the physical conditions in the [O~III] 
emission-line gas. At this ionization, Cloudy predicts $\mathcal{M}$ = 3300 at the ionized face of an illuminated 
slab and we use this value to solve for $v(r)$. 

\begin{figure}[!htbp]
\centering

\includegraphics[width=0.45\textwidth]{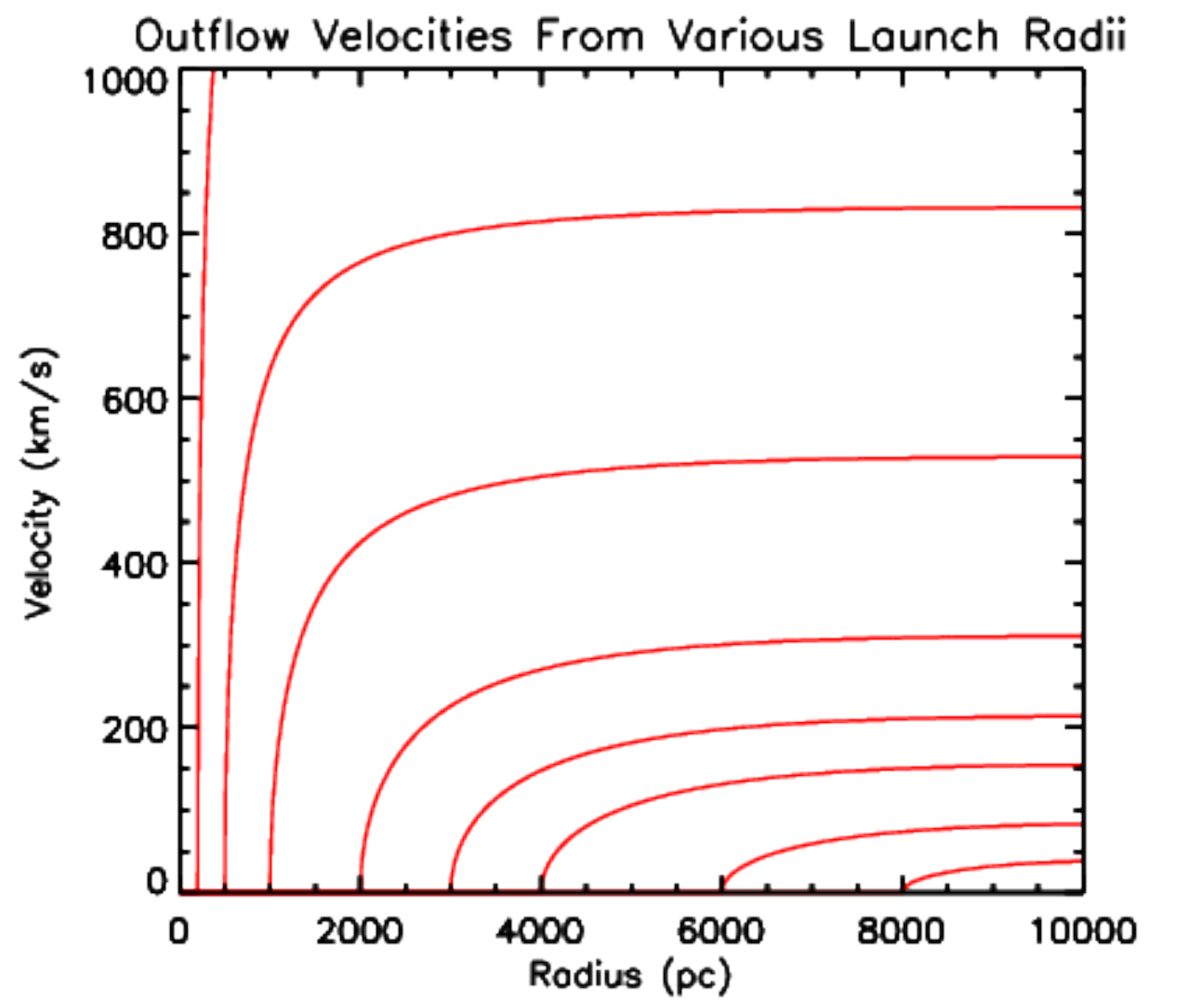}

\caption{Velocity profiles for knots of [O~III] gas in 2MASX~J0423 launched from various 
radii (all generated assuming $\mathcal{M}$ = 3300), in the absence of interaction with 
an ambient medium. Based on these results, radiatively accelerated gas can be driven 
throughout the inner 10\,kpc.}
\label{fig:vel_profile}

\end{figure}

Using our kinematic model to generate velocity distributions for knots of emission line gas 
launched from various radii, we find that radiative driving should be successful throughout 
the inner 10\,kpc of the system, which encompasses the entirety of the observed optical emission.
However, while radiative driving may explain the kinematic offsets between emission lines 
throughout the system, the high FWHM gas and adjacent gas exhibiting large velocity offsets observed near 
the ends of the optical gas lanes cannot be produced by radiative driving at these distances. As such, 
an additional mechanism must contribute to the observed kinematics.


\subsection{Mechanically Driven Kinematics}
One possibility that may explain the observed kinematics at large radii is that AGN winds driven from 
smaller radii are colliding with high density gas lanes further out. As shown in Figure \ref{fig:vel_profile}, 
continuous acceleration of optical clouds across the inner few hundreds of parsecs from the nucleus 
should result in knots of ionized-gas traveling with velocities v $>$ 1000 km s$^{-1}$, however such large 
velocities are not observed in this system. This may be due to the observed optical ionized gas outflows only tracing 
the surface of molecular gas reservoirs adjacent to the AGN ionizing radiation field. Unbound from the molecular gas 
reservoirs, outflows of optically emitting gas expand thermally and are further ionized into outflows of X-ray emitting gas.  
Possessing a large filling factor, this X-ray wind is likely be diffuse and difficult to detect until it comes in contact 
with high-density molecular gas lanes at larger radii. The X-ray winds then impart their energy into the gas lanes, which 
compress and shock the lane material, driving it laterally out of the plane of the disk, i.e. pancaking, producing the 
observed large velocity gradients, localized high FWHM kinematics, and LINER-like ionization diagnostics in regions near 
the ends of the observed gas lanes.

From our observations, the X-ray morphology across 2MASX~J0423 also mimics the gas lane structures observed in our optical 
observations and is concentrated near the high FWHM kinematics observed in the optical. This suggests that the observed 
X-ray emission is likely continuum emission produced thermally via shocks. This is supported by our SED fit to the X-ray 
emission in the southern lobe, where we found the emission to best resemble a soft thermal plasma model with 
kT $= 0.97^{+0.106}_{-0.133}$~keV. Emission peaking at $\approx$ 1 keV is consistent with a shock velocity of 
approximately 1000 km s$^{-1}$ (e.g. \citealt{Kra00b}) and the FWHM of the optical gas near the point of impact in the 
southern gas lane.

We model the constant energy injection rate going into the energy-conserving pancaked region following 
work by \citet{Nes06}, assuming the mechanically driven kinematics are expanding into a low-density medium, $n_0$.  
This provides as a serviceable upper bound for an energy injection rate into the gas lane from the X-ray 
wind. Assuming minimal radiative losses, their Equation 3 defines a constant energy injection rate as

\begin{equation}
\dot{E} \approx 1.5 \times (\Omega/4\pi) \times 10^{46} R^{2}_{10}v^{3}_{1000}n_{0} \,erg\,s^{-1}
\end{equation}

\noindent where the extent of the mechanically driven gas $R_{10}$ is in units of 10 kpc, $v_{1000}$ 
is in units of 1000 km s$^{-1}$, and $\Omega/4\pi$ is the covering factor of the driven gas in steradians. 
From the approximate measurement of FWHM$\sim$1000 km s$^{-1}$ for the broad emission line gas in the southern 
field of view, we measure a maximum outflow velocity defined as 1/2 $\times$ the full width at zero maximum (FWZM), 
approximately the 3$\sigma$ velocity offset from the centroid of the broad FWHM region, of $v = 1275$\,km s$^{-1}$.
Using $R_{10}$ = 2.4, $v_{1000}$ = 1.275, $\Omega/4\pi$ = 1.0, and $n_0$ = 0.5, we calculate an energy 
injection rate of $\dot{E}$ = 8.95$\times$10$^{44}$\,erg s$^{-1}$. This is likely an upper limit because 
covering factor and surrounding medium density are likely overestimated.

Similar calculations for the X-ray wind impacting the northern gas lane produce an upper limit 
kinetic energy injection rate of $\dot{E} = 6.7\times$10$^{43}$\,erg s$^{-1}$, using $R_{10}$ = 1.25 kpc, 
$v_{1000}$ = 0.83, derived by converting an average FWHM of the broader line region in the north 
of 650 km s$^{-1}$ to a FWZM of 830 km s$^{-1}$, and $n_0$ = 0.5.

Interestingly, we also note that the brightest knots of radio emission are also colocated with the impact 
regions and resultant thermal X-ray emission in 2MASX~J0423. For example, bifurcated radio knots surround the 
southern optical gas lane as shown in Figure \ref{fig:optxrad}, existing adjacent but laterally exterior to 
thermal X-ray emission and shocked, high FWHM optical emission. Following \citet{Zak14}, this may be evidence 
that radio emission observed in this radio-quiet AGN is largely a by-product of relativistic particles accelerated 
in the shocks caused by the AGN driven outflows. Although {\it HST} imaging does not trace gas lanes to larger 
radii, potentially thermal X-ray structures continue to trace the general path of the gas lanes, with cospatial, 
localized pockets of additional radio structure. This suggests that pressure from the X-ray winds colliding 
with gas lanes at larger radii continues to compress the molecular gas, forming shocks which in turn produce 
cosmic rays. The combination of cosmic rays and enhanced magnetic field lines in the compressed gas allow for 
the in situ formation of the observed radio structure, forming the observed pseudo-jet morphology and filling up 
the evacuated cavity at lower frequencies.

This production of radio emission is likely to be similar to processes in supernova remnants,
where shocks initiated by the supernova wind accelerates particles and then produces synchotron radiation 
(\citealt{Zak14} and references therein). Therefore the efficiency of converting the kinetic energy of 
the outflow wind into radio synchrotron emission is assumed to be similar between quasar-driven and 
starburst-driven winds, where an efficiency of 3.6$\times$10$^{-5}$ is required to convert kinetic energy 
into 1.4\,GHz radio luminosity.

From the archival {\it VLA} 1.4\,GHz observations, we measure fluxes of the radio knots adjacent 
to the high FWHM optical regions north and south of the AGN and calculate luminosities of 
$\nu L_{\nu}$[1.425\,GHz] = 2.65 $\times$10$^{39}$\,erg s$^{-1}$ and 2.94 $\times$10$^{40}$\,erg s$^{-1}$, 
respectively. Comparing these values to the calculated energy injection rates required to produce the 
laterally expanding optical kinematics described above, we measure efficiencies of 3.95$\times$10$^{-5}$ and 
3.28$\times$10$^{-5}$ for the north and south regions, respectively, which are consistent with that 
required to produce synchrotron emission via starburst-driven winds.

The differences in flux and FWHM in the north and south gas lanes is likely due to the interaction 
between AGN radiation and host disk material immediately north of the AGN nucleus where the largest velocity 
offsets in Figure \ref{fig:vel_diff} are observed. Here, we see large concentrations of optical, X-ray, and 
radio emission which suggests that the mechanical processes observed at large radii are also occurring here. 
However, as the impact of the wind onto the host gas lane is likely at a more oblique angle than the impacts 
observed at larger radii, it is difficult to derive observed optical energy injection rates to test the 
relationship between injected energy and observed 1.4\,GHz radio luminosity

\section{Discussion}
\label{sec:dis}

From our analysis, we attribute the observed optical kinematics to three distinct distributions: 

1) A general rotation pattern, more easily observed in lower ionization potential emission lines (i.e. [O~II]). 
The northern lane is redshifted, rotating into the plane of the sky, and the southern lane is blueshifted, rotating 
out of the plane of the sky.

2) Radially outflowing kinematics traveling away from the nuclear continuum peak via radiative driving, observed as velocity 
differences between emission lines with different ionization potentials. Velocities for [O~III] are redshifted compared 
to other emission lines both north and south of the AGN nucleus.

3) Laterally outflowing kinematics approximately perpendicular to the radial outflows via mechanical driving at radial 
distances greater than 5$''$, observed both as velocity gradients across lane structures and high FWHM emission lines. 
The largest gradients in both the north and south fields are radially colocated with high FWHM gas regions at the end of 
the optical gas lane structures, which suggests the velocity shifts are the expanding edges of the high FWHM region.

With continuous outflows traveling radially outward potentially through the entire observed optical system, 2MASX~J0423 has 
outflows of radiatively-driven optical ionized gas, defined as radial velocities offset from systemic, larger than 
those observed in a majority of nearby Type 2 AGN in spatially resolved kinematic studies which typically exhibit an maximum 
outflow distances of 1 - 2\,kpc \citep{Fis13,Fis18}. Our measurements suggest that the bolometric luminosity of this system 
is not exceptional. However, the relatively small stellar mass within the host for this system allows outflows to exist 
throughout the system, assuming an absence of interactions with the ambient medium (i.e. lanes of high density molecular gas). 
As such, we suggest that the mass ratio between the stellar bulge of a galaxy and its SMBH is critical in determining if 
feedback works. Extending this to systems near Cosmic Noon, where bulges were typically defined, AGN are likely to be more 
successful in the smaller gravitational potentials of their host galaxies. 

Our analysis also suggests that the extent of optical, radially driven, high-velocity outflows does not define 
the maximum radius of impact for AGN feedback in a host galaxy. In previous high-resolution studies of [O~III] 
kinematics in luminous QSO2s \citep{Fis18}, ionized gas kinematics exhibited high-velocity radial outflows at small 
radii, and returned to systemic velocity or followed a rotation pattern at larger radii. However, this gas at large radii
is likely experiencing some influence by the AGN as kinematics at these distances often exhibited large FWHM values 
indicative of some kinematic disturbance \citep{Bel13,Ram17}. If these targets follow a similar scenario to  
2MASX~J0423, the optical ionized, radially outflowing gas at small radii in these systems is likely further ionized 
to become an X-ray wind, which continues out to further distances than what is observed via optical and near-infrared 
measurements and impacts the previously rotating, photoionized host disk gas. As it is currently extremely difficult to 
obtain spatially resolved X-ray spectroscopy and measure the kinematics of the diffuse outflows in the X-ray, we can 
instead use signatures of the wind disrupting ambient host material, in the form of large FWHM optical 
and infrared emission-lines, and by-products of thermal X-ray emission, and extended synchotron radio emission, 
to act as a proxy for the extent of outflowing AGN winds interacting with their hosts in radio-quiet AGN.


It is notable that when radio structures are well aligned with optical ionized gas structures, they must be approximately 
aligned with high density gas lanes in the system, as the optical structures have been shown to be the ionized surfaces 
of molecular gas reservoirs (i.e. gas lanes; \citealt{Fis17}). If the observed radio emission was from a radio jet, 
to strongly interact with the southern gas lane of 2MASX~J0423, as observed at a distance of $\approx$ 4.65 kpc from 
the nucleus, and assuming a lane scale height of 500 pc, it would need be aligned within 3$^{\circ}$ of the lanes. Given a 
random alignment between jet axis and plane, the probability of successful alignment within 3$^{\circ}$~is $\sim$ 5\%. 
Previous analysis of radio and optical alignment \citep{Fal98,Kuk99} show that extended radio emission is often intertwined 
or aligned with the optical structure in a similar fashion to 2MASX~J0423. As such, having radio emission and adjacent gas 
ionization and non-rotating kinematics be due to a radio jet would require the axis of the jets observed in these targets 
to consistently be near perfectly aligned with the plane of the disk or gas lanes. 

We find several targets share similarities to 2MASX~J0423, including SDSS~J1356+1026, a merging system housing a Type 2 AGN 
at z = 0.123. Optical spectroscopy for the system largely consists of emission from the two nuclei and a kinematic 
$"$bubble$"$ which produces a ring-like [O~III] structure in the two-dimensional spectrum, extending 8-12\,kpc from the AGN 
nucleus, and exhibits velocities $\sim\pm$230 km s$^{-1}$ from systemic \citep{Gre12}. These kinematics are attributed 
to a shell of gas expanding into the intergalactic medium, similar to the mechanical driving that we observe in the 
optical kinematics of the southern gas lane of 2MASX~J0423, where [O~III] gas is fainter near the X-ray impact point and 
brighter in the expanding mechanically driven gas. Additionally, {\it Chandra} X-ray observations show similar 
relationships between extended structures in the optical and soft X-ray \citep{Gre14,Com15}, such that the bright 
optical bubble feature is colocated with the brightest extended X-ray emission. These similarities 
suggest that the kinematics at multi-kpc distances in SDSS~J1356+1026 are likely due to shocked gas associated with 
AGN driven outflows. High-resolution, high-sensitivity {\it VLA} radio continuum observations do not yet exist for this 
target, however detecting extended radio emission intertwined with the extended X-ray and optical structure would 
confirm similar processes to those in 2MASX~J0423 are in play. 

Another target with striking similarities to 2MASX~J0423 which does have optical, X-ray, and radio imaging 
is NGC 4258; a Seyfert galaxy believed to possess large radio jets driven by the AGN that lie within the plane of the star-forming 
spiral host galaxy and have therefore interacted strongly with the interstellar medium \citep{Van72}. 
The radio structures have been referred to as $“$anomalous arms$”$ and several studies of this source discuss 
the jet-gas interaction, how matter in the host galaxy can be entrained, heated, or shocked by the jet, and how 
the galaxy's spiral stellar structure responds to such a disruption (e.g., \citealt{Mar89,Cec00,Lai10}). 
Imaging of this target exhibits two prominent arms north and south of the nucleus in the optical which spiral outward 
and are dominated by young, bright stars, with the anomalous arms visible in H$\alpha$, soft X-ray, and radio 
continuum \citep{Cec92,Yan07}. Spatially resolved X-ray spectroscopy reveals the extended soft X-ray emission for the 
entire galaxy to be well described by thermal components, suggesting that the gas is heated by shocks traveling at 
velocities similar to those implied by optical kinematic measurements. Figure 7b in \citet{Cec95} orients the radio 
structure to a position that is remarkably similar to 2MASX~J0423 and exhibits a very similar radio morphology; a bright 
radio structure immediately above the nucleus with a fainter structure at large radii and a bright radio structure below the nucleus 
at large radii, with large FWHMs present in H$\alpha$ when not extinguished by the host disk. The northwest radio arm in NGC 4258 is 
also bifurcated, similar to the south radio arm of 2MASX~J0423. We found the radio structure to separate around the 
optical gas lane structure, beginning near the X-ray impact region that exhibits high FWHM optical lines. Assuming a similar 
scenario in NGC 4258, forking in the northwest lane of NGC 4258 also begins near an apparent X-ray impact region that exhibits 
high FWHM optical lines, which suggests that the AGN feedback processes in NGC 4258 may also be related to winds. 

Findings from 2MASX~J0423 also provide further understanding in our previous kinematic analysis of Mrk 573 in \citet{Fis17}, 
where we initially decoded non-rotating ionized-gas kinematics along the axis of the projected NLR as radiative, in situ 
acceleration of material residing in the host disk. At the furthest extents of non-rotating gas, however, there are 
also co-radial knots of ionized gas which continue to follow rotation as measured by stellar kinematics. Applying the 
kinematic explanation we derive for 2MASX~J0423, the non-rotating kinematics at distances $>$ 500 pc are likely the 
by-product of winds launched from smaller radii. These winds run into dust lanes, shocking the dense medium, forming 
in situ X-ray and radio emission, and creating the same pancaking optical kinematic profile we observe in 2MASX~J0423 
which disturbs the intrinsic, rotating ENLR gas. This scenario is supported through several observations across similar 
wavebands to those in this analysis. {\it HST} [O~III] long-slit spectroscopy of the arcs in Mrk 573 from \citet{Fis10} 
exhibit velocity gradients across each of the three arcs illuminated by the AGN, similar to the gradients we see 
in 2MASX~J0423. {\it Chandra} X-ray flux distributions in \citet{Pag12} are aligned such that photoionized line 
emission traces the linear, optical outflowing region and the interior side of the optical arcs. Radio structures 
are also aligned with the optical gas, where high-resolution 8.49\,GHz observations show emission located along the 
a linear feature with the largest adjacent H$_2$ reservoir (i.e. densest spiral arm) and in between the linear outflows 
and pancaking arcs \citep{Fal98}. Lower resolution 5\,GHz emission forms a reservoir interior to the arcs, filling the 
area between the outflows and gas lane arcs \citep{Pag12}. Ionized gas kinematics in several other AGN (e.g. Mrk 34, 
NGC 3516, NGC 3393; \citealt{Fis13}) also exhibit velocity gradients centered on the rotational velocity kinematics at 
large radii that are likely similar examples of X-ray winds pancaking into molecular gas lanes.

\section{Conclusions}
\label{sec:conclusion}
We analyzed SNIFS optical IFU spectroscopy and imaging from {\it HST}, {\it Chandra}, and {\it VLA} observatories 
to measure optical ionized gas kinematics and ionization sources, X-ray spectral diagnostics, radio spectral indices, 
and the morphological relationship between the three wavebands. Our conclusions are as follows:

1) Radiative driving of optical, AGN-ionized gas is possible throughout the inner 10\,kpc of this system, encompassing 
the entirety of the observed optical emission. Successful driving of [O~III] gas out to multi-kpc scales in this system is due 
to a less concentrated mass profile in the host galaxy than other AGN of similar luminosity. As more luminous AGN with more 
massive hosts exhibit less productive radial outflow, we suggest lower host masses may be critical for galaxies to regularly 
quench star formation.

2) Optical high FWHM gas near systemic or following rotational patterns  observed at large radii in this system cannot be produced 
by radiative driving and therefore do not trace the maximum extent at which optical gas can be driven by the AGN. Instead, 
such a signature likely measures the extent of outflows of X-ray emitting gas, launched from smaller radii, driving into high density 
gas lanes at large radii and laterally pushing the ambient lane gas.

3) Outflows driving into these dense lanes also produce shocks that heat and compress the lane gas, producing a significant 
amounts of thermal X-rays and cosmic rays, which in turn allow for the in situ formation of radio knots. Therefore observed 
extended soft X-ray and radio continuum emission in this target, and likely other radio-quiet AGN, may often be by-products 
of radiatively driven winds originating at smaller radii.

\acknowledgments
The authors thank the anonymous referee for their helpful comments that improved the clarity of this work.

TCF thanks Mitchell Revalski, Beena Meena, Francisco Martinez, and Garrett Polack for insightful discussions, and 
Nathan Secrest for guidance in astrometric alignment between images. 

KLS is grateful for support from the National Aeronautics and Space Administration through Einstein Postdoctoral Fellowship Award
Number PF7-180168, issued by the Chandra X-ray Observatory Center, which is operated by the Smithsonian Astrophysical 
Observatory for and on behalf of the National Aeronautics Space Administration under contract NAS8-03060.

Based on observations made with the NASA/ESA Hubble Space Telescope, obtained from the data archive at the Space Telescope 
Science Institute. STScI is operated by the Association of Universities for Research in Astronomy, Inc. under NASA contract 
NAS 5-26555. Support for this work was provided by NASA through the Space Telescope Science Institute, 
which is operated by AURA, Inc., under NASA contract NAS 5-26555. 

The scientific results reported in this article are based in part on observations made by the Chandra X-ray Observatory.

The National Radio Astronomy Observatory is a facility of the National Science Foundation operated under cooperative 
agreement by Associated Universities, Inc.

\bibliographystyle{apj}             
\bibliography{apj-jour,main}       


\clearpage

\appendix
\section{Additional Optical Emission Line Measurements With SNIFS}
\label{sec:appendix}

\begin{figure*}[!htbp]
\centering
\includegraphics[width=0.31\textwidth]{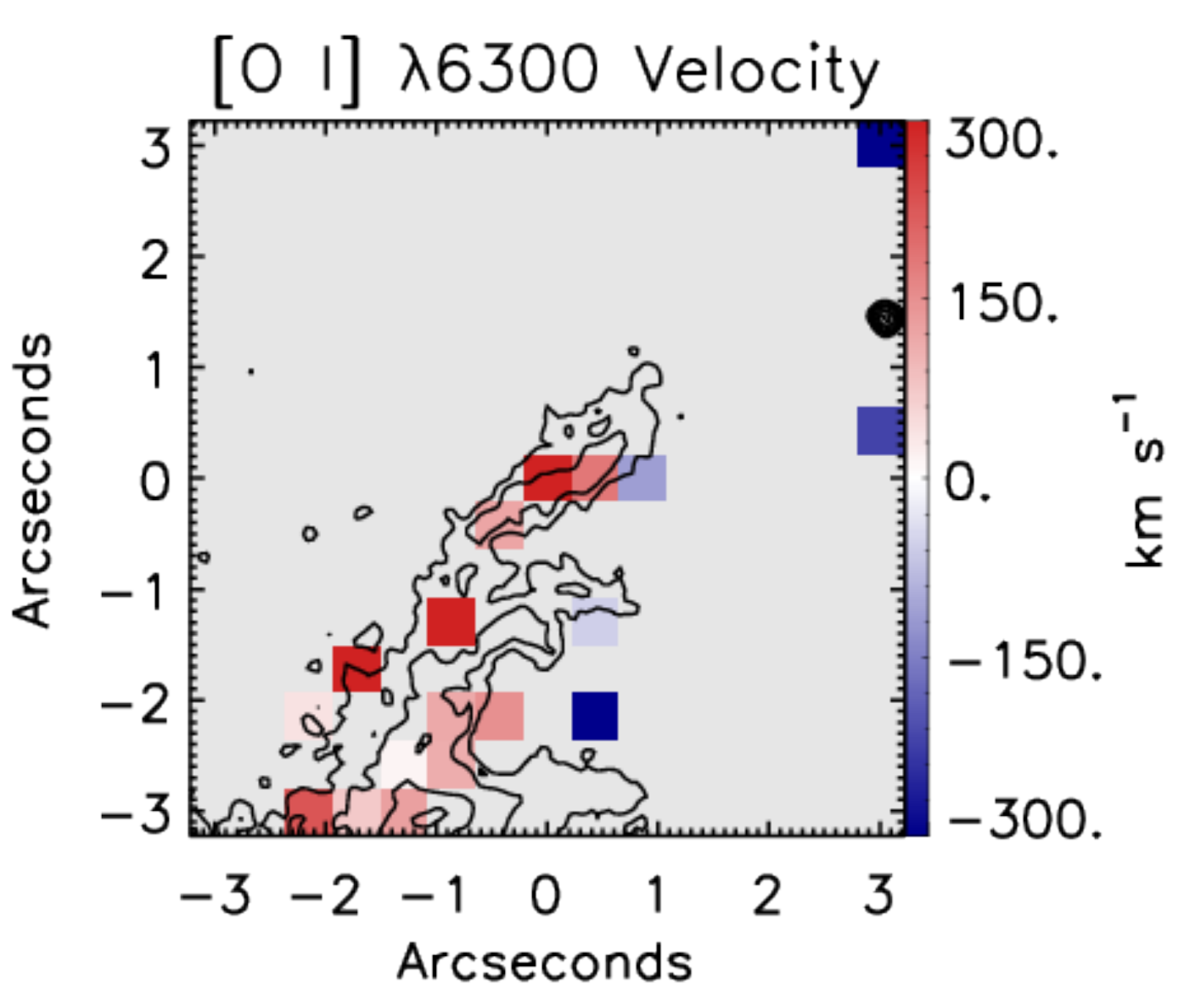}
\includegraphics[width=0.31\textwidth]{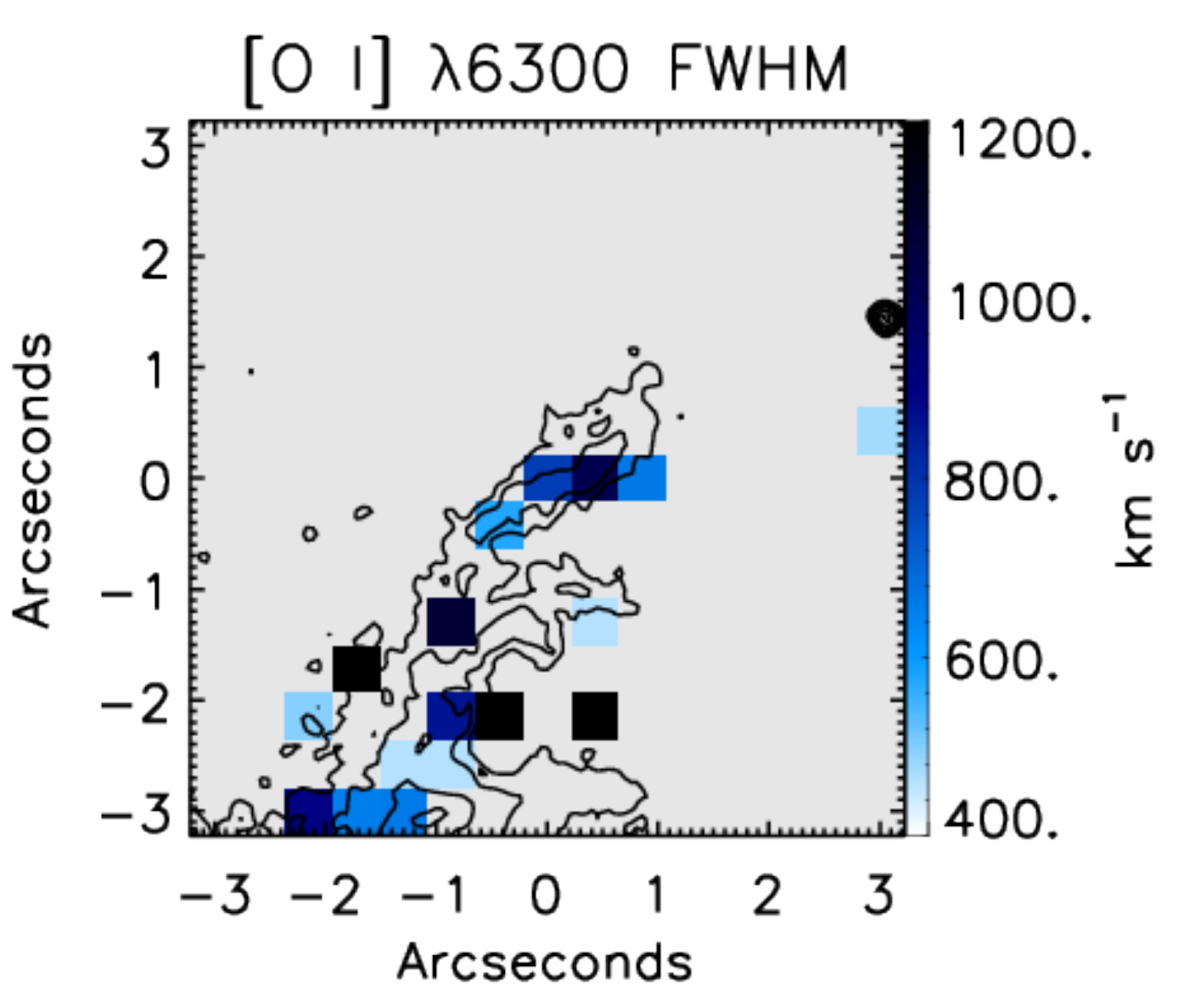}
\includegraphics[width=0.31\textwidth]{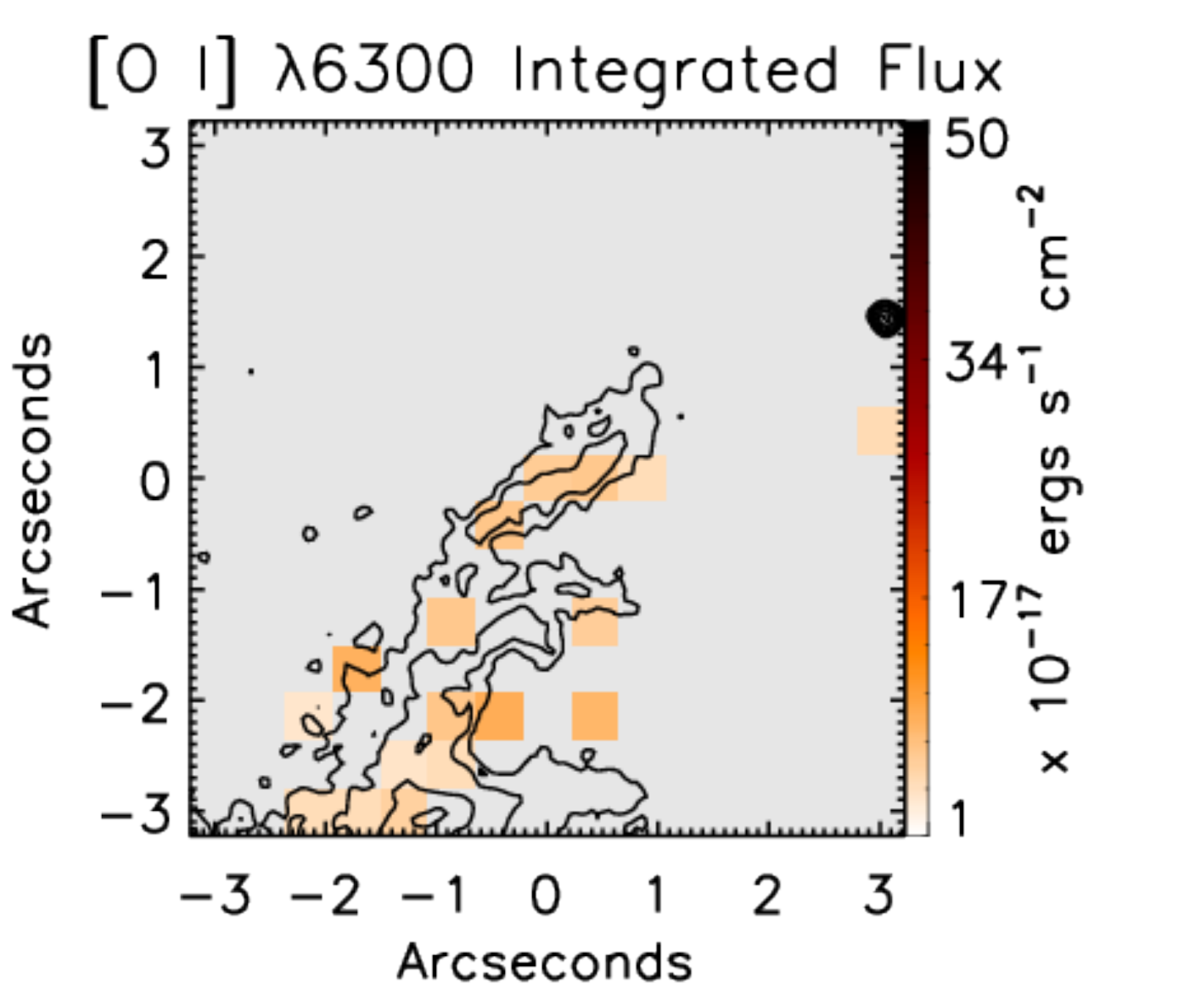}\\
\includegraphics[width=0.31\textwidth]{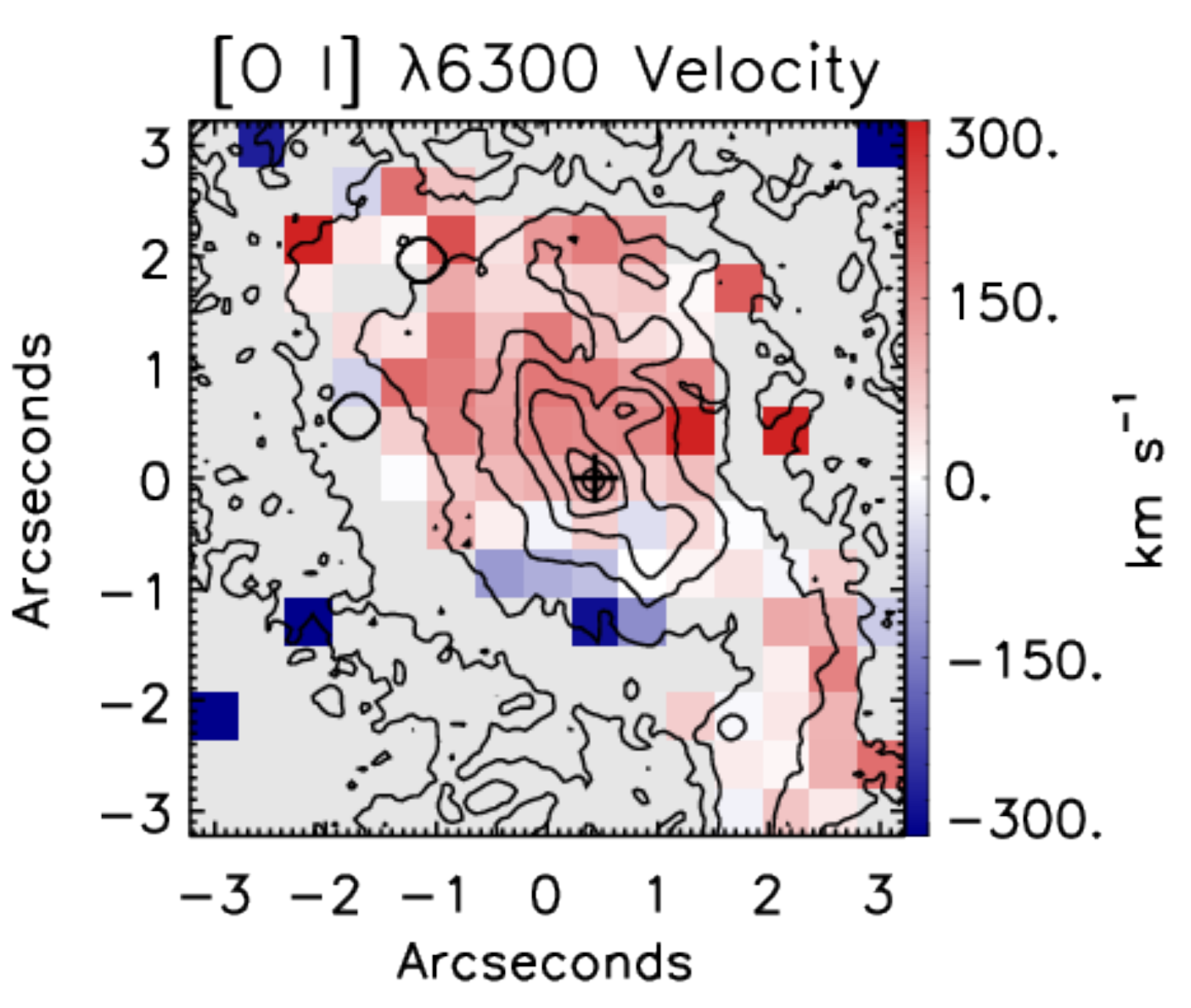}
\includegraphics[width=0.31\textwidth]{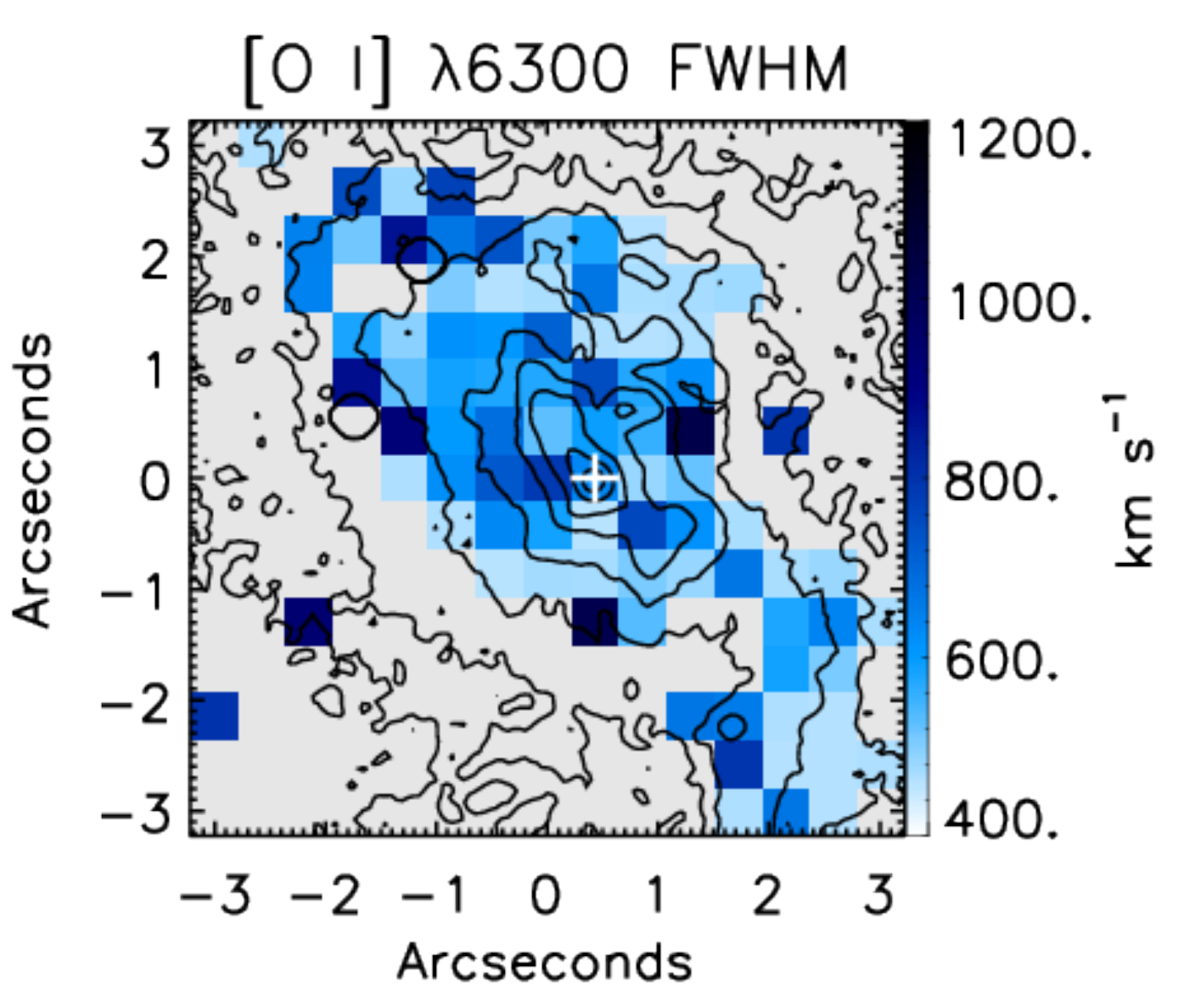}
\includegraphics[width=0.31\textwidth]{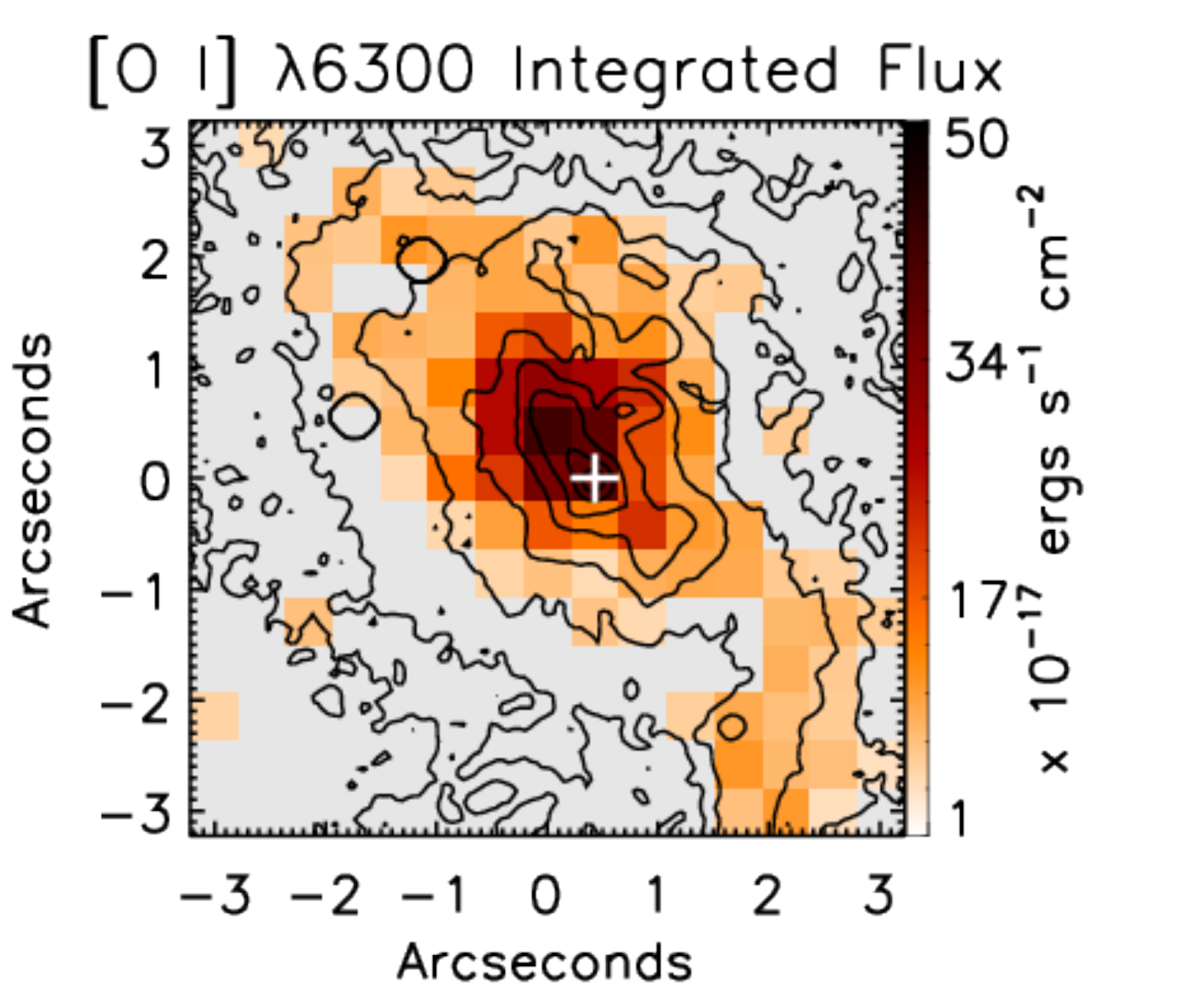}\\
\includegraphics[width=0.31\textwidth]{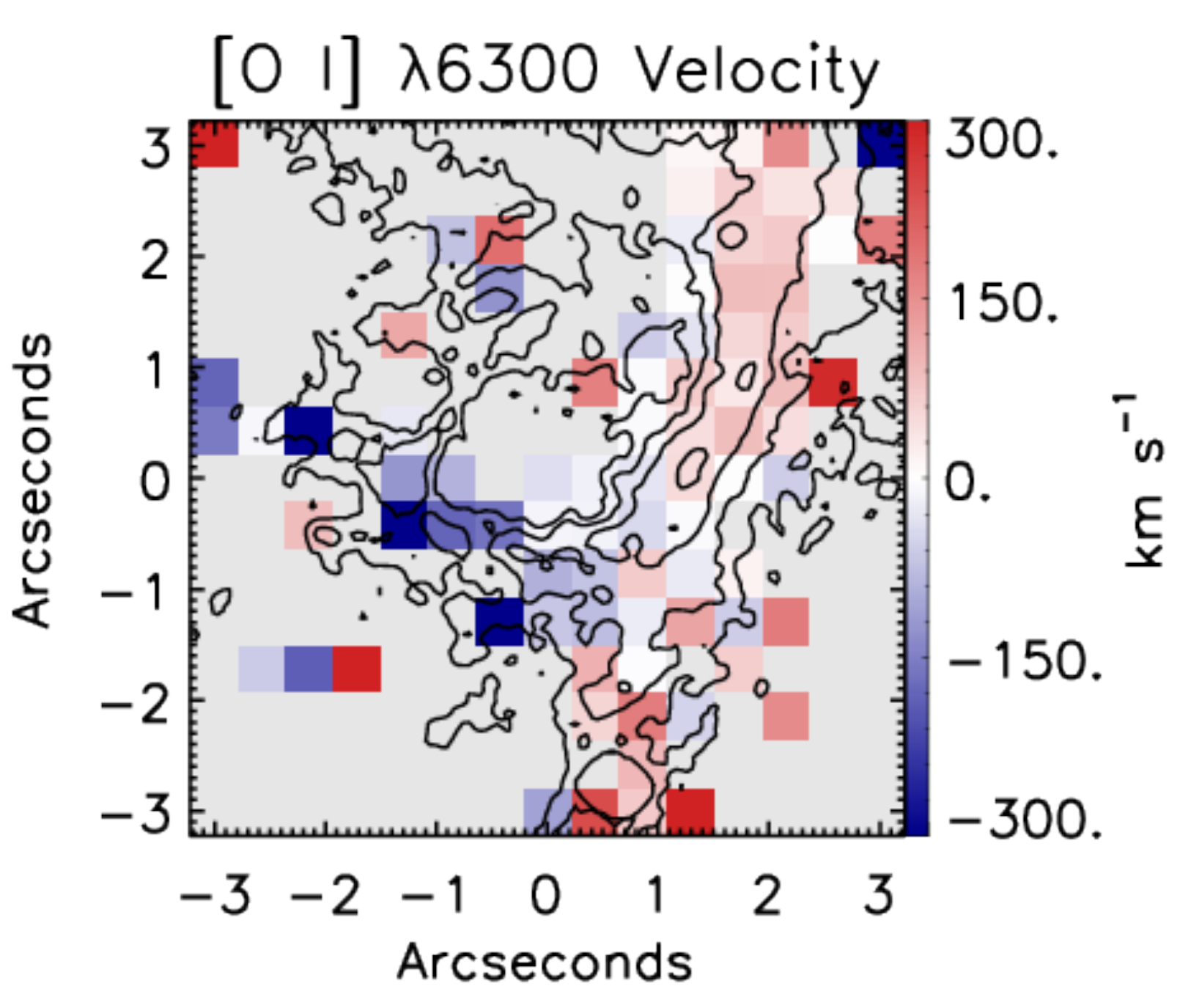}
\includegraphics[width=0.31\textwidth]{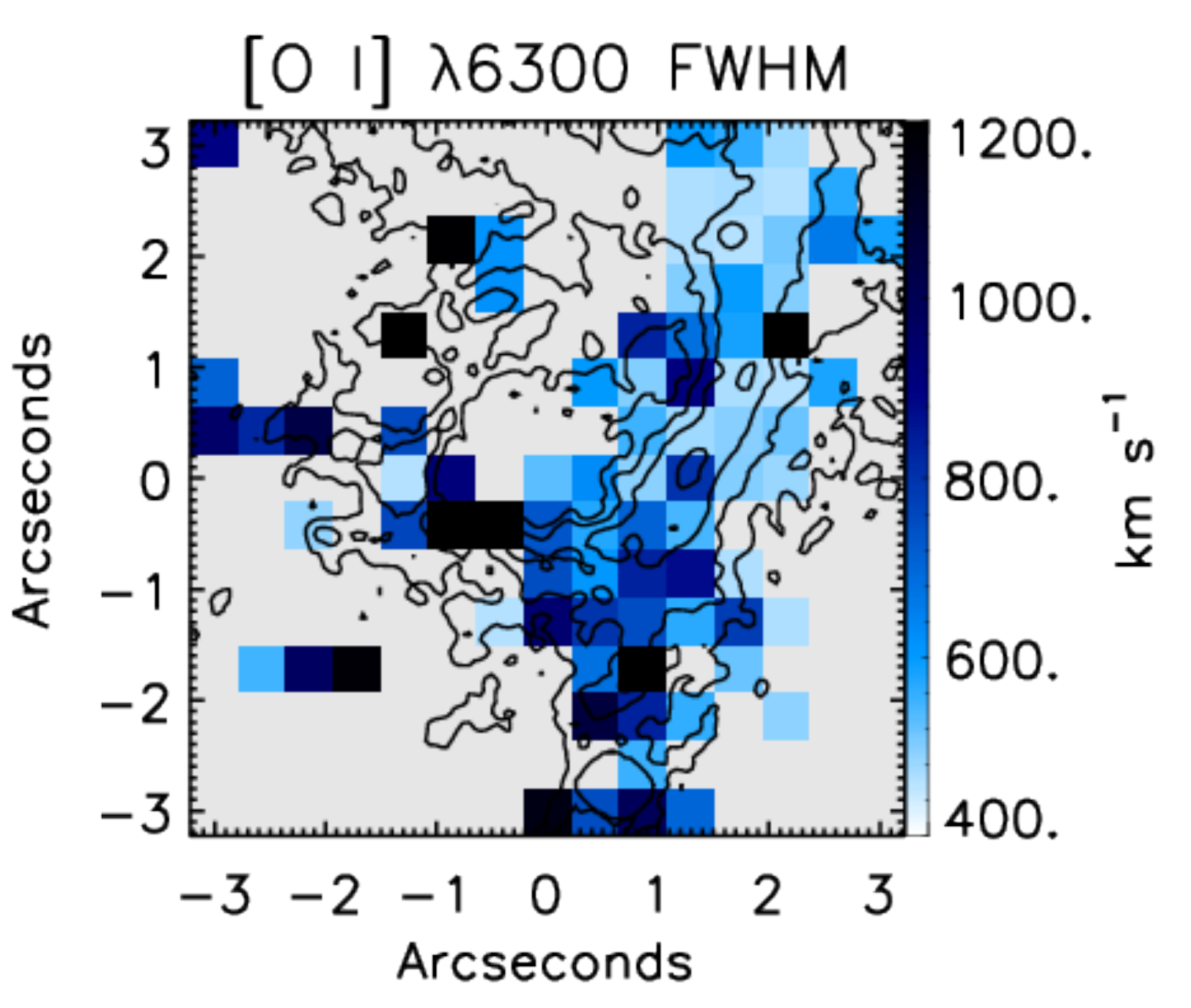}
\includegraphics[width=0.31\textwidth]{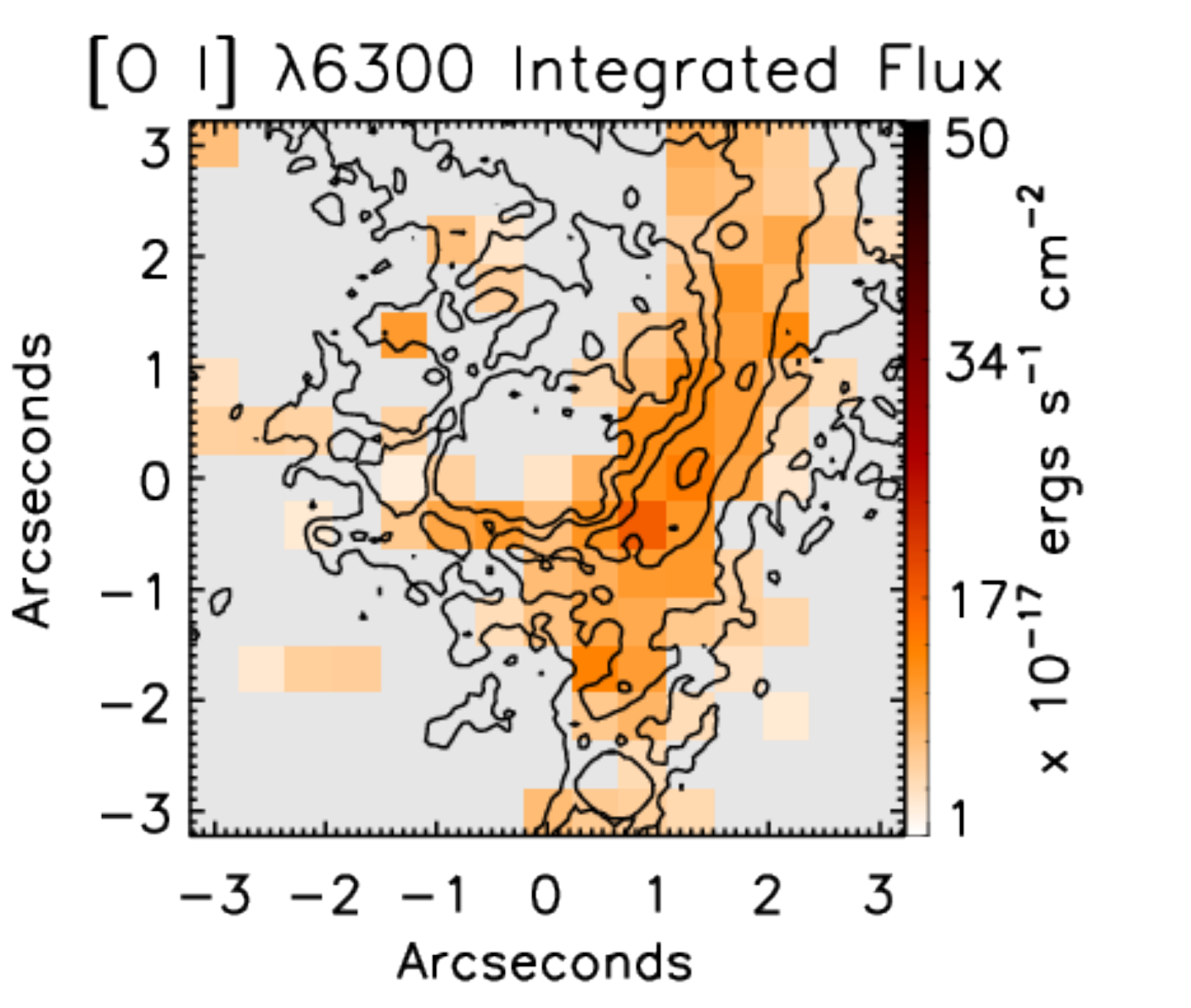}\\

\caption{[O~I] $\lambda$6300 kinematic measurements in 2MASX~J0423 from SNIFS IFU observations. First, 
second, and third columns display emission-line profile centroid velocity, FWHM, and integrated 
flux maps, respectively. First, second, and third rows display measurements for the top, 
center, and bottom fields of view, respectively. Black contours represent \emph{HST}/WFPC2 F675W 
imaging. The optical continuum flux peak is depicted by a cross. One 0.43$" \times$ 0.43$"$ spaxel 
samples approximately 380\,pc $\times$ 380\,pc.}
\label{fig:oimaps}

\end{figure*}

\begin{figure*}[!htbp]
\centering
\includegraphics[width=0.31\textwidth]{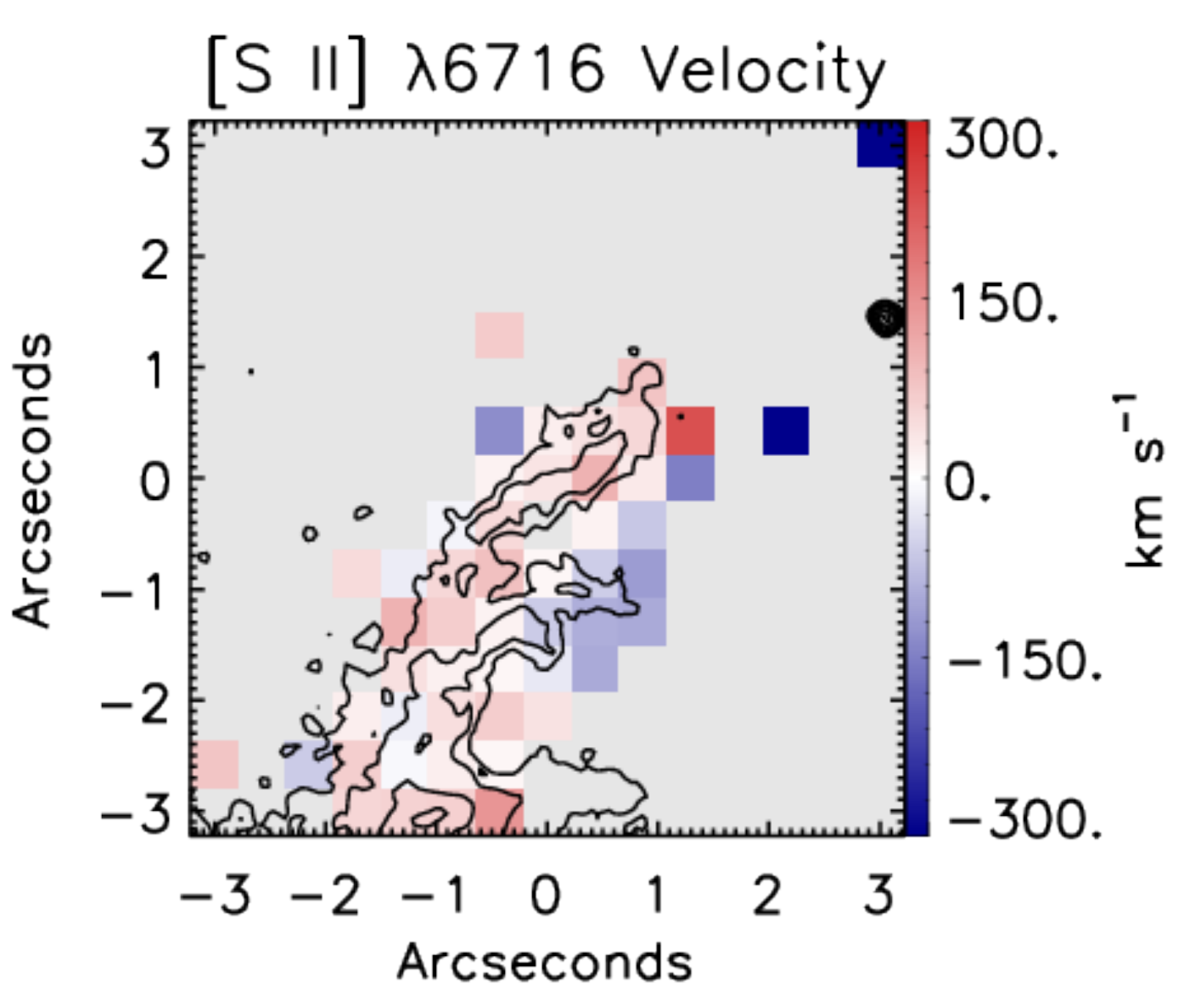}
\includegraphics[width=0.31\textwidth]{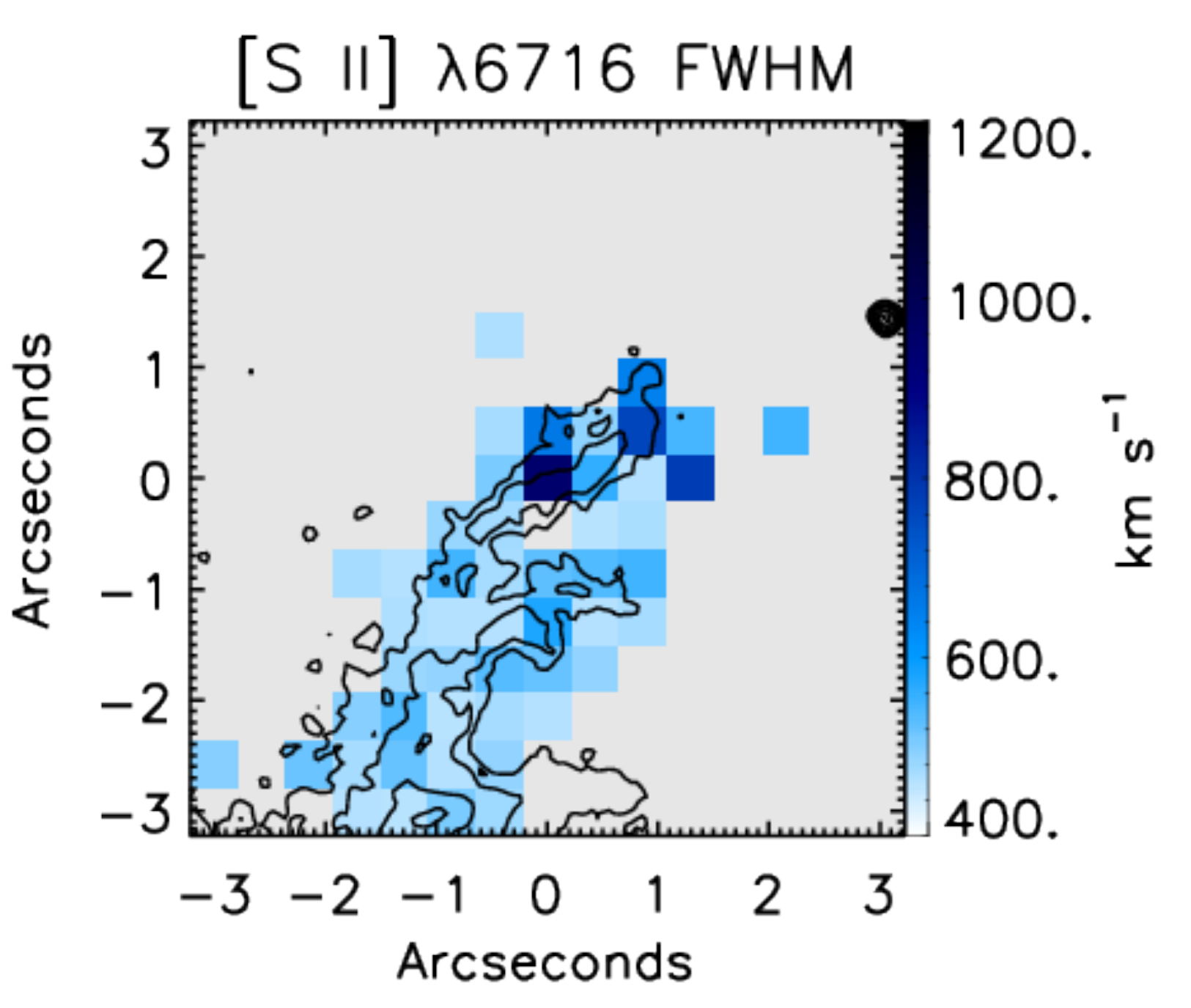}
\includegraphics[width=0.31\textwidth]{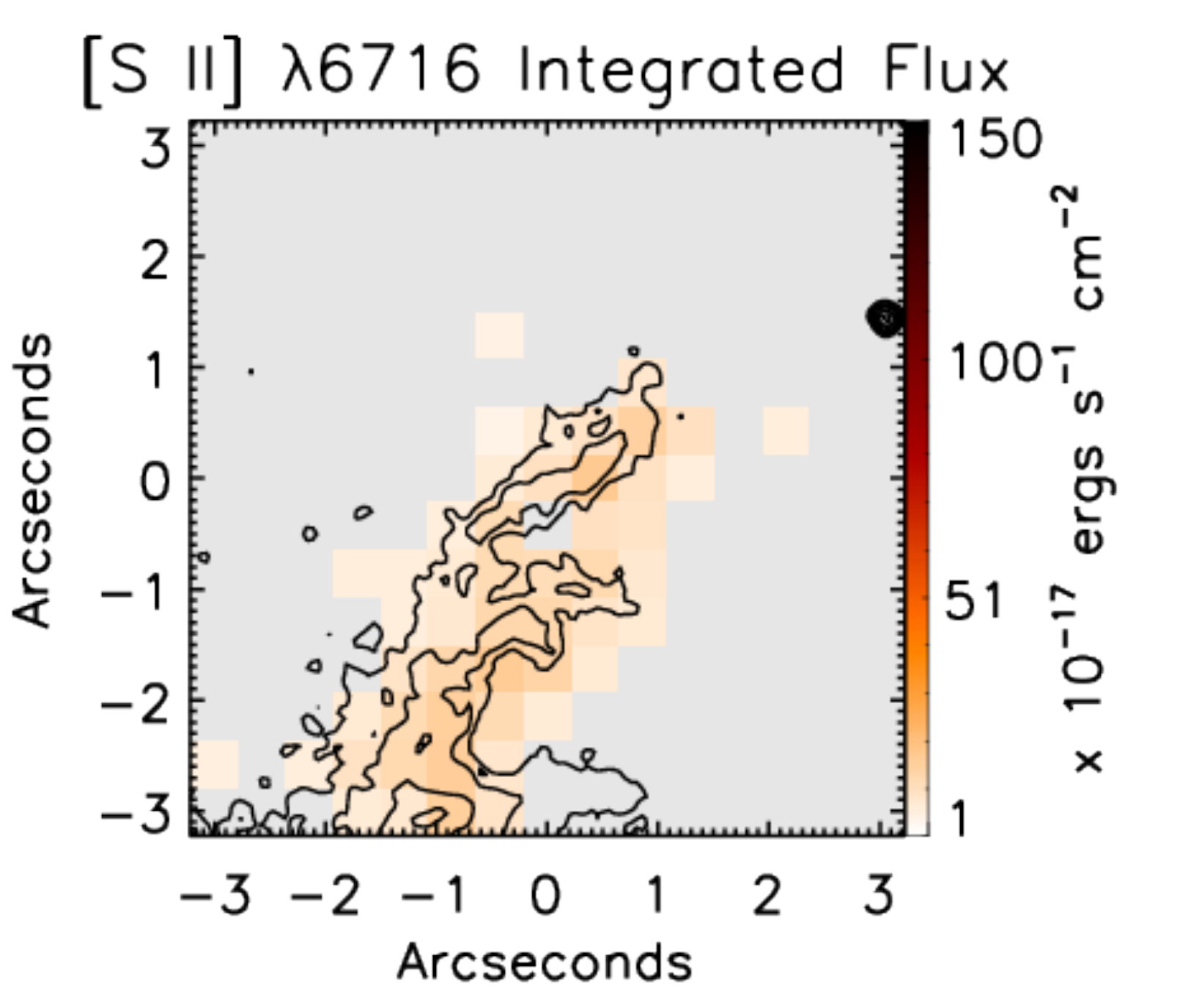}\\
\includegraphics[width=0.31\textwidth]{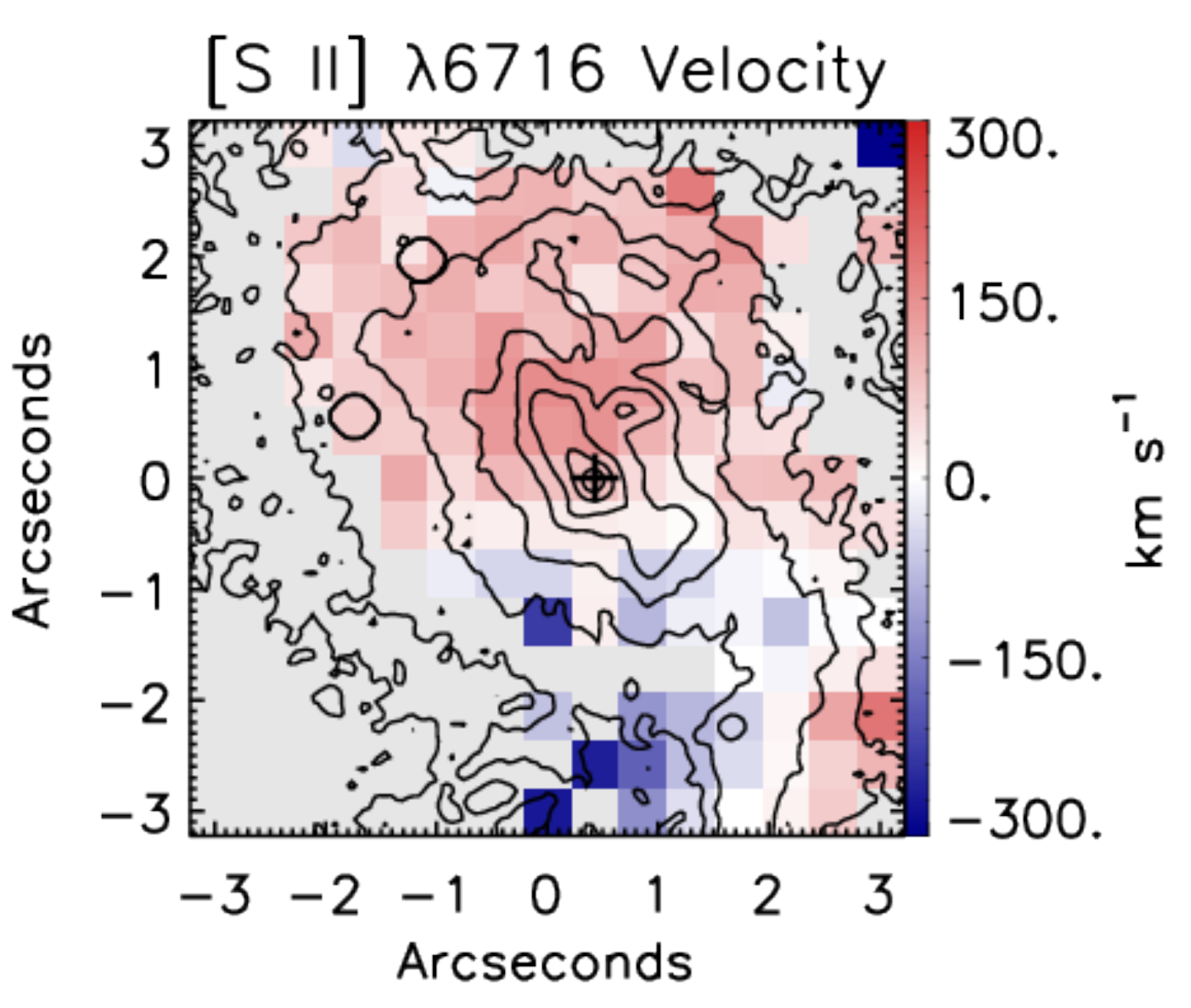}
\includegraphics[width=0.31\textwidth]{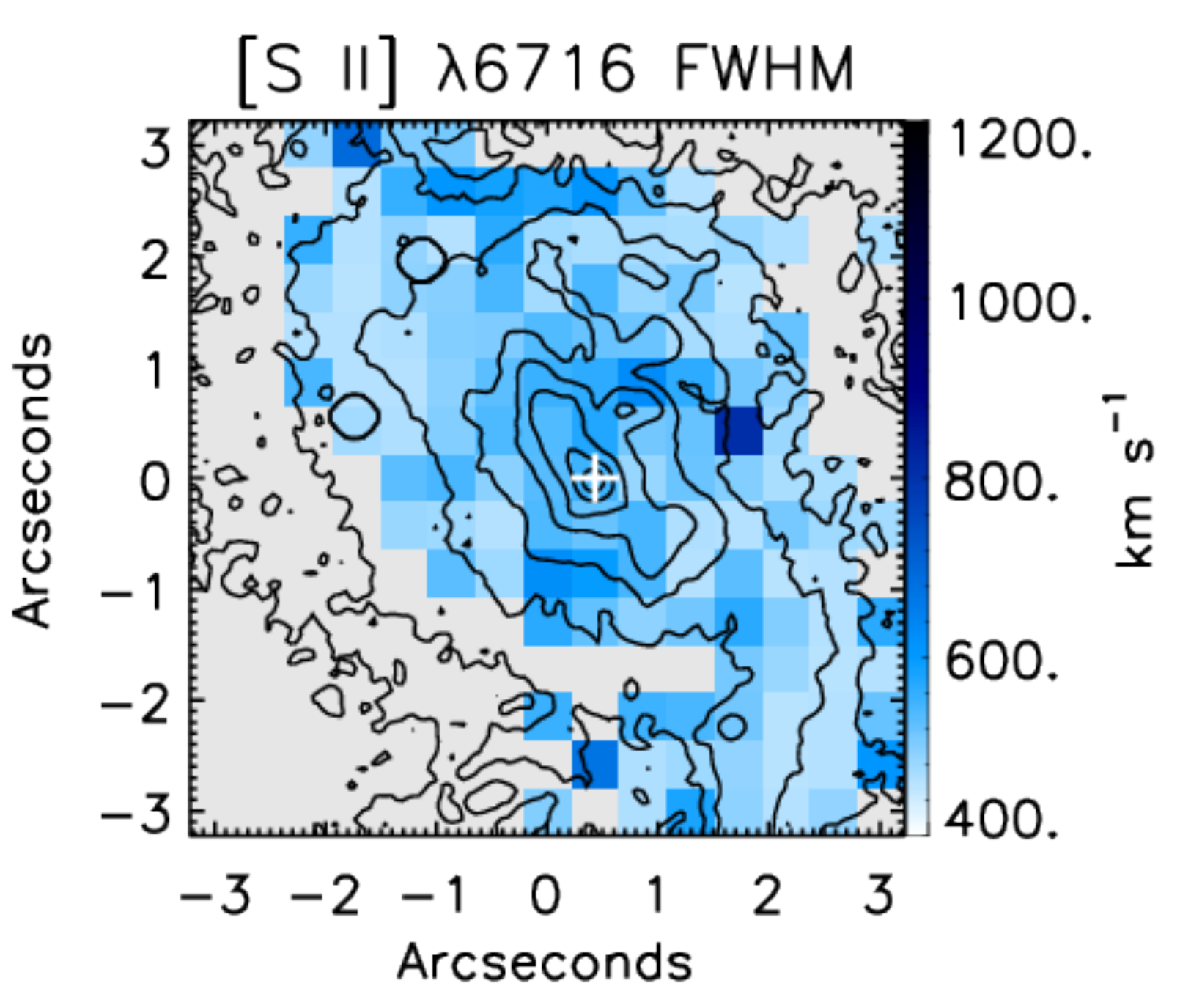}
\includegraphics[width=0.31\textwidth]{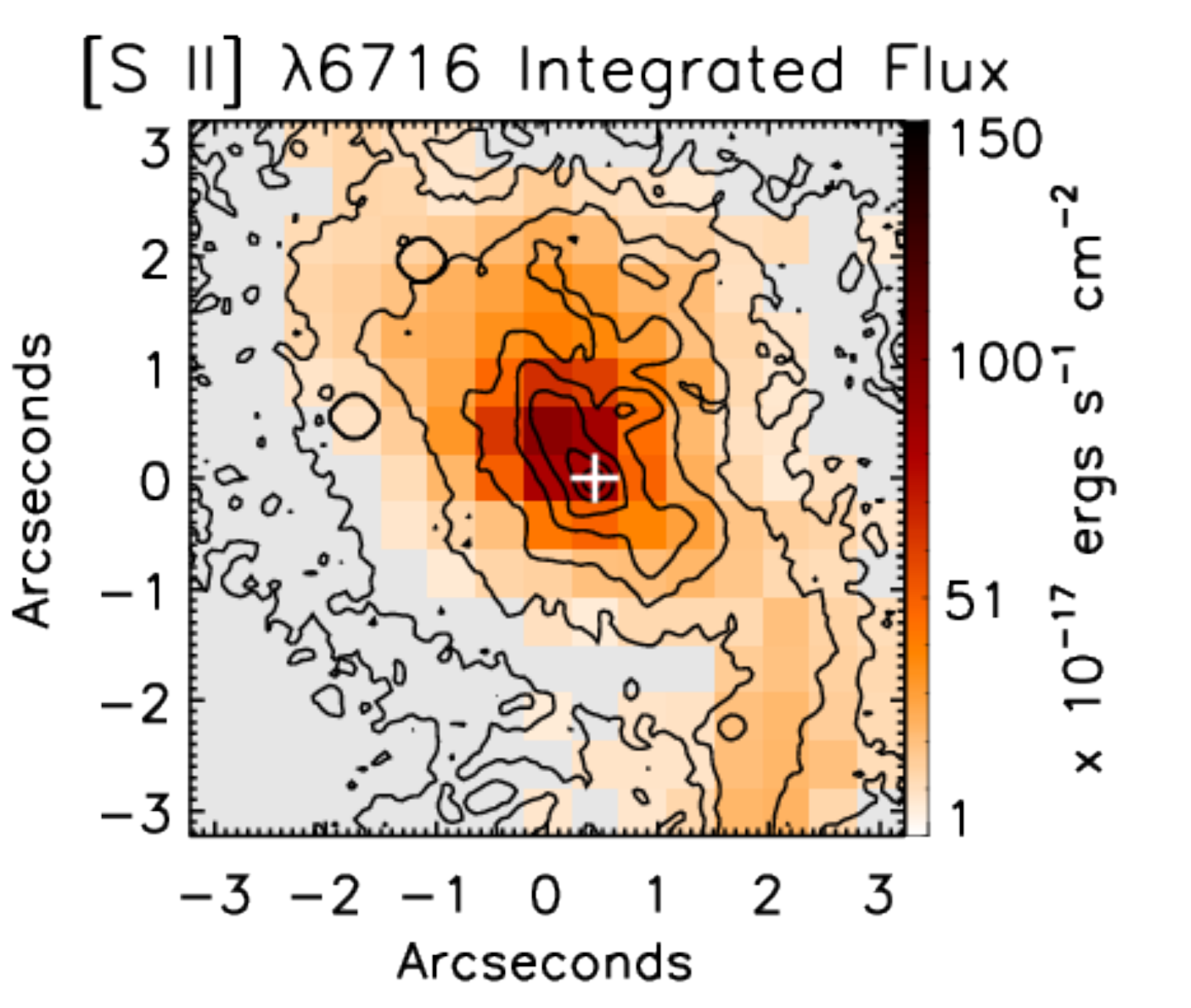}\\
\includegraphics[width=0.31\textwidth]{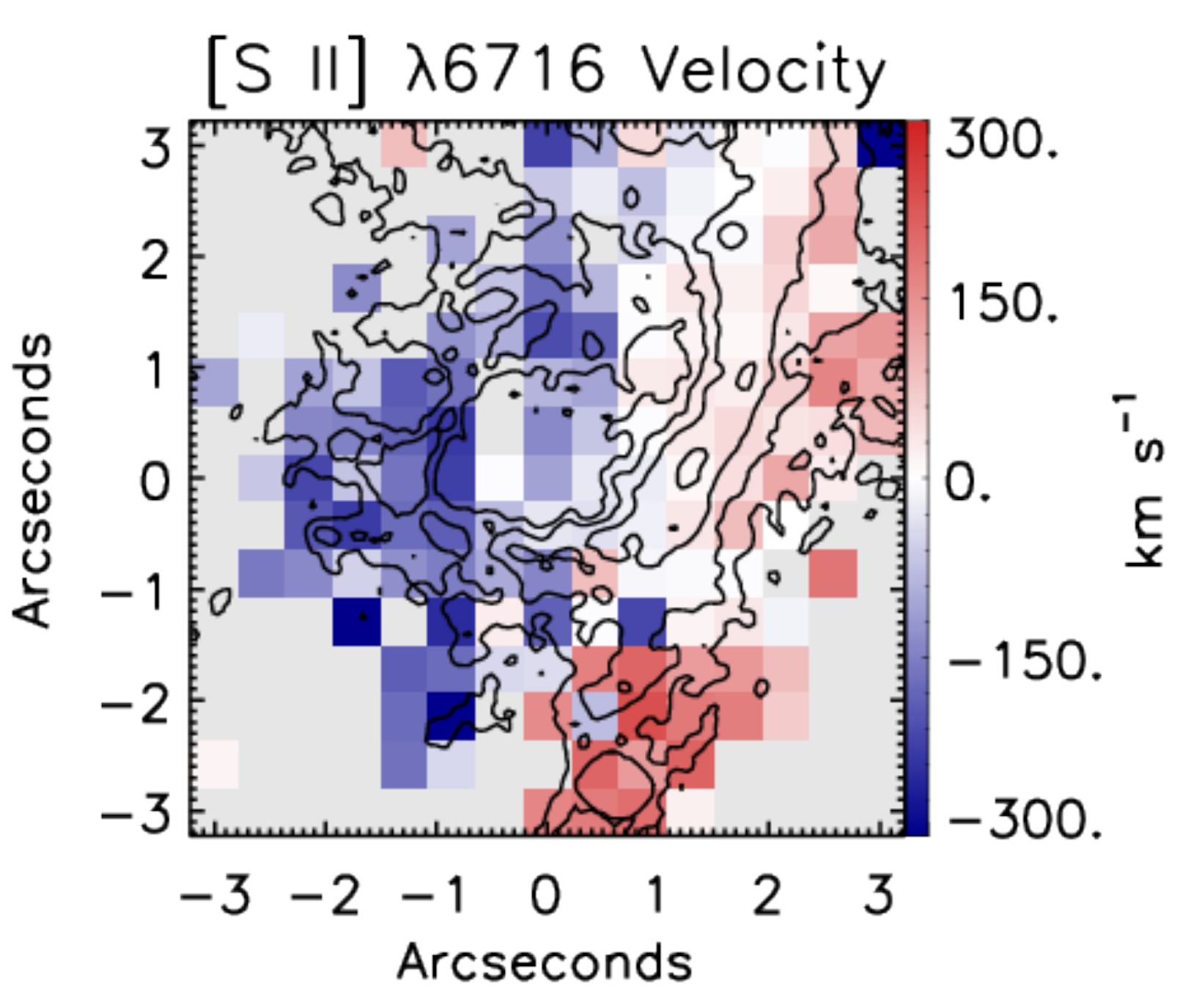}
\includegraphics[width=0.31\textwidth]{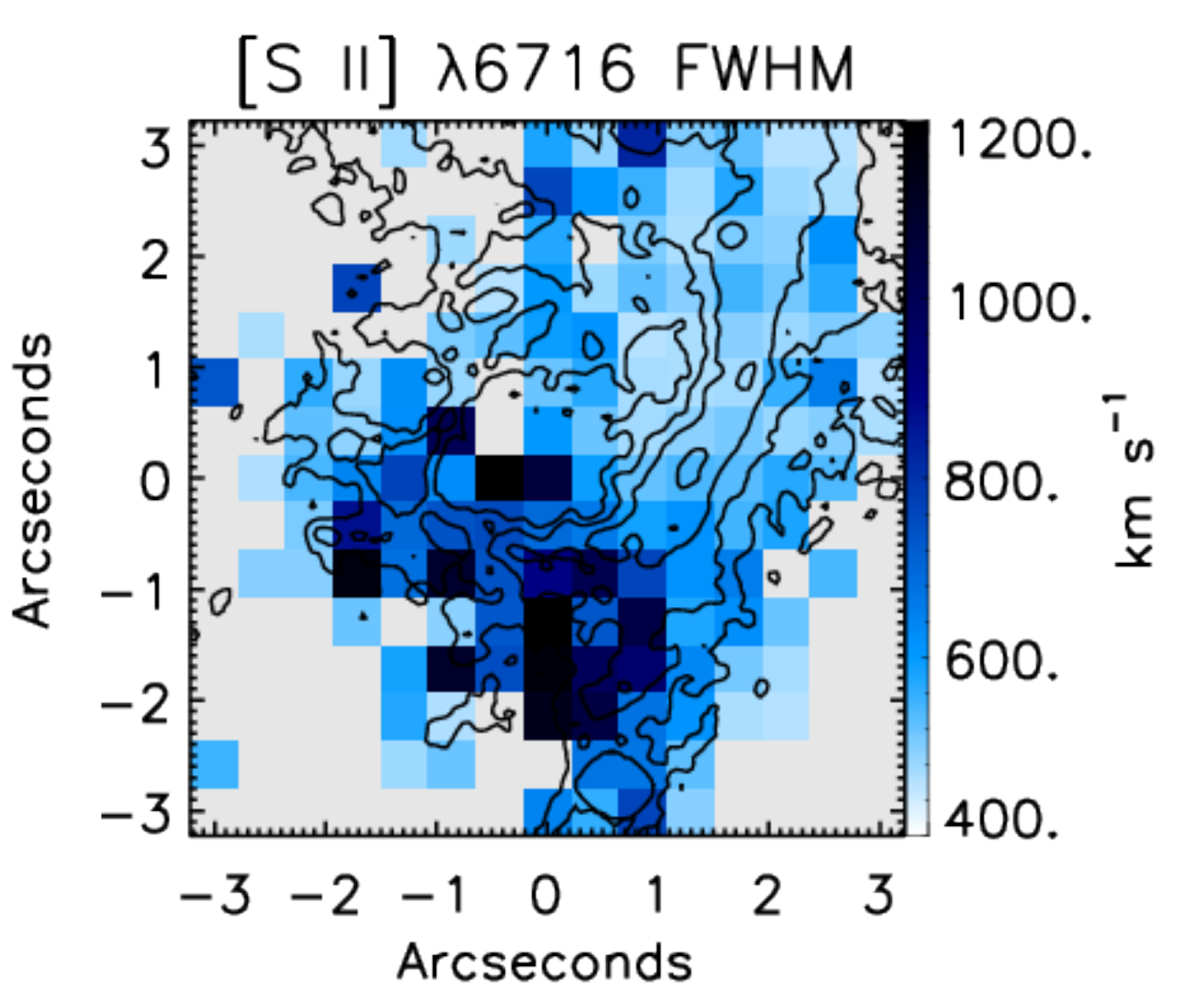}
\includegraphics[width=0.31\textwidth]{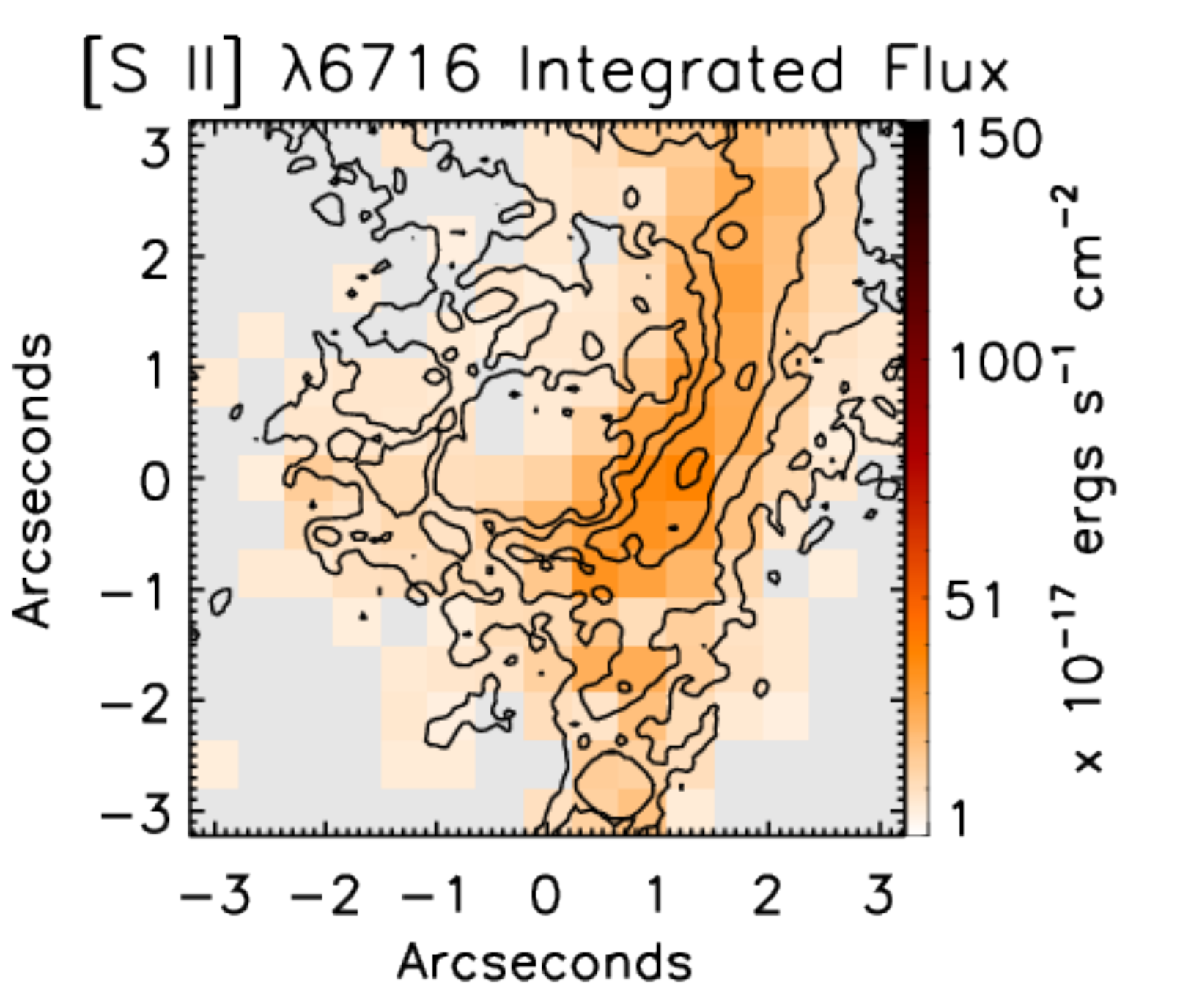}\\

\caption{[S~II] $\lambda$6716 kinematic measurements in 2MASX~J0423 from SNIFS IFU observations. First, 
second, and third columns display emission-line profile centroid velocity, FWHM, and integrated 
flux maps, respectively. First, second, and third rows display measurements for the top, 
center, and bottom fields of view, respectively. Black contours represent \emph{HST}/WFPC2 
F675W imaging. The optical continuum flux peak is depicted by a cross. One 0.43$" \times$ 0.43$"$ spaxel 
samples approximately 380\,pc $\times$ 380\,pc.}
\label{fig:siimaps}

\end{figure*}

\begin{figure*}[!htbp]
\centering
\includegraphics[width=0.31\textwidth]{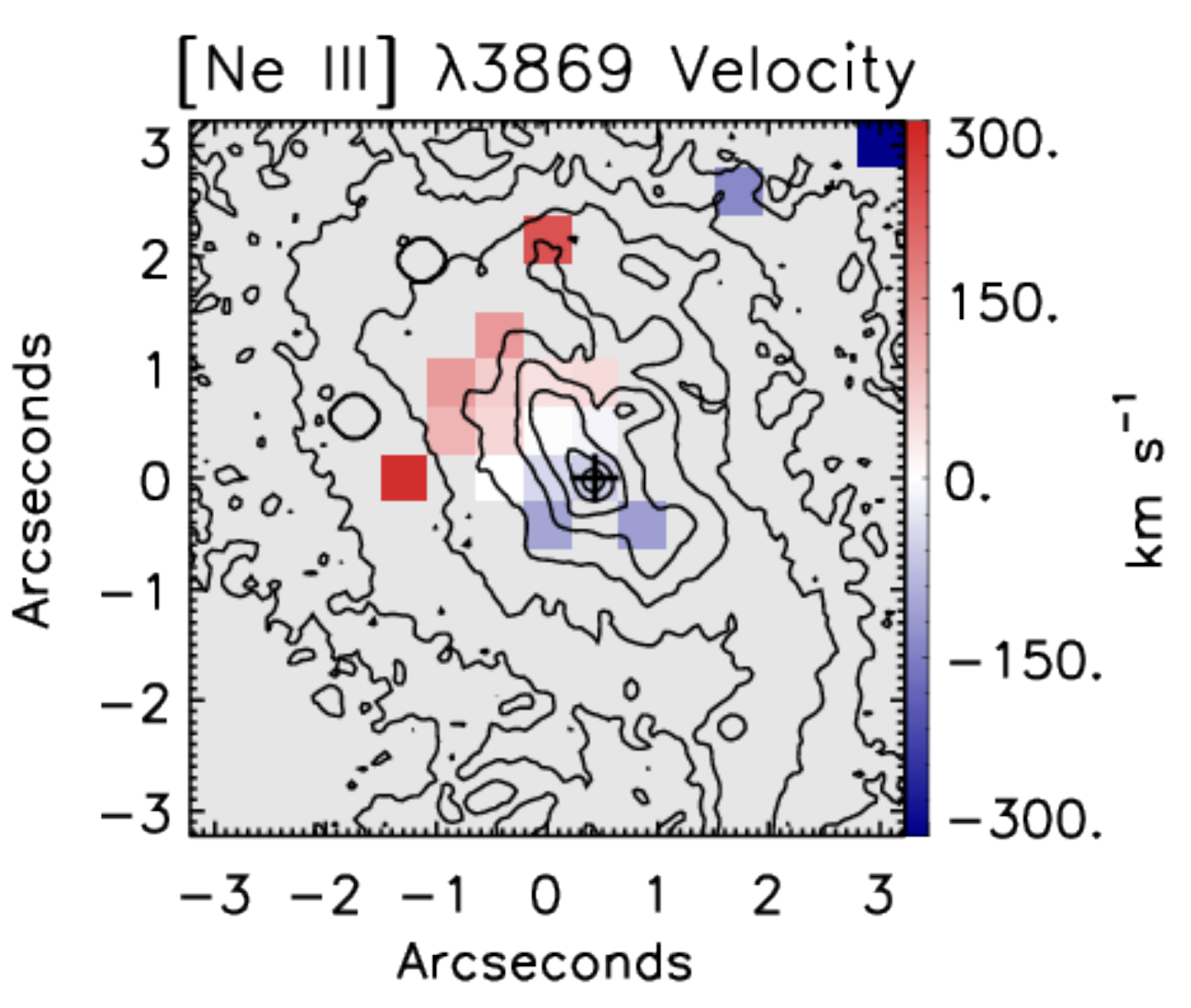}
\includegraphics[width=0.31\textwidth]{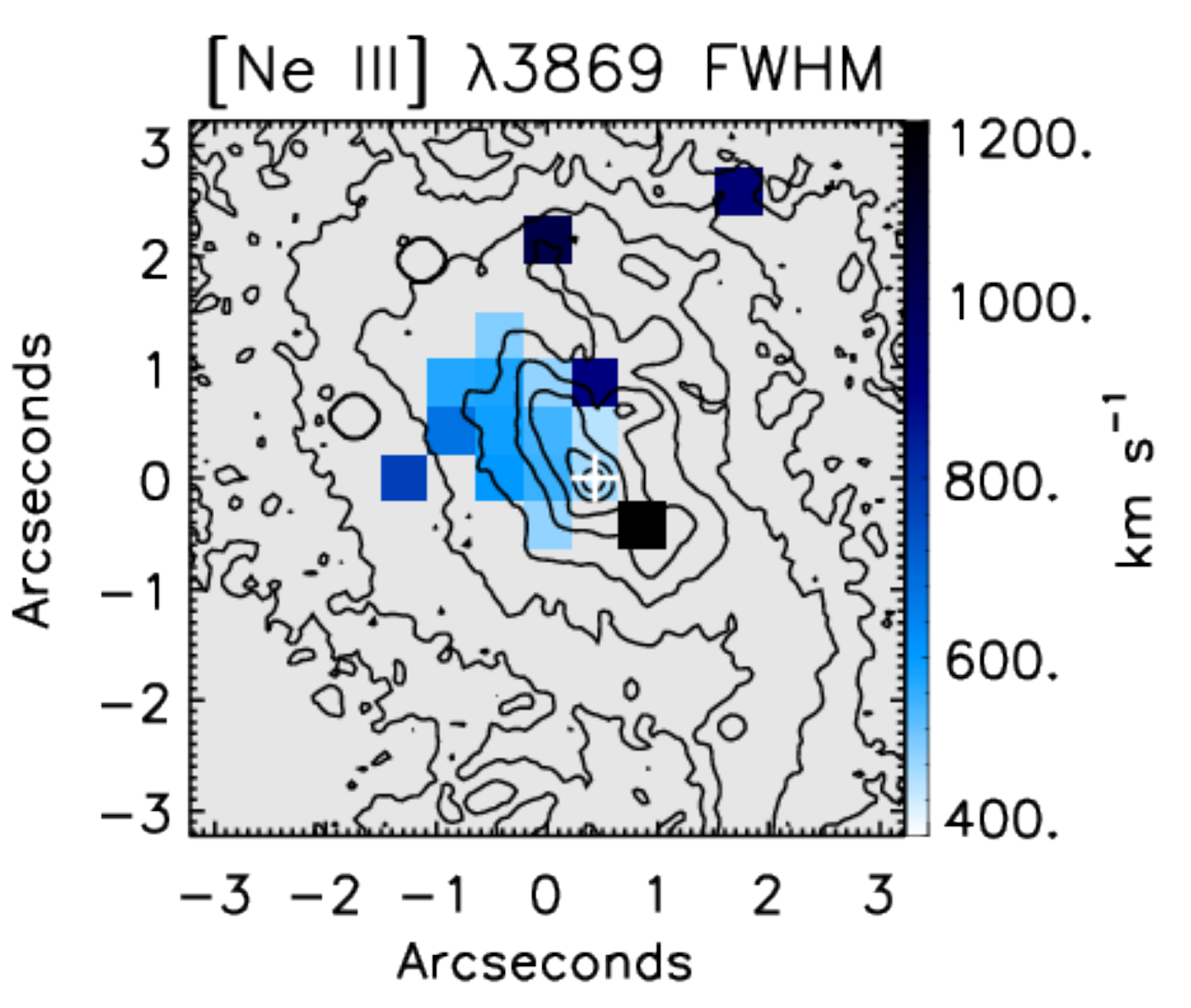}
\includegraphics[width=0.31\textwidth]{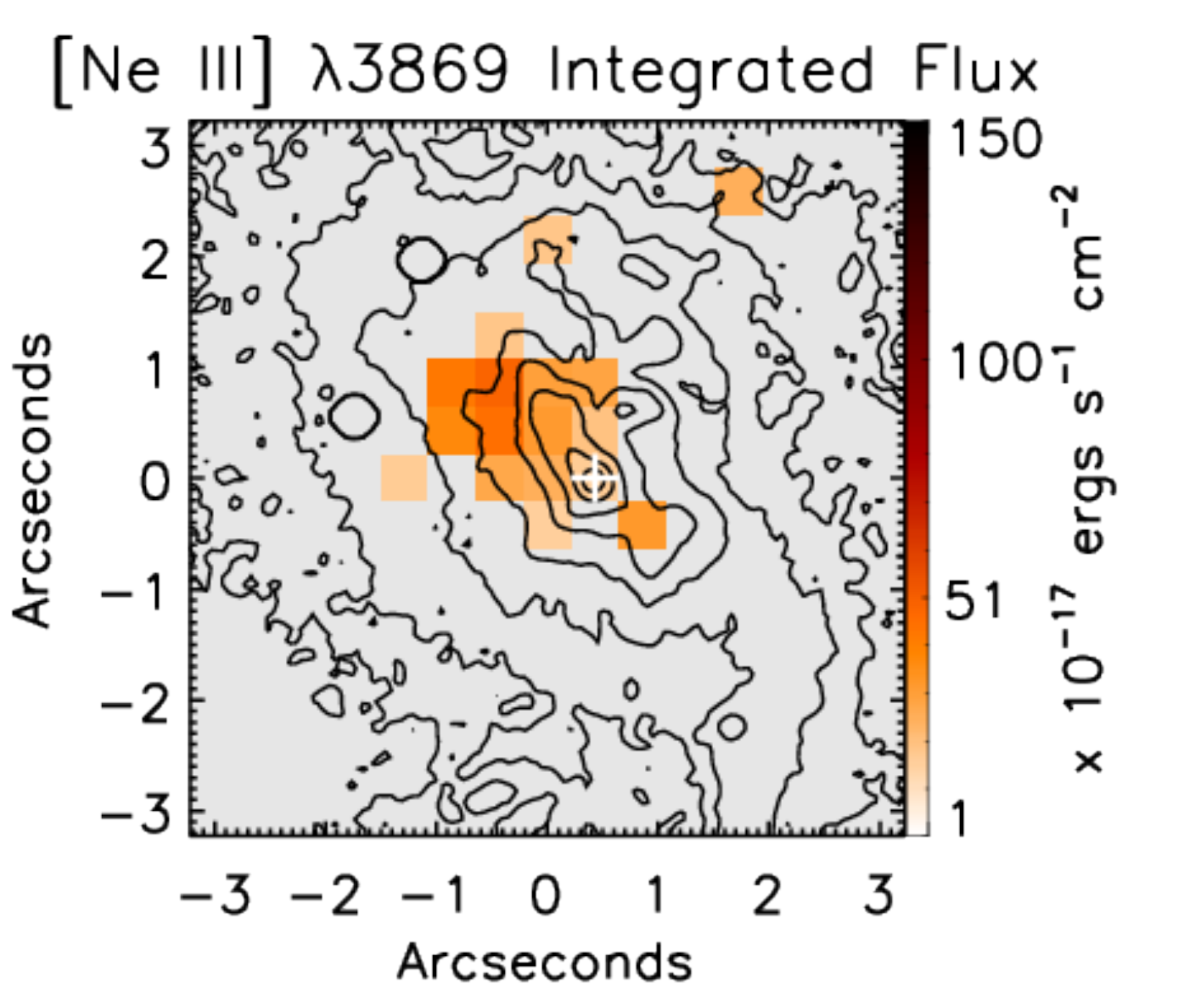}\\
\includegraphics[width=0.31\textwidth]{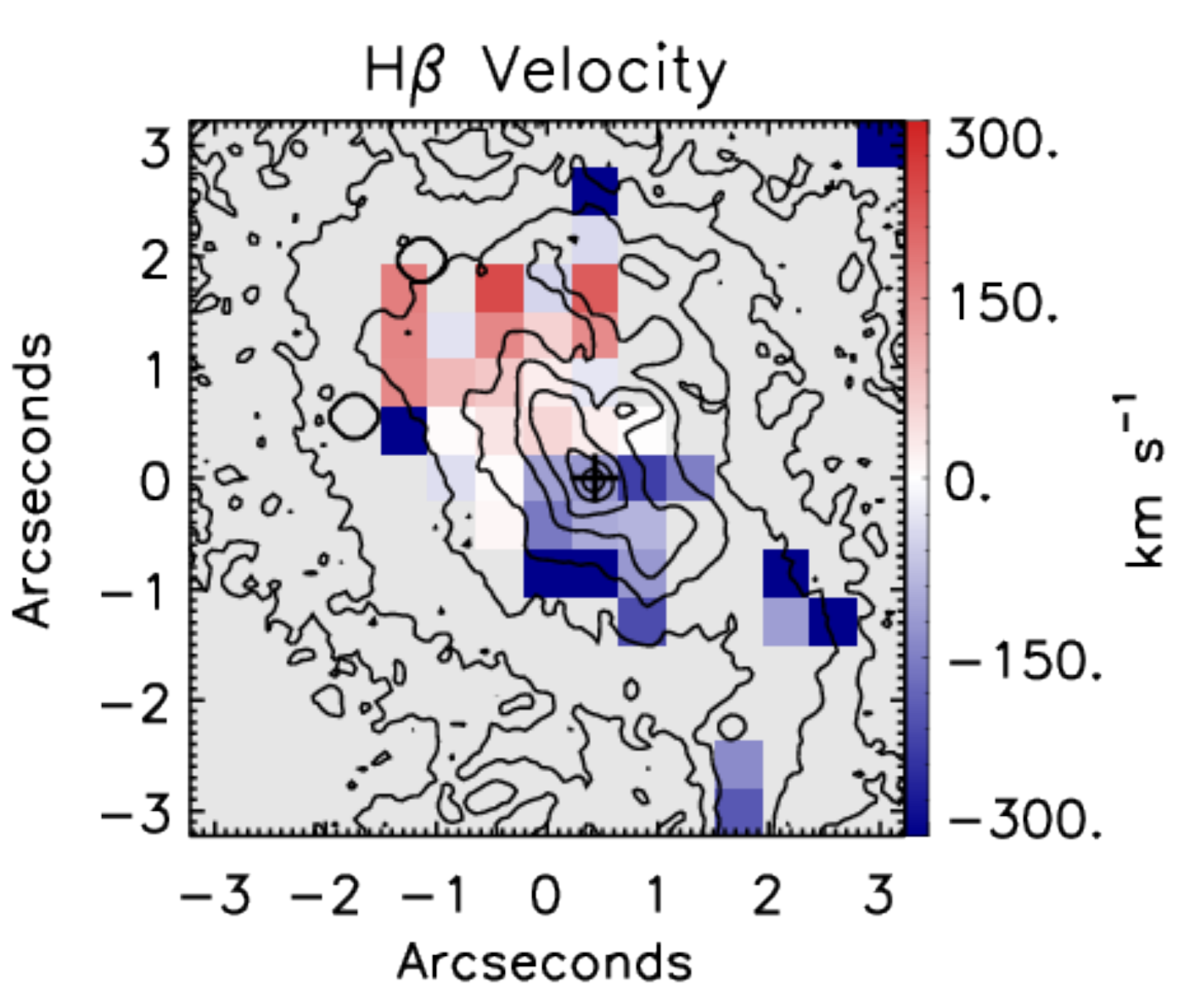}
\includegraphics[width=0.31\textwidth]{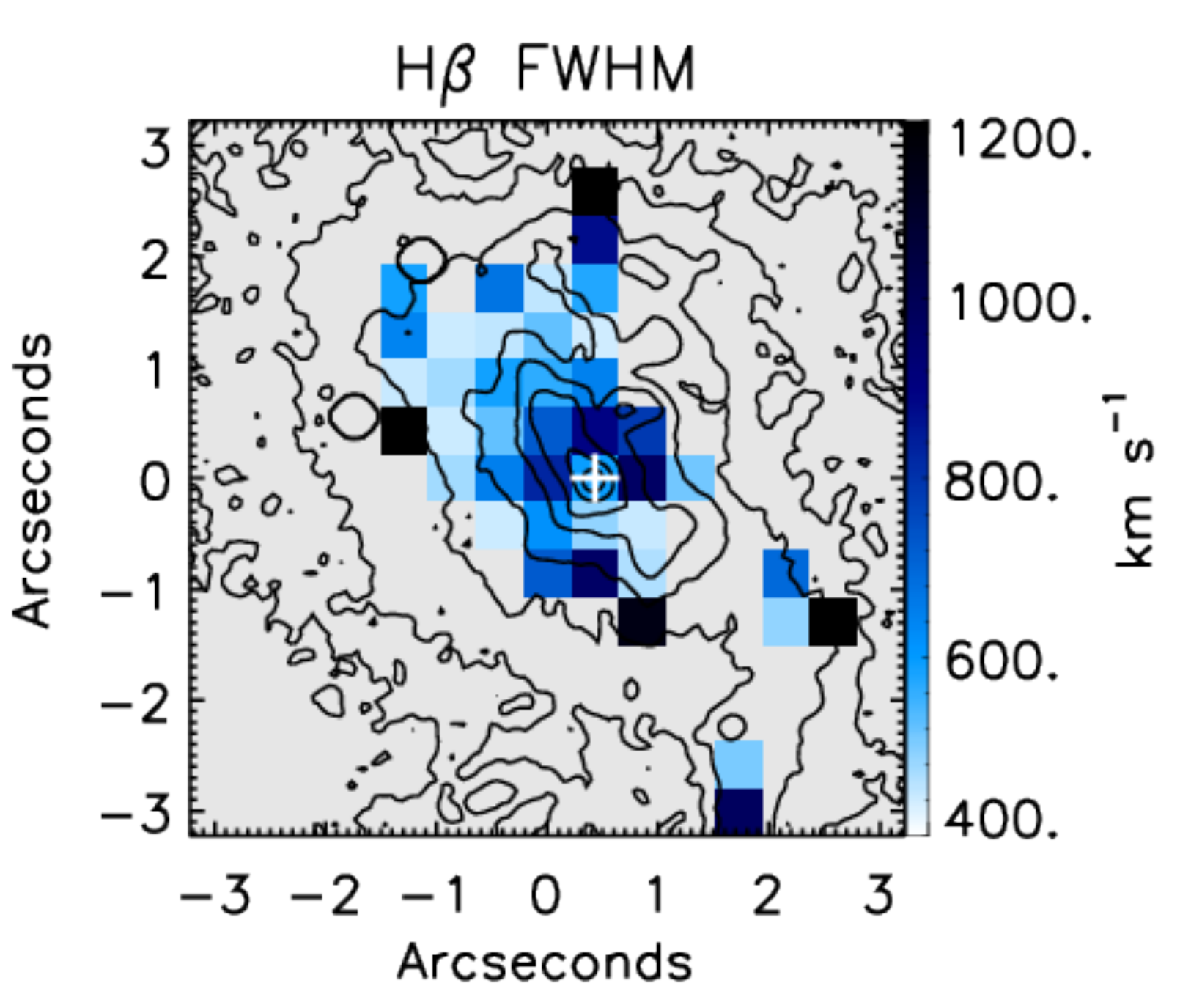}
\includegraphics[width=0.31\textwidth]{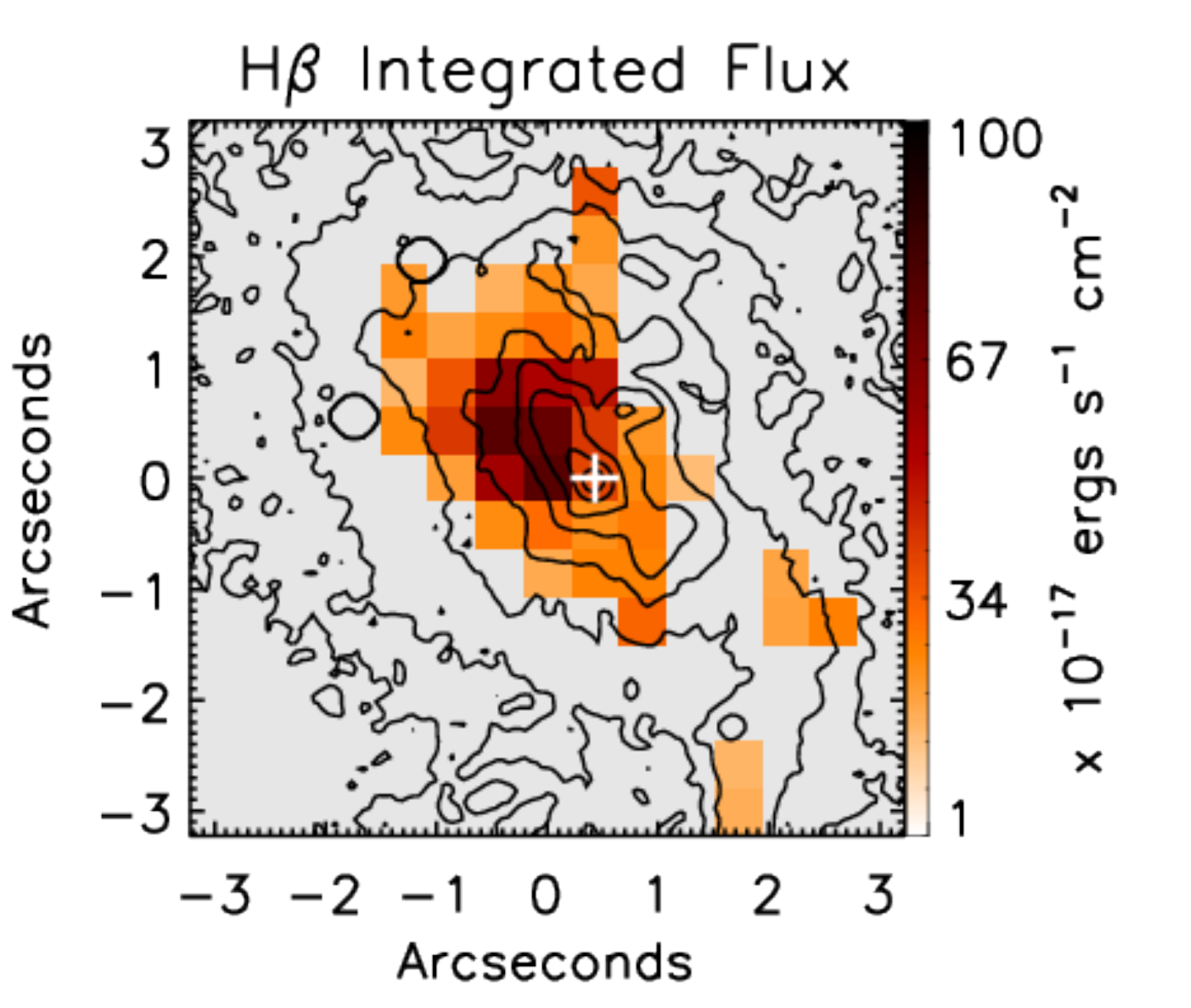}\\
\includegraphics[width=0.31\textwidth]{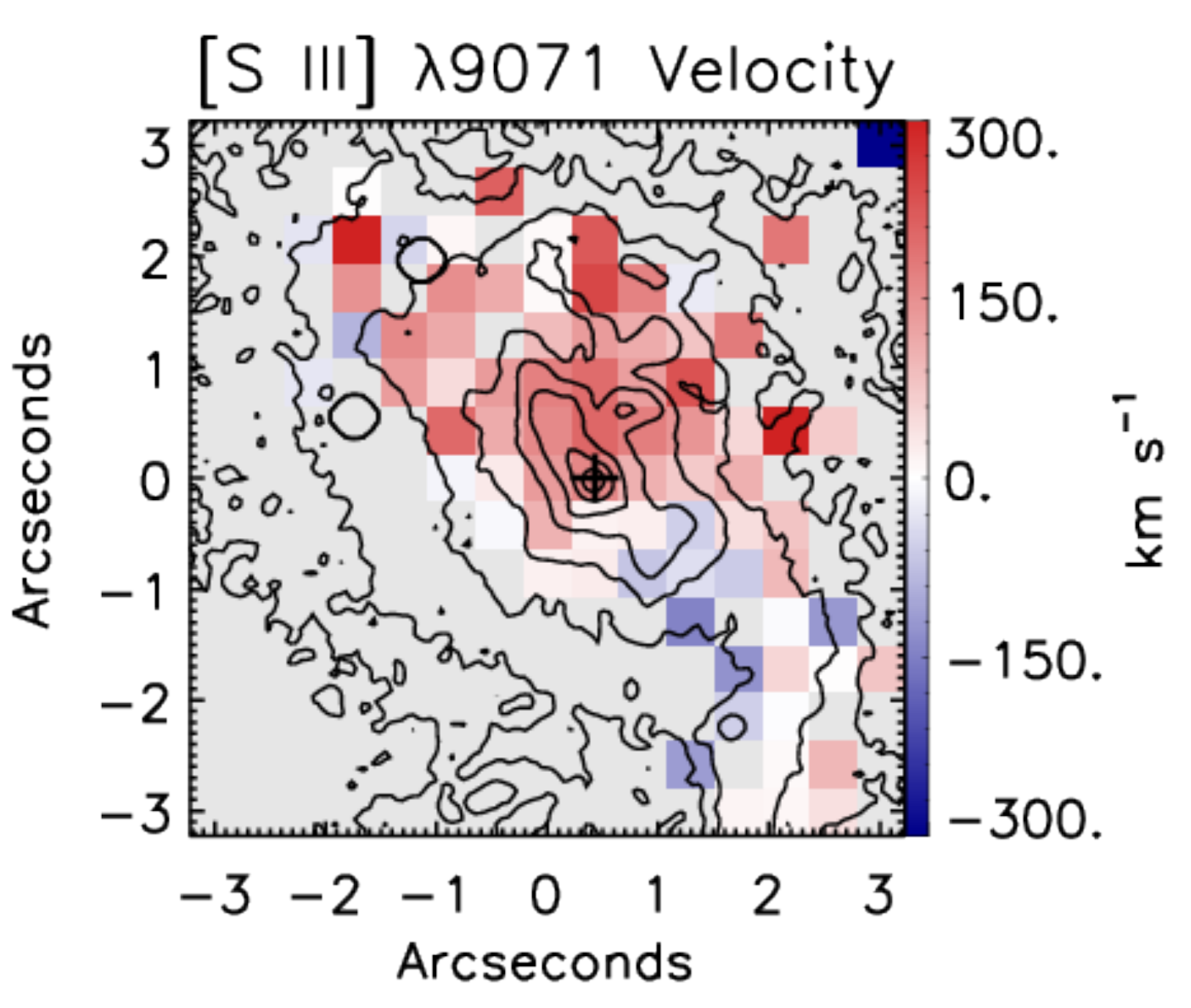}
\includegraphics[width=0.31\textwidth]{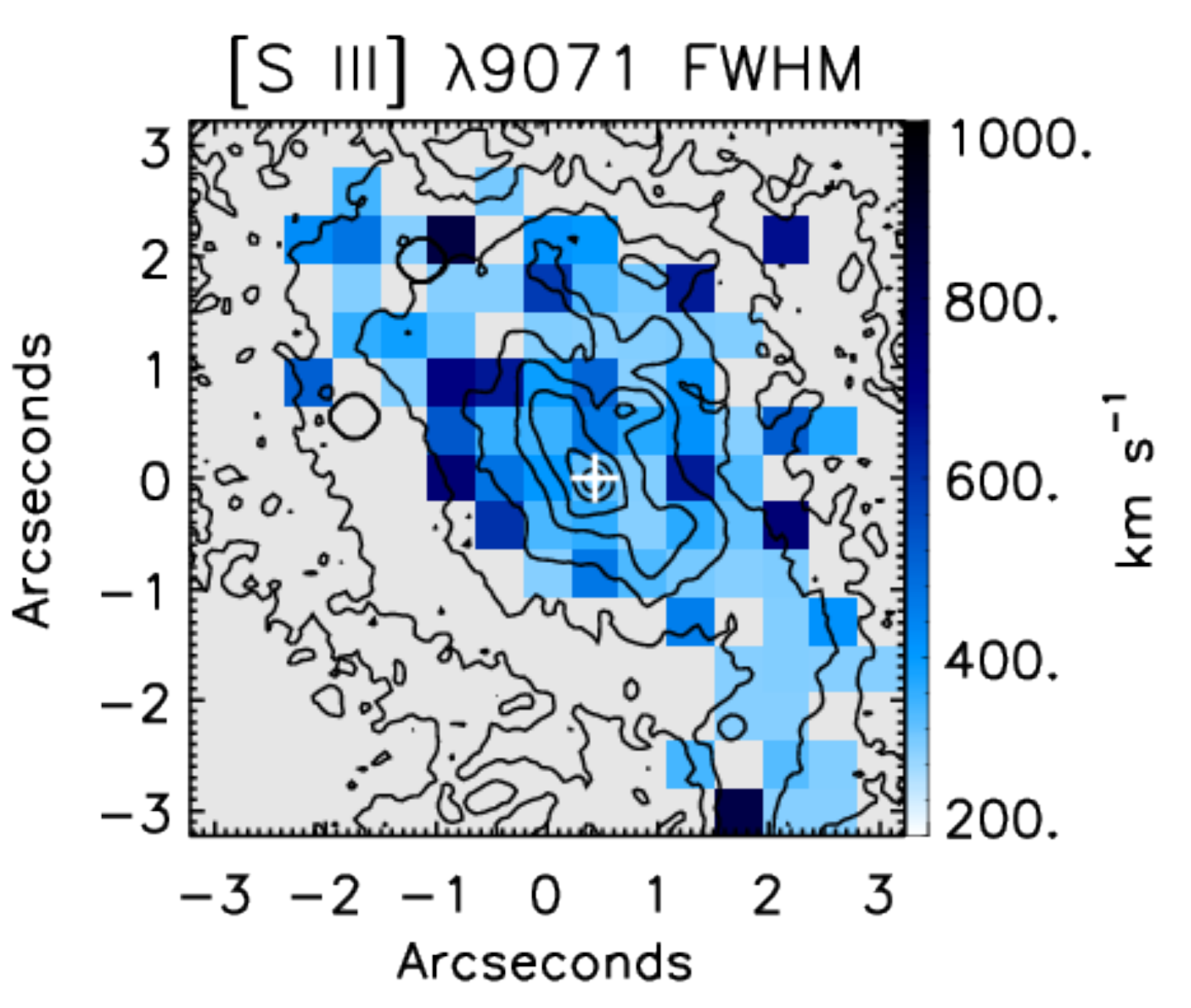}
\includegraphics[width=0.31\textwidth]{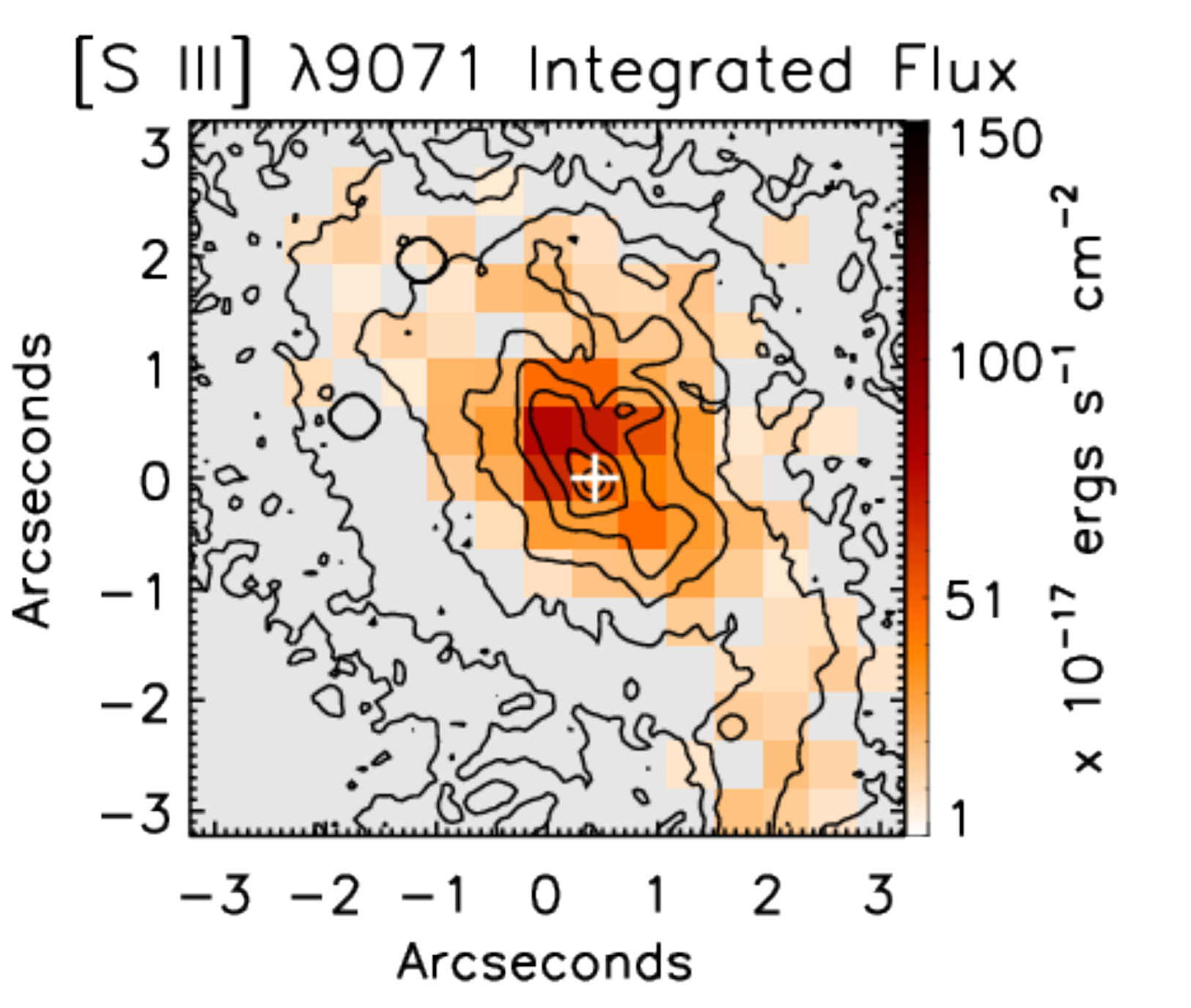}\\

\caption{Nuclear [Ne~III]$\lambda$3869 (top row), H$\beta$ (middle row), and [S~III] $\lambda$9071 (bottom row) 
kinematic measurements in 2MASX~J0423 from SNIFS IFU observations in the central field of view. First, 
second, and third columns display emission-line profile centroid velocity, FWHM, and integrated 
flux maps, respectively. Black contours represent \emph{HST}/WFPC2 F675W 
imaging. The optical continuum flux peak is depicted by a cross. One 0.43$" \times$ 0.43$"$ spaxel 
samples approximately 380\,pc $\times$ 380\,pc.}
\label{fig:altmaps}

\end{figure*}

\end{document}